\documentclass[12pt,a4paper,hypertex]{article}
\usepackage{jheppub}
\allowdisplaybreaks[1]
\usepackage{bm}
\usepackage{subfigure}

\title{Turbulent strings in AdS/CFT}

\author[1]{Takaaki Ishii}
\author[2]{and Keiju Murata}
\affiliation[1]{Crete Center for Theoretical Physics, Department of Physics, University of Crete,\\
PO Box 2208, 71003 Heraklion, Greece}
\affiliation[2]{Keio University, 4-1-1 Hiyoshi, Yokohama 223-8521, Japan}
\emailAdd{ishii@physics.uoc.gr}
\emailAdd{keiju@phys-h.keio.ac.jp}

\abstract{
We study nonlinear dynamics of the flux tube between an external quark-antiquark pair in $\mathcal{N}=4$ super Yang-Mills theory 
using the AdS/CFT duality. In the gravity side, the flux tube is realized by a fundamental string whose endpoints are attached to the AdS boundary. We perturb the endpoints in various ways and numerically compute the time evolution of the nonlinearly oscillating string. As a result, cusps can form on the string, accompanied by weak turbulence and power law behavior in the energy spectrum. When cusps traveling on the string reach the boundary, we observe the divergence of the force between the quark and antiquark. Minimal amplitude
of the perturbation below which cusps do not form is also investigated. No cusp formation is found when the string moves in all four AdS space directions, and in this case an inverse energy cascade follows a direct cascade.
}

\preprint{CCTP-2015-10 \\ \hspace*{\fill} CCQCN-2015-80}


\begin{document}

\maketitle

\section{Introduction}

The gauge/gravity duality~\cite{Maldacena:1997re,Gubser:1998bc,Witten:1998qj} is 
successfully applied to investigating strongly coupled gauge theories. 
Through this duality, it is hoped that one can access their nontrivial aspects 
that are hard to be handled because of the strong coupling. 
In particular, among advantages in using gravity duals, it is worth notifying that 
dynamics in time-dependent systems can be powerfully computed from time evolution in classical gravity. 
Applications range over far-from-equilibrium dynamics governed by nonlinear equations, 
and using numerical techniques for solving them attracts much attention. 
For instance, physics of strongly coupled plasma of quarks and gluons 
at RHIC and LHC brings motivations to numerically study dual gravitational dynamics; 
a series of seminal works is in Refs.~\cite{Chesler:2008hg,Chesler:2009cy,Chesler:2010bi,Chesler:2013lia,Chesler:2015wra}.

Far-from-equilibrium processes in the D3/D7 brane system dual to
$\mathcal{N}=2$ supersymmetric QCD have been recently
studied in
Refs.~\cite{Ishii:2014paa,Hashimoto:2014yza,Hashimoto:2014xta,Hashimoto:2014dda,Ali-Akbari:2015bha},
where partial differential equations for time evolution were solved numerically.
As a phenomena characteristic in non-linear dynamics, 
it has been found that long time evolution of the D7-brane generates a singularity on the brane,
and this  is understood from the viewpoint of weak turbulence on the
D7-brane: The energy in the spectrum is transferred from large to small
scales~\cite{Hashimoto:2014yza,Hashimoto:2014xta,Hashimoto:2014dda}.
Small scale fluctuations there correspond to excited
heavy mesons in the dual gauge theory. The turbulent behavior of the
D7-brane can be interpreted as production of many heavy mesons in
the dual gauge theory, and the singularity formation is interpreted as deconfinement of such mesons.

\begin{figure}
\begin{center}
\includegraphics[scale=0.6]{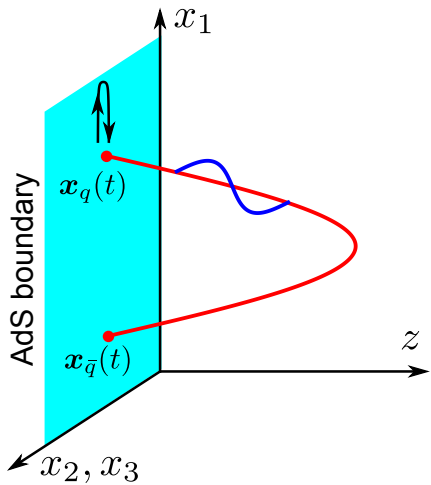}
\end{center}
\caption{
A schematic picture of our setup. 
Perturbing the endpoints induces fluctuations on the string.
}
\label{ponchi}
\end{figure}

To gain a deep insight into this kind of nonlinear dynamics in the gauge/gravity duality, in this paper, we will consider a string in AdS dual to the flux tube between a quark-antiquark pair in $\mathcal{N}=4$ super Yang-Mills theory, and fully solve its nonlinear time evolution with a help of numerical techniques. This setup corresponds to focusing on a Yang-Mills flux tube compared with the collective mesons described in the D3/D7 system. In addition, working in this setup is simpler than using the D3/D7 system and will provide a clear understanding of the turbulent phenomena and instabilities in probe branes in the gauge/gravity duality. To initiate the time evolution, we will perturb the endpoints of the string for an instant. Our setup is schematically illustrated in Fig.~\ref{ponchi}. The endpoints are forced to move momentarily and then brought back to the original locations. We will loosely use the terminology ``quench'' for expressing this process. This action introduces waves propagating on the string, and we will be interested in their long time behavior where nonlinearity in the time evolution of the string plays an important role. When we perform numerical computations, we will use the method developed in~\cite{Ishii:2014paa}, which turned out to be efficient for solving the time evolution in probe brane systems.
We are also motivated by the weakly turbulent instability found by
Bizo\'{n} and Rostworowski~\cite{Bizon:2011gg}.
Our setup may give a simple playground to study that kind of phenomenon.

The string hanging from the AdS boundary is one of the most typical probes in the gravity dual. This gives the gravity dual description of a Wilson loop corresponding to the potential between a quark and an antiquark~\cite{Rey:1998ik,Maldacena:1998im}. Although the potential in a conformal theory is different from that in real QCD, linear confinement can be realized in nonconformal generalization~\cite{Witten:1998zw}. In finite temperatures, a string extending to the AdS black hole corresponds to a deconfined quark~\cite{Rey:1998bq,Brandhuber:1998bs}, and this has been utilized for studying the behavior of moving quarks in Yang-Mills plasma~\cite{Herzog:2006gh,Gubser:2006bz,CasalderreySolana:2006rq,Gubser:2006nz,CasalderreySolana:2007qw,Liu:2006ug,Liu:2006he}. Moving quark-antiquark pairs were also considered~\cite{Liu:2006nn,Chernicoff:2006hi}. In far-from-equilibrium systems, holographic Wilson loops have also been used as probes for thermalization~\cite{Balasubramanian:2010ce,Balasubramanian:2011ur}. Nevertheless, veiled by these applications to QGP, nonlinear (and non-dissipative) dynamics of the probe hanging string in AdS has not been shed light on so much.
Some analytic solutions of non-linear waves on an extremal surface in AdS have been studied in Ref.~\cite{Mikhailov:2003er}. Notice that that configuration corresponds to a straight string in the Poincare coordinates.

When it comes to nonlinear dynamics of a string, formation of cusps
would be primarily thought of. In fact, it is well known in flat space
in the context of closed cosmic strings that cusp formation is
ubiquitous~\cite{Turok:1984cn}. 
Cusp formation of fundamental strings ending on D-branes has been also
found in Ref.~\cite{Davis:2008kg}.
We will turn our attention to whether there is such formation of cusps also
in AdS.\footnote{A development of a cusp in a decelerating trailing string was discussed in Ref.~\cite{Garcia:2012gw}.}

The organization of the rest of this paper is as follows. We start from reviewing the static solution and linear perturbations in Section~\ref{sec:static} and~\ref{sec:lin}, where we introduce a parametrization convenient for our use. In Section~\ref{sec:nlin}, we explain the setup for our time dependent computations. We introduce four patterns of quenches that we consider, derive the evolution equations and the boundary conditions, prepare initial data, and explain measures for evaluating the time evolution. Sections~\ref{sec:longi}, \ref{sec:transv}, and \ref{sec:trancirc} are reserved for numerical results: In Section~\ref{sec:longi}, we discuss two of the four quenches where the oscillations of the boundary flux tube are restricted to compression waves and therefore we call them longitudinal. We evaluate cusp formation, turbulent behavior in the energy spectrum, and the forces acting on the quark endpoints. In Section~\ref{sec:transv}, we show results of a quench where the flux tube oscillates in one of its transverse directions. Finally, in Section~\ref{sec:trancirc}, we examine the last quench where the motion of the flux tube is in all three spatial directions of the boundary 3+1 dimensions. Section~\ref{sec:sumdis} is devoted to summary and discussion. In appendices, we explain details for the numerical computations and derive some formulae used in the main text.

\section{A review of the static solution}
\label{sec:static}

We briefly review the holographic calculation of
the static quark-antiquark potential in the near horizon limit of extremal
D3-branes~\cite{Rey:1998ik,Maldacena:1998im}. The background metric is
$\mathrm{AdS}_5 \times \mathrm{S}^5$, 
\begin{align}
ds^2 &= \frac{\ell^2}{z^2} \left( -dt^2 + dz^2 + d\bm{x}^2 \right) + \ell^2 d\Omega_5^2\ ,
\label{stat_AdS5S5metric}
\end{align}
where $\ell^4 = 4\pi g_s N_c\alpha'^2$ and $\bm{x}\equiv(x_1,x_2,x_3)$. 
We consider a rectangular Wilson loop with the quark-antiquark
separation $L$ along the $x_1$-direction where the quark and antiquark
are located at $x_1=\pm L/2$. 
The dynamics of the string is described by Nambu-Goto action,
\begin{equation}
 S=-\frac{1}{2\pi \alpha'}\int d\tau d\sigma \sqrt{-\gamma}\ ,
\label{NG0}
\end{equation} 
where $\gamma=\textrm{det}(\gamma_{ab})$ and $\gamma_{ab}$ is the induced metric on the string.
It is convenient to take a static gauge where the worldsheet coordinates $(\tau,\sigma)$
 coincide with target space coordinates as $(\tau, \sigma) = (t,z)$.
The static solution is then described by a single function 
$x_1=X_1(z)$.\footnote{
We will use small letters for coordinates and 
capital letters for functions specifying the string position.
}
The Nambu-Goto action becomes
\begin{align}
S &=-\frac{\sqrt{\lambda}}{2\pi} \int dt dz \, \frac{1}{z^2} \sqrt{1+X_1'{}^2} \ ,
\label{stat_NG_action}
\end{align}
where $\lambda =  4 \pi g_s N_c$ is the 't Hooft coupling.

\begin{figure}
\begin{center}
\includegraphics[scale=0.45]{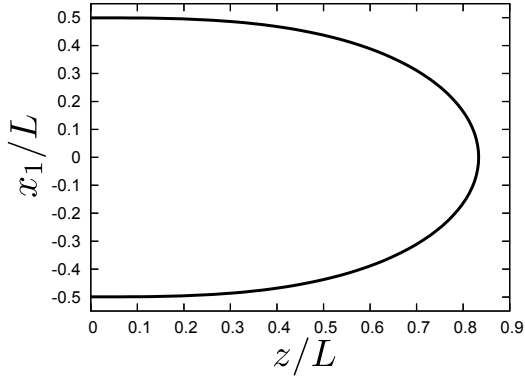}
\end{center}
\caption{
The string profile of the static solution.
}
 \label{stat_prof}
\end{figure}

Solving the equation of motion of $X_1(z)$ gives the bulk string
configuration. 
Let $z=z_0$ specify the bulk
bottom of the string reached at $x=0$, where the regular boundary
condition $\partial_x z=0$ is imposed. The embedding solution is 
given by
\begin{equation}
X_1(z) = \pm z_0 \int_{z/z_0}^1 dw \frac{w^2}{\sqrt{1 - w^4}}
= \pm z_0 \big[ \Gamma_0 + F(z/z_0;i) - E(z/z_0;i) \big] \ ,
\label{staticsol}
\end{equation}
where $\Gamma_0 \equiv \sqrt{2}\pi^{3/2}/\Gamma(1/4)^2 \simeq 0.599$, and $F$ and $E$ are 
the incomplete elliptic integrals of the first and second kinds defined as
\begin{equation}
F(x;k)=\int^x_0 dt \frac{1}{\sqrt{(1-t^2)(1-k^2t^2)}} \ , \quad
E(x;k)=\int^x_0 dt \sqrt{\frac{1-k^2t^2}{1-t^2}} \ .
\label{ellip}
\end{equation}
Setting $z=0$ in \eqref{staticsol}, we obtain $L/2 = z_0 \Gamma_0$. 
In Fig.~\ref{stat_prof}, we show the profile of the static string in the $(z,x_1)$-plane.
We will use this solution as the initial configuration for our
time evolution.

It is also known that the dependence of the potential energy on $L$ is Coulomb. The energy is evaluated from the on-shell action, which in general diverges at the boundary, but this divergence can be regulated by comparing with the diverging energy of two strings straightly extending to the Poincare horizon. The regularized energy is then given by
\begin{align}
E_\mathrm{reg} = - \frac{4 \pi^2 \sqrt{\lambda}}{\Gamma(1/4)^4 L} \ .
\end{align}
In this sense, the quark-antiquark potential in the AdS background does not correspond to the confining potential of real QCD. This, however, is considered as a handy playground for testing nonlinear evolution in the gauge/gravity duality.

\section{Linear perturbation theory}
\label{sec:lin}

\begin{figure}
  \centering
  \subfigure[Profiles of special functions]
  {\includegraphics[scale=0.45]{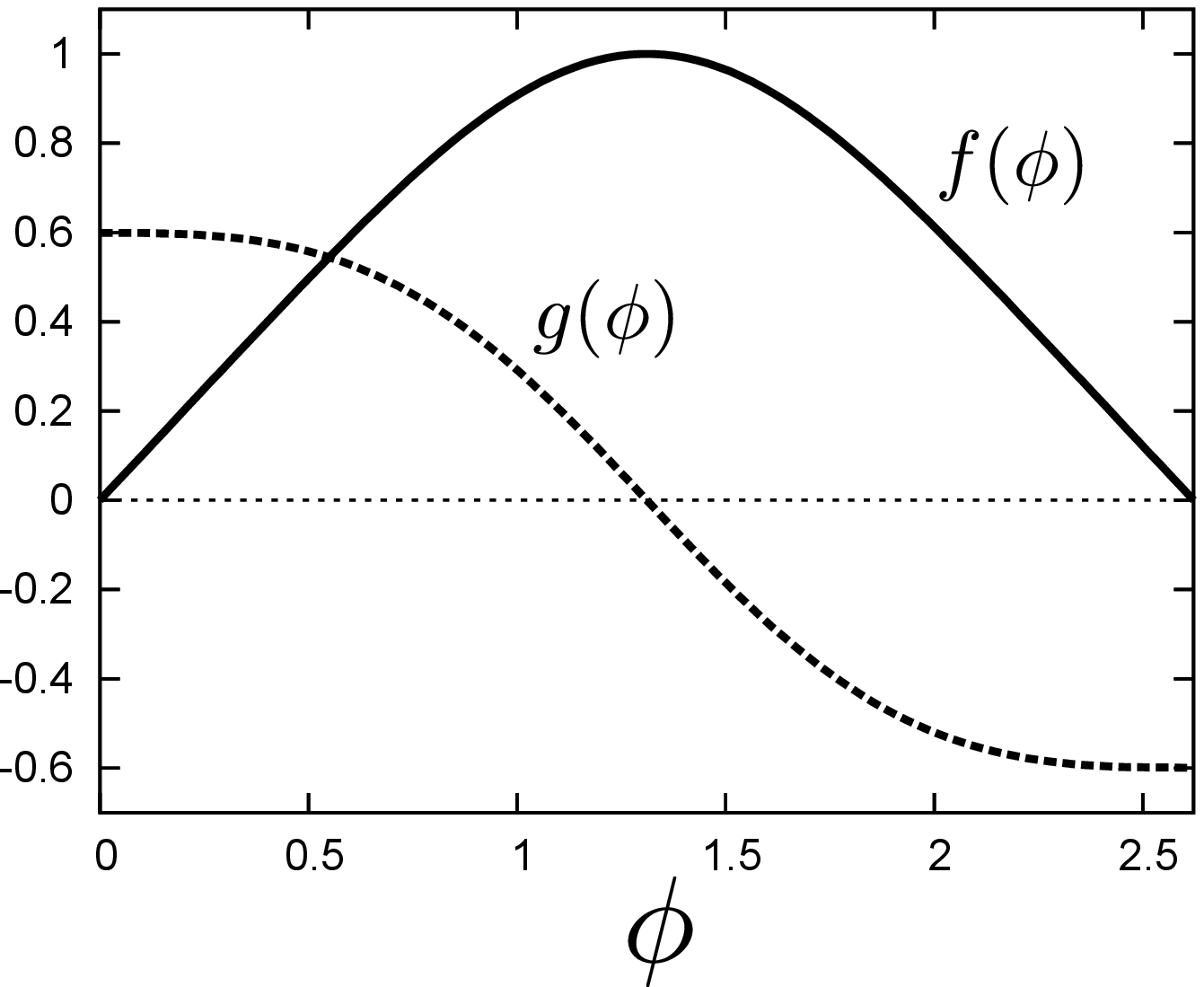}\label{fg}
  }
  \subfigure[Polar-like coordinates]
  {\includegraphics[scale=0.45]{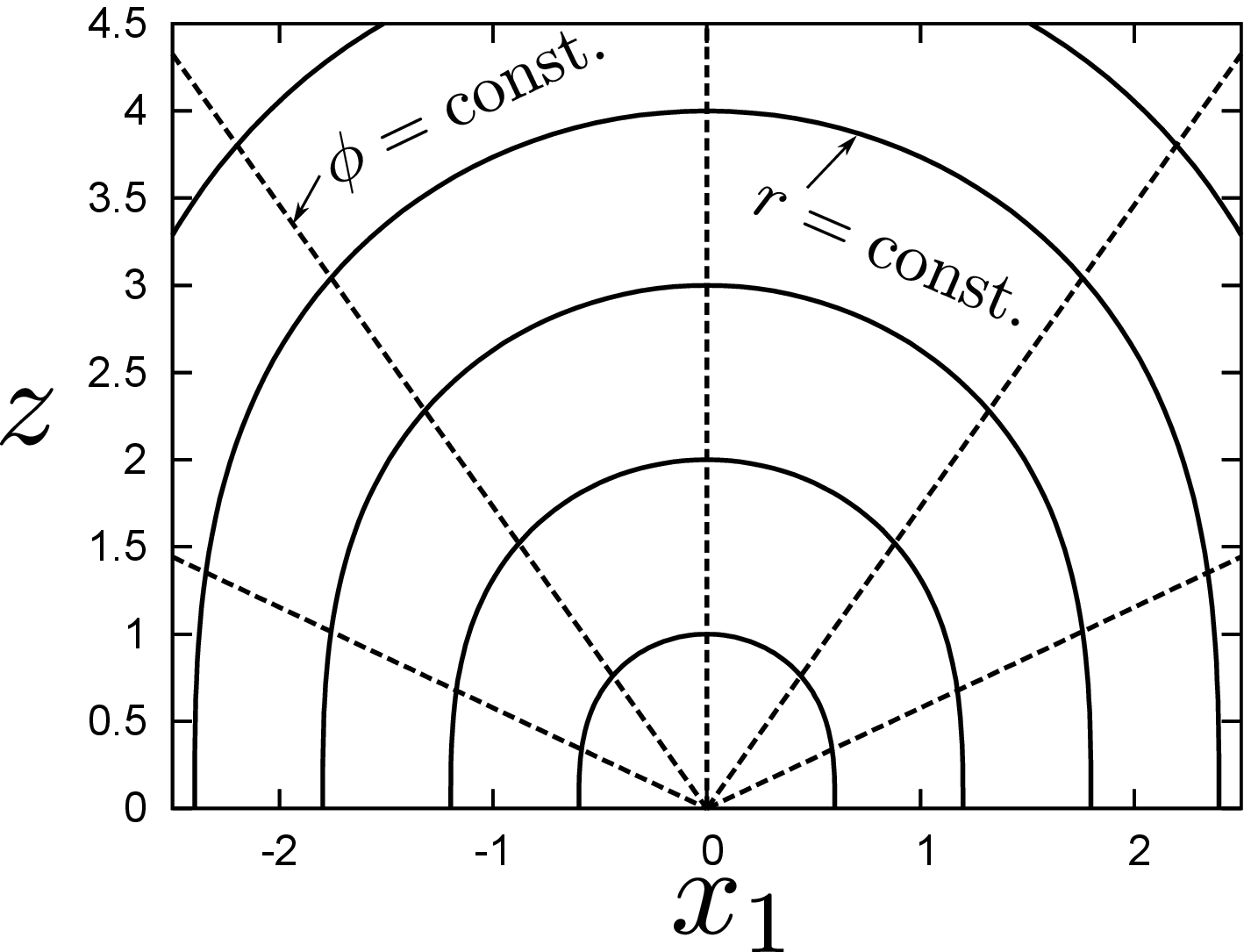} \label{rphi}
  }
  \caption{
(a) The profiles of the functions $f(\phi)$ and $g(\phi)$.
(b) $r=\textrm{const.}$ and $\phi=\textrm{const.}$ surfaces are shown in 
the $(z,x_1)$-plane.
}
\end{figure}

Linearized fluctuations and stability of holographic quark-antiquark potentials
 have been studied in Refs.~\cite{Callan:1999ki,Klebanov:2006jj,Avramis:2006nv,Arias:2009me}. 
Here, we solve the linearized fluctuations of the static solution in coordinates convenient for our use.
In~(\ref{staticsol}), we parametrized the location
of the string in terms of the $z$-coordinate.
In this coordinate, however, the static solution $X_1(z)$ becomes
multi-valued, and this may not be suitable for considering linear perturbations.
Instead, we introduce new polar-like coordinates $(r,\phi)$ in which 
the static embedding is expressed by single-valued
functions,
\begin{equation}
 z=r f(\phi)\ ,\quad x_1=r g(\phi)\ ,
\label{static_para}
\end{equation}
where we define the functions $f$ and $g$ as
\begin{align}
 f(\phi) &\equiv \textrm{sn}(\phi;i)\ ,
\label{fdef} \\
g(\phi) & \equiv-\int^\phi_{\beta_0/2}d\phi'\,\, f(\phi')^2
=
\begin{cases}
\phi-E(\textrm{sn}(\phi;i);i)+\Gamma_0 & (\phi\leq\beta_0/2)\\
\phi+E(\textrm{sn}(\phi;i);i)-\Gamma_0-\beta_0 & (\phi>\beta_0/2)
\end{cases}
\ ,
\label{gdef}
\end{align}
where 
$\textrm{sn}(x;k)$ is a Jacobi elliptic function defined
as the inverse function of $F(x;k)$ given in Eq.~(\ref{ellip}): $F(\textrm{sn}(x;k);k)=x$.
For $k=i=\sqrt{-1}$, the Jacobi elliptic function has roots at $x=\beta_0 n$
($n\in\mathbb{Z}$), where 
$\beta_0=\pi/(2\Gamma_0)\simeq2.622$.
We find that there is a nice relation between $f$ and $g$,
\begin{equation}
 f'(\phi)^2+g'(\phi)^2=1\ .
\label{fgrel}
\end{equation}
With these functions $f$ and $g$, the static
embedding~(\ref{staticsol}) is simply given by $r=z_0$.
The profiles of the functions $f(\phi)$ and $g(\phi)$ are shown in
Fig.~\ref{fg}.
We also depict the $r=\textrm{const.}$ and $\phi=\textrm{const.}$
surfaces in the $(z,x_1)$-plane in Fig.~\ref{rphi}.

\begin{figure}
  \centering
  \subfigure[Longitudinal modes]
  {\includegraphics[scale=0.45]{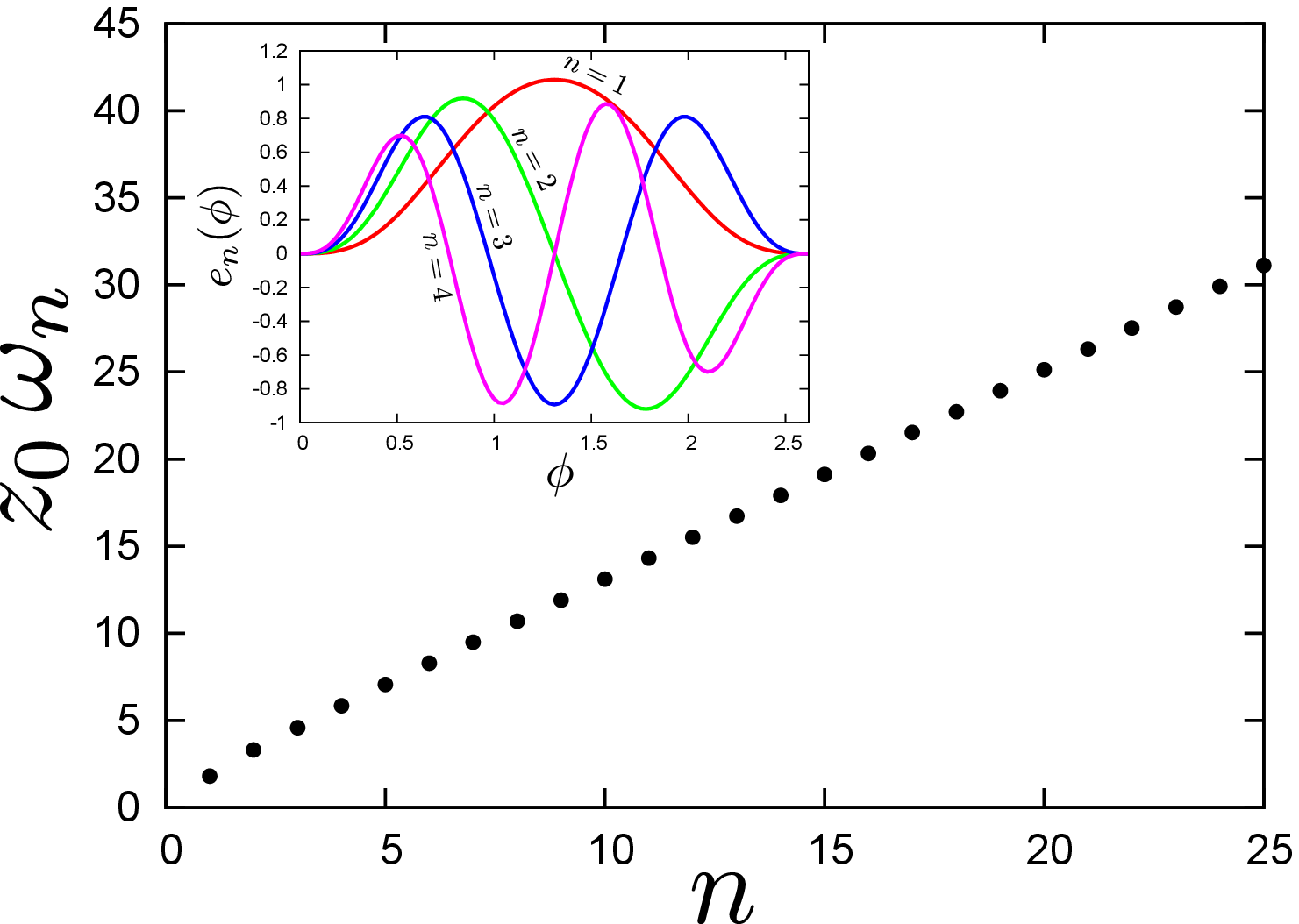}\label{eigenfig_longi}
   }
  \subfigure[Transverse modes]
  {\includegraphics[scale=0.45]{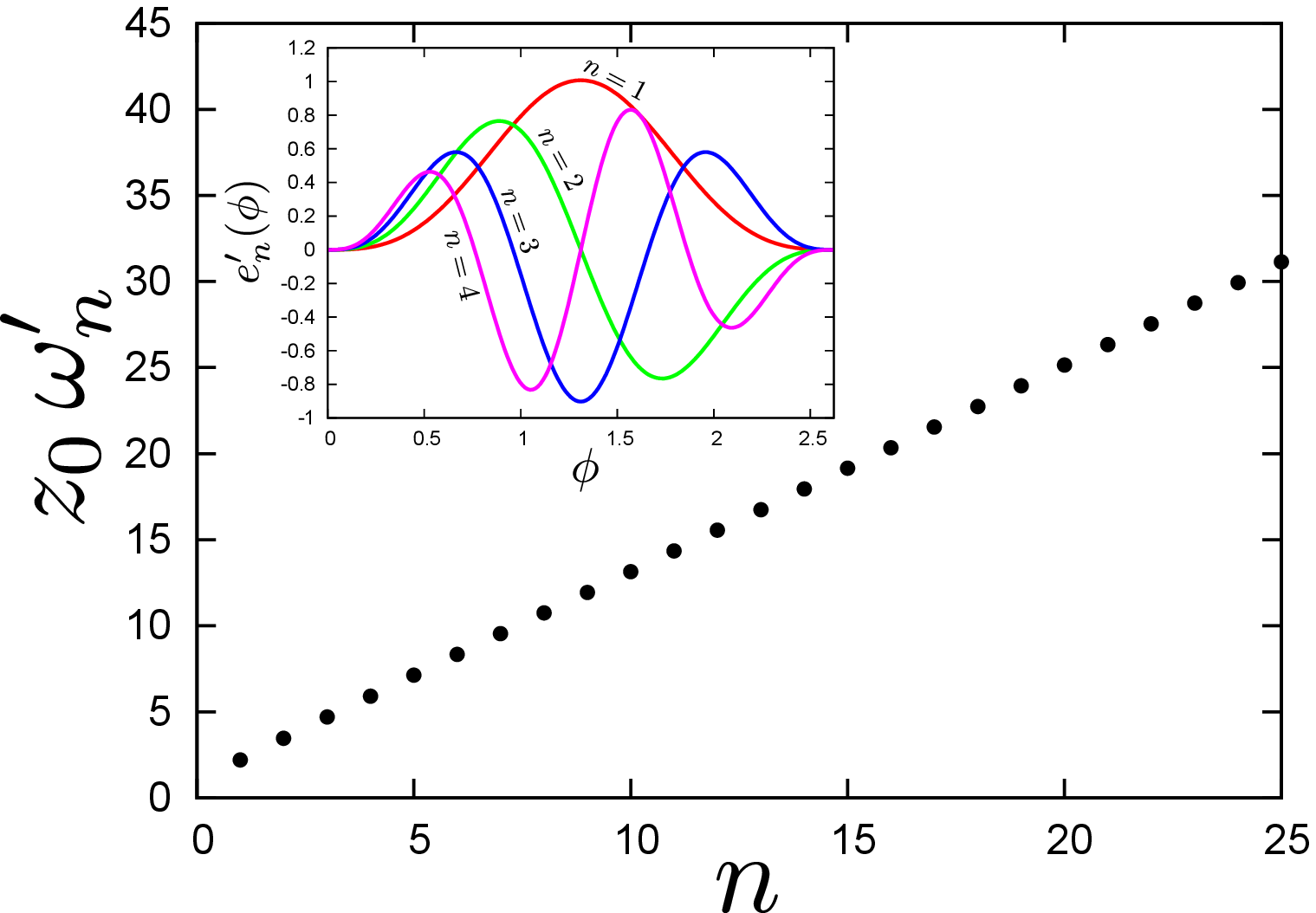}\label{eigenfig_trans}
  }
  \caption{
Normal mode frequencies of the longitudinal and transverse modes.
The eigenfunctions for $n=1,2,3,4$ are also shown in the inset in each figure.
}
\label{evef}
\end{figure}

Using $t$ and $\phi$ as the worldsheet coordinates, 
we can describe the dynamics of the string in terms of three functions,
\begin{equation}
r=R(t,\phi),
\quad 
x_2=X_2(t,\phi), 
\quad
x_3=X_3(t,\phi)\ .
\end{equation}
We consider perturbations around the static solution, $R=z_0, X_2=X_3=0$, as
\begin{equation}
R(t,\phi)=z_0\{1+\chi_1(t,\phi)\}\ ,\quad
X_2(t,\phi)=z_0\,\chi_2(t,\phi), \quad
X_3(t,\phi)=z_0\,\chi_3(t,\phi)\ , 
\label{chidef}
\end{equation}
where $\chi_i$ ($i=1,2,3$) are dimensionless perturbation variables.
We will refer to $\chi_1$ as the
longitudinal mode and $\chi_i$ ($i=2,3$) as the transverse modes.
Then, in the second order in $\chi$, the Nambu-Goto action becomes
\begin{equation}
 S=\frac{\sqrt{\lambda}}{4\pi z_0}\int dt d\phi\,\left[
  h(\phi)\,(z_0^2\,  \dot{\chi}_1^2-\chi'_1{}^2)
+\sum_{i=2,3}\frac{1}{f^2(\phi)}(z_0^2\,
  \dot{\chi}_i^2-\chi_i'{}^2)\right]\ ,
\end{equation}
where we define $\dot{}\equiv \partial_t$ and ${}'\equiv\partial_\phi$,
and introduce $h(\phi)\equiv [(g/f)'f]^2$.
To derive the above expression, we used the relation of $f$ and
$g$~(\ref{fgrel}) and 
omitted the total derivative terms.
The equations of motion for $(\chi_1,\chi_2,\chi_3)$ are
\begin{equation}
\begin{split}
&(\partial_t^2 + \mathcal{H})\chi_1=0\ ,\qquad
\mathcal{H}\equiv -\frac{1}{z_0^2h}\partial_\phi h \partial_\phi\ ,\\
&(\partial_t^2 + \mathcal{H}')\chi_i=0\ ,\qquad
\mathcal{H}'\equiv -\frac{f^2}{z_0^2}\partial_\phi \frac{1}{f^2}
 \partial_\phi\ 
\quad (i=2, \, 3)\ .
\end{split}
\label{chieq}
\end{equation}
Operators $\mathcal{H}$ and $\mathcal{H}'$ are Hermitian under the inner
products
\begin{equation}
 (\alpha,\beta)\equiv \int^{\beta_0}_0 d\phi \,h(\phi)\,
\alpha(\phi)\beta(\phi)\ ,\qquad
 (\alpha,\beta)'\equiv \int^{\beta_0}_0 d\phi \,\frac{1}{f(\phi)^2}\,
\alpha(\phi)\beta(\phi)\ ,
\end{equation}
respectively.
We denote the eigenvalues and eigenfunctions of $\mathcal{H}$ and
$\mathcal{H}'$ as $\{\omega_n^2,e_n(\phi)\}$ and
$\{\omega'_n{}^2,e_n'(\phi)\}$, respectively. 
These are labeled by integers $n\geq 1$ in ascending
order of $\omega_n$ and $\omega_n'$. 
The eigenfunctions are orthonormalized as $(e_n,e_m)=\delta_{nm}$ and $(e'_n,e'_m)'=\delta_{nm}$.

It is easy to check the linear stability of the static embedding
against the longitudinal perturbation as 
\begin{equation}
 \omega_n^2 = (e_n,\mathcal{H}e_n) 
=\frac{1}{z_0^2}\int^{\beta_0}_0 d\phi \,h\, (\partial_\phi e_n)^2\geq 0\ .
\end{equation}
In the same way, the stability 
against the transverse perturbations can be also checked, $\omega_n'{}^2\geq 0$.

We numerically determine normal mode frequencies and eigenfunctions.
These are plotted in Fig.~\ref{evef}. 
The spectra are asymptotically linear: 
$\omega_n, \omega_n' \propto n$ for $n\to\infty$.
In fact, from the WKB analysis, we can obtain 
$z_0 \omega_n\simeq 2\Gamma_0 (n+1)$ 
for $n\to \infty$~\cite{Callan:1999ki,Klebanov:2006jj}. 
Our numerical results are consistent with the WKB approximation.

\section{Non-linear dynamics of fundamental strings}
\label{sec:nlin}

The main focus of this paper is to study nonlinear dynamics of the string, where we make use of numerical techniques for solving the time evolution. In this section, the setup for this is prepared.

\subsection{Setup}

We consider AdS$_5 \times \mathrm{S}^5$~(\ref{stat_AdS5S5metric}) as the background
spacetime, and
take the static solution~(\ref{staticsol}) as the initial configuration.
We then consider ``quench'' on the endpoints of the string: 
We move their positions momentarily and put them back to the original positions.
A schematic picture of this setup is depicted in Fig.~\ref{ponchi}.

\begin{figure}
\begin{center}
\includegraphics[scale=0.8]{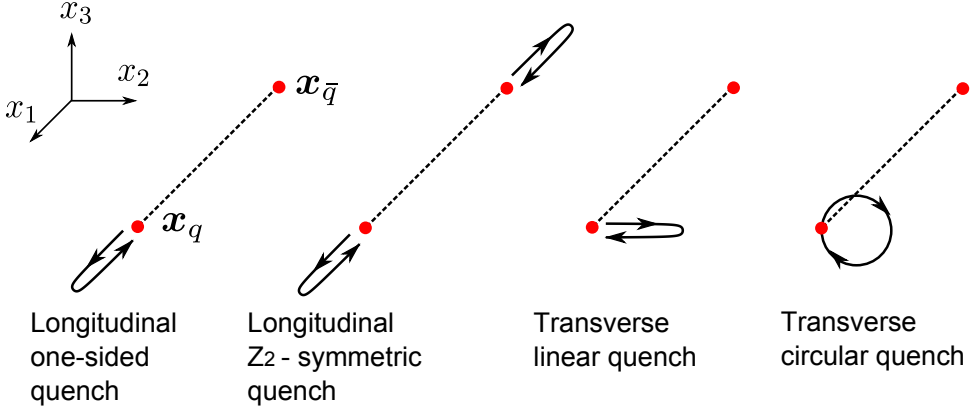}
\end{center}
\caption{
Schematic pictures of the quenches we will consider.
}
 \label{quench_ponchi}
\end{figure}

Let us denote the two endpoints of the string as $\bm{x}_q(t)$ and 
$\bm{x}_{\bar{q}}(t)$,
corresponding to the locations of the quark and antiquark, respectively.
In this paper, we consider the following four kinds of quenches on $\bm{x}_q(t)$ and 
$\bm{x}_{\bar{q}}(t)$:
\begin{enumerate}
\renewcommand{\labelenumi}{\textbf{(\roman{enumi})}}
\item \textbf{Longitudinal one-sided quench}:
\begin{equation}
 \bm{x}_q(t)=\left(\frac{L}{2}+\epsilon L\alpha(t),0,0\right)\ ,\quad
 \bm{x}_{\bar{q}}(t)=\left(-\frac{L}{2},0,0\right)\ , 
\label{quench1}
\end{equation}
where $\alpha(t)$ is a compactly supported $C^\infty$ function defined by
\begin{equation}
 \alpha(t)=
\begin{cases}
\exp\left[2\left(\frac{\Delta t}{t-\Delta t}-\frac{\Delta t}{t}+4
\right)\right]\qquad &(0<t<\Delta t)\\
0\qquad &(\textrm{else})
\end{cases} \ .
\end{equation}
The profile of this function is shown in Fig.~\ref{comp}.
The flux tube vibrates in its longitudinal direction, and 
motions are not induced in the transverse directions
by this quench.
Thus, in this case, 
the motion of the string is restricted in (2+1)-dimensions spanned by
$(t,z,x_1)$.
\begin{figure}
\begin{center}
\includegraphics[scale=0.45]{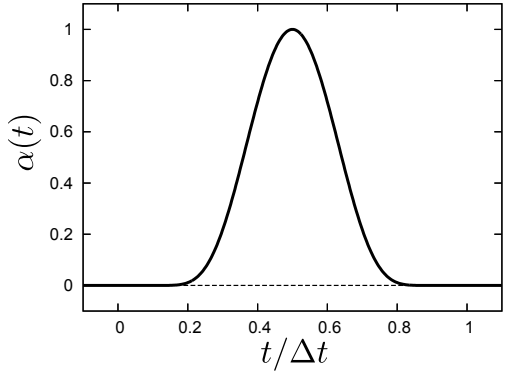}
\end{center}
\caption{
A compactly supported $C^\infty$ function for quench.
}
 \label{comp}
\end{figure}
\item \textbf{Longitudinal $\mathbb{Z}_2$-symmetric quench}:
\begin{equation}
 \bm{x}_q(t)=\left(\frac{L}{2}+\epsilon L\alpha(t),0,0\right)
\ ,\quad
 \bm{x}_{\bar{q}}(t)=\left(-\frac{L}{2}-\epsilon L\alpha(t),0,0\right)
\ .
\label{quench2}
\end{equation}
We simultaneously quench both endpoints in the opposite directions along the flux tube.
The string motion induced by this quench is invariant under $x_1\to -x_1$ and
restricted in the same (2+1)-dimensions as (i).
\item \textbf{Transverse linear quench}:
\begin{equation}
 \bm{x}_q(t)=\left(\frac{L}{2},\epsilon L \alpha(t),0\right)\ ,\quad
 \bm{x}_{\bar{q}}(t)=\left(-\frac{L}{2},0,0\right)\ .
\label{quench3}
\end{equation}
We shake one of the endpoints along $x_2$-direction.
String fluctuations in this direction are induced by the quench but not in $x_3$-direction.
Thus, the string oscillates in (3+1)-dimensions spanned by $(t,z,x_1,x_2)$.
\item \textbf{Transverse circular quench}:
\begin{equation}
 \bm{x}_q(t)=\left(\frac{L}{2},\epsilon L \alpha(t),\pm\epsilon L \sqrt{\alpha(t)(1-\alpha(t))}\right)\ ,\quad
 \bm{x}_{\bar{q}}(t)=\left(-\frac{L}{2},0,0\right)\ ,
\label{quench4}
\end{equation}
where we choose the upper and lower signs for $t\leq \Delta t/2$ and
$t> \Delta t/2$, respectively.
The orbit of $\bm{x}_q$ is a circle given by 
$(x_2-\epsilon L/2)^2+x_3^2=(\epsilon L/2)^2$.
String fluctuations along both $x_2$- and $x_3$-directions are induced
by this quench, and hence
the string moves in all (4+1)-dimensions spanned by $(t,z,x_1,x_2,x_3)$.
\end{enumerate}
In Fig.~\ref{quench_ponchi}, we show schematic pictures of the quenches (i)-(iv).
These patterns are chosen to represent typical string motions, particularly with different dimensionality.
String dynamics is specified by two parameters $\epsilon$ and  $\Delta t$ once we choose a quench type.

In this paper, we will take modest values for the amplitude of the quench 
($\epsilon\sim 0.01$)
since we are interested in nonlinear evolution starting from small deviation from the linear theory 
and focus on weak turbulence and cusp formation driven by the nonlinearity.
For a large value of $\epsilon$, we expect some effect similar to the overeager effect
found in Ref.~\cite{Ishii:2014paa}: 
The string will be able to plunge into the Poincare
horizon $z=\infty$ because of the strong perturbation.
Such strong quenches will be studied in detail elsewhere~\cite{work_in_progress}.

\subsection{Basic equations}

To calculate the time evolution on the string worldsheet, 
we find it convenient to use double null coordinates.
With worldsheet coordinates $(u,v)$, 
the string position is parametrized as
\begin{equation}
 t=T(u,v)\ ,\quad z=Z(u,v)\ ,\quad 
\bm{x}=\bm{X}(u,v)\ .
\end{equation}
Using these expressions with Eq.~\eqref{stat_AdS5S5metric}, we obtain the induced
metric as
\begin{equation}
\begin{split}
&\gamma_{uu}=\frac{\ell^2}{Z^2}(-T_{,u}^2+Z_{,u}^2+\bm{X}_{,u}^2)\ ,\quad
\gamma_{vv}=\frac{\ell^2}{Z^2}(-T_{,v}^2+Z_{,v}^2+\bm{X}_{,v}^2)\ ,\\
&\gamma_{uv}=\frac{\ell^2}{Z^2}(-T_{,u}T_{,v}+Z_{,u}Z_{,v}+\bm{X}_{,u}\cdot\bm{X}_{,v})\ .
\end{split}
\label{gammas}
\end{equation}
The reparametrization freedom of the worldsheet
coordinates allows us to impose the double null condition on the induced metric as
\begin{equation}
C_1\equiv\gamma_{uu}=0\ ,\qquad
C_2\equiv\gamma_{vv}=0\ .
\label{CON}
\end{equation}
Notice that these conditions do not fix the coordinates completely:
There are residual coordinate freedoms, 
\begin{equation}
 u=u(\bar{u})\ ,\qquad v=v(\bar{v})\ .
\label{resg}
\end{equation}
These will be fixed by boundary conditions and initial data.

In the double null coordinates, the Nambu-Goto action~(\ref{NG0}) becomes
\begin{equation}
\begin{split}
 S&=-\frac{1}{2\pi\alpha'}\int dudv
 \sqrt{\gamma_{uv}^2-\gamma_{uu}\gamma_{vv}}=\frac{1}{2\pi\alpha'}\int dudv
 \gamma_{uv}\\
&=\frac{\ell^2}{2\pi\alpha'}\int dudv\,
\frac{1}{Z^2}(-T_{,u}T_{,v}+Z_{,u}Z_{,v}+\bm{X}_{,u}\cdot\bm{X}_{,v})\ ,
\end{split}
\label{action1}
\end{equation}
where at the second equality we eliminate the square root in the action
using the double null conditions~(\ref{CON}).
(Note that $\gamma_{uv}$ is negative.)
From the action, we obtain the evolution equations of the string as
\begin{equation}
\begin{split}
&T_{,uv}=\frac{1}{Z}(T_{,u}Z_{,v}+T_{,v}Z_{,u})\ ,\\
&Z_{,uv}=\frac{1}{Z}(T_{,u}T_{,v}+Z_{,u}Z_{,v}-\bm{X}_{,u}\cdot
 \bm{X}_{,v})\ ,\\
&\bm{X}_{,uv}=\frac{1}{Z}(\bm{X}_{,u}Z_{,v}+\bm{X}_{,v}Z_{,u})\ .
\end{split}
\label{evol}
\end{equation}
Using these, we find that the constraints~(\ref{CON})
are preserved under the time evolution:
\begin{equation}
\partial_v C_1=\partial_u C_2=0\ .
\end{equation}
Hence if we impose $C_1=C_2=0$ on the initial surface and the boundaries, 
the constraints~(\ref{CON}) are automatically satisfied in the whole computational domain.

Our numerical method for solving the evolution equations is briefly
summarized in Appendix~\ref{sec:numerical} and 
its numerical error is estimated in Appendix~\ref{sec:err}.
For more details of the numerical method, see also Appendix A in~\cite{Ishii:2014paa}.

\subsection{Boundary conditions}

On the worldsheet, there are two time-like boundaries that correspond
to the two endpoints of the string attaching on the AdS boundary.
We need boundary conditions there.
Using the residual coordinate freedoms~(\ref{resg}), 
we can fix the locations of the boundaries to $u=v$ and $u=v+\beta_0$. 
The boundary conditions for the spatial parts of the target space coordinates are given by 
\begin{equation}
Z|_{u=v}=0\ ,\quad \bm{X}|_{u=v}=\bm{x}_q\ ,\quad
Z|_{u=v+\beta_0}=0\ ,\quad \bm{X}|_{u=v+\beta_0}=\bm{x}_{\bar{q}}\ .
\end{equation}
We also need boundary conditions for $T$.
To derive them, we solve Eq.~(\ref{evol}) and
(\ref{CON}) near the boundaries. 
Defining $\tau=u+v$ and $\sigma=u-v$ ($\tau\in (-\infty,\infty)$,
$\sigma\in [0,\beta_0]$), 
we obtain the asymptotic solutions around $\sigma=0$ as
\begin{equation}
\begin{split}
&T=t_0(\tau)+\left[\frac{1}{2}\ddot{t}_0
-\gamma^2\bm{v}\cdot\bm{a}\,\dot{t}_0^2\right]\sigma^2+\bm{v}\cdot\bm{x}_3\sigma^3+\cdots\
 ,\\
&Z=\frac{\dot{t}_0}{\gamma}\sigma
+\left[
\frac{\dddot{t}_0}{6\gamma}-\frac{1}{2}\gamma \bm{v}\cdot \bm{a}\,
 \ddot{t}_0 \dot{t}_0
-\frac{\gamma}{6}(3\gamma^2 \bm{a}^2+\bm{v}\cdot \bm{j})\,\dot{t}_0^3
\right]\sigma^3+\cdots \ , \\
&\bm{X}=\bm{x}_q(\tau)+\left[
\frac{1}{2}\bm{v}\,\ddot{t}_0-\left(\gamma^2-\frac{1}{2}\right)\bm{a}\,\dot{t}_0^2\right]\sigma^2
+\bm{x}_3(\tau) \sigma^3 +\cdots\ .
\end{split}
\label{expand}
\end{equation}
where $\dot{}=d/d\tau$, 
$\bm{v}=\dot{\bm{x}}_q/\dot{t}_0$, $\bm{a}=\dot{\bm{v}}/\dot{t}_0$, $\bm{j}=\dot{\bm{a}}/\dot{t}_0$ and
$\gamma=1/\sqrt{1-\bm{v}^2}$.
The asymptotic solutions near $\sigma=\beta_0$ can be 
given by replacing 
$\sigma\to \beta_0-\sigma$ and $\bm{x}_q\to \bm{x}_{\bar{q}}$ in the above
expressions. 
From the second equation in Eq.~(\ref{expand}), 
we have $\partial_\sigma
Z|_{\sigma=0}=\sqrt{\dot{t}_0^2-\dot{\bm{x}}_q^2}$.
Using this, we obtain 
\begin{equation}
 T_{,\tau}|_{\sigma=0}=\sqrt{(Z_{,\sigma}|_{\sigma=0})^2 +
  \dot{\bm{x}}_q^2}\ ,\quad
 T_{,\tau}|_{\sigma=\beta_0}=\sqrt{(Z_{,\sigma}|_{\sigma=\beta_0})^2 +
  \dot{\bm{x}}_{\bar{q}}^2}\ .
\label{bdry_T_evo}
\end{equation}
These equations determine the time evolution of $T$ at the boundaries $\sigma=0, \, \beta_0$.
Their numerical implementation is explained in Appendix~\ref{sec:numerical}.

For consistency, the speed of the quark endpoints during the quench should be slower than light, $|\bm{v}|<1$.
Otherwise, the Lorentz factor $\gamma$ becomes imaginary.
Solving this condition, we obtain constraints for the quenches (i)-(iii),
\begin{equation}
\frac{\epsilon L}{\Delta t} 
< \frac{(19-5\sqrt{13}) e^{\sqrt{13}-3}}{12
\sqrt{6 \sqrt{13}-21}} 
\simeq 0.1865 \ ,
\end{equation}
and for the transverse circular quench (iv),
\begin{equation}
\frac{\epsilon L}{\Delta t} < \frac{1}{4 \sqrt{2}} \simeq 0.1768 \ .
\end{equation}
The parameter values examined in this paper satisfy these conditions.

\subsection{Initial data}
Before the quenches are applied, we assume that the string is static, 
namely, we use the static solution~(\ref{staticsol}) as the initial configuration.
For numerical computations, we need initial data written in the $(u,v)$-coordinates.
Substituting Eq.~(\ref{staticsol}) into Eqs.~(\ref{CON}) and (\ref{evol}),
we can express the static solution in terms of $(u,v)$ as
\begin{equation}
\begin{split}
&T(u,v)=z_0\,[\phi_1(u)+\phi_2(v)]\ ,\quad
Z(u,v)=z_0\,f(\phi_1(u)-\phi_2(v))\ ,\\
&X_1(u,v)=z_0\,g(\phi_1(u)-\phi_2(v))\ ,\quad
X_2(u,v)=X_3(u,v)=0\ ,
\end{split}
\label{staticuv}
\end{equation}
where $\phi_1$ and $\phi_2$ are arbitrary functions associated
with the residual coordinate freedom~(\ref{resg}), and
the functions $f$ and $g$ are defined in Eqs.~(\ref{fdef}) and (\ref{gdef}).

Locating the initial surface at $v=0$ on the worldsheet,
we parametrize the initial configuration as
\begin{equation}
\begin{split}
&T(u,0)=z_0\,u\ ,\quad
Z(u,0)=z_0\,f(u)\ ,\\
&X_1(u,0)=z_0\,g(u)\ ,\quad
X_2(u,0)=X_3(u,0)=0\ ,
\end{split}
\label{inidata}
\end{equation}
where we set the free functions $\phi_1(u)=u$ and $\phi_2(0)=0$ so that the boundaries are at $u=0, \, \beta_0$.
If the string endpoints are not perturbed,
our numerical calculations describe the static evolution of the exact
solution~(\ref{staticuv}).

\subsection{Quantities for evaluation}

In the non-linear dynamics of the string, we expect to observe formation of cusps.
For detailed analyses, we will in particular use the following quantities for evaluation.

\subsubsection{Cusp formation}
\label{sec:cusp_cond}

Let us consider the profile of the string on a surface with $t=$constant.
There is a cusp if the string does not change its target space position when the
parameter on the string is varied. Using the $u$-coordinate as a parameter
on the string in the $t=$constant surface, we obtain
the conditions for the cusp formation as
$\partial_u X_I|_{t=\textrm{const}}=0$ 
where 
$X_I\equiv (Z,X_i)$ and $i=1,\,2,\,3$.
These conditions are rewritten as
\begin{equation}
 J_I\equiv T_{,u}X_{I,v}-T_{,v}X_{I,u}=0\ .
\label{cuspformJ}
\end{equation}
In our numerical computations, we monitor the roots of $J_I$, 
which typically appear as curves on the $(u,v)$-plane. 
If these curves overlap at a point, we find cusp
formation, and if the overlap continues in the time evolution,
it is implied that the cusps continue to exist.

As an obvious corollary, if \eqref{cuspformJ} is satisfied, there is a necessary condition that
\begin{equation}
X_I{}_{,u} X_J{}_{,v} - X_I{}_{,v} X_J{}_{,u} =0
\label{cuspneceJ}
\end{equation}
is also satisfied.
This condition is conveniently utilized for a consistency check of the cusp formation detected by \eqref{cuspformJ}.

\subsubsection{Energy spectrum in the non-linear theory}
\label{Enspec}

We will also study the
energy spectrum of the non-linear fluctuations of the string
because, from the spectrum, it is expected to find weak turbulence on the string.
Once a dynamical solution $(T(u,v),Z(u,v),\bm{X}(u,v))$ is calculated,
we can convert it to the polar-like coordinates introduced in
Eq.~(\ref{static_para}): 
$r=R(u,v)$ and $\phi=\Phi(u,v)$.\footnote{
Solving $Z(u,v)/X_1(u,v)=f(\phi)/g(\phi)$ for $\phi$,
we have $\phi=\Phi(u,v)$. Then, we can obtain $R(u,v)$ from  
$R(u,v)=Z(u,v)/f(\Phi(u,v))$.
}
Eliminating the worldsheet coordinates $(u,v)$ from 
$(T(u,v)$, $R(u,v)$, $\Phi(u,v)$, $X_2(u,v)$, $X_3(u,v))$, 
we can express the dynamical solution using target space coordinates
$(t,\phi)$ as
\begin{equation}
 r=R(t,\phi),\qquad
 x_2=X_2(t,\phi),\qquad
 x_3=X_3(t,\phi) \ .
\end{equation}
As in Eq.~(\ref{chidef}), we define the non-linear version of the ``perturbation'' variables
$\hat{\chi}_1,\hat{\chi}_2,\hat{\chi}_3$ as
\begin{equation}
\hat{\chi}_1(t,\phi)=\frac{R(t,\phi)}{z_0}-1\ ,\quad
\hat{\chi}_i(t,\phi)=\frac{X_i(t,\phi)}{z_0}\quad (i=2,3)\ .
\label{pertvar}
\end{equation}
We then decompose $\hat{\chi}_1$, $\hat{\chi}_2$ and $\hat{\chi}_3$ with
the eigenfunctions of the linear theory $e_n(\phi)$ 
and $e_n'(\phi)$, which were introduced below Eq.~(\ref{chieq}), as
\begin{equation}
\hat{\chi}_1=\sum_{n=1}^\infty c_n(t) e_n(\phi)\ ,\quad
\hat{\chi}_i=\sum_{n=1}^\infty c_n^i(t) e_n'(\phi) \quad (i=2,3)\ .
\end{equation}
Using the mode coefficients $c_n$ and $c_n^i$, we define the
energy contribution from the $n$-th mode $\varepsilon_n(t)$ and the total energy in terms of the linear theory $\varepsilon(t)$
as
\begin{equation}
 \varepsilon_n(t)=\frac{\sqrt{\lambda} z_0}{4\pi}\left[\dot{c}_n{}^2 +
						  \omega_n^2 c_n{}^2 + 
\sum_{i=2,3}(
  \dot{c}_n^i{}^2 + \omega_n'{}^2 c_n^i{}^2)\right]\ ,\quad
\varepsilon(t)=\sum_{n=1}^\infty \varepsilon_n\ .
\label{eps_n}
\end{equation}
The quantify $\varepsilon_n$ is conserved in linear theory.
Therefore, if we find time dependence in $\varepsilon_n$, it is a fully
non-linear effect. 
Note that the total energy $\varepsilon$ defined in the linear theory is also time
dependent in the non-linear theory although 
the time dependence is suppressed by the amplitude of the quench: 
$\dot{\varepsilon}/\varepsilon=\mathcal{O}(\epsilon)$.
Since we consider only small $\epsilon$ in this paper 
($\epsilon\sim 0.01$), we do not put emphasis on the time dependence of the total
energy. In our actual numerical calculations, we take a cutoff at
$n=50$ for evaluating $\varepsilon$ in Eq.~(\ref{eps_n}).
Its cutoff dependence is also not essential for our following arguments on the energy spectrum on each $t$-slice.

\subsubsection{Forces acting on the heavy quarks}
\label{sec:force}

From the dynamical solution of the string, 
we can read off the time dependence of the forces acting on the quark and antiquark
located at $\bm{x}=\bm{x}_q$ and $\bm{x}_{\bar{q}}$.
The force acting on the quark can be evaluated from the on-shell Nambu-Goto action as
\begin{equation}
 \langle F_i(t) \rangle = \frac{\sqrt{\lambda}}{4\pi}
\gamma^{-1}(\delta_{ij}+\gamma^2 v_i v_j) \partial_z^3 X_j|_{z=0}\ ,
\end{equation}
where $\gamma=1/\sqrt{1-(d\bm{x}_q/dt)^2}$. The same formula can be applied to
the force acting on the antiquark $\langle \bar{\bm{F}}(t) \rangle$ by replacing $\bm{x}_q\to \bm{x}_{\bar{q}}$.
The derivation of this expression is summarized in
Appendix~\ref{app:force}.
For our numerical analysis, it is convenient to rewrite this expression 
in terms of the $(\tau,\sigma)$-coordinates.
Using the asymptotic expansions~(\ref{expand}), we obtain 
\begin{equation}
\langle \bm{F}(t) \rangle =
\frac{3\sqrt{\lambda}}{2\pi}\frac{\gamma^2 \bm{x}_3}{\dot{t}_0^3}\ .
\label{Force2}
\end{equation}
We need to extract the third order coefficient $\bm{x}_3$ from our
numerical data. 
For this purpose, it is convenient to define
\begin{equation}
 \bm{Y}\equiv \bm{X}-\bm{x}_q-(T-t_0)\bm{v}-\gamma^2\left[
\gamma^2(\bm{v}\cdot\bm{a})\bm{v}-\left(\gamma^2-\frac{1}{2}\right)\bm{a}
\right]Z^2\ .
\end{equation}
Using Eq.~(\ref{expand}), we find that this function behaves as 
$\bm{Y}\simeq [\bm{x}_3-(\bm{v}\cdot\bm{x}_3)\bm{v}]\sigma^3$ near the boundary.
To evaluate $\bm{x}_3$, we fit numerical data for $\bm{Y}$ by a function $\bm{c}\, \sigma^3$ in 
$\sigma\in [0,0.1]$.

\section{Results for the longitudinal quenches}
\label{sec:longi}

From this section to Section~\ref{sec:trancirc}, 
we discuss results of numerical computations for our quenches (i)-(iv).
In this section, we firstly treat the longitudinal one-sided quench (i)
and then discuss the longitudinal $\mathbb{Z}_2$-symmetric quench (ii).

\subsection{Cusp formation}
\label{sec:Longi_cusp}

\begin{figure}
  \centering
  \subfigure[Just after the quench]
  {\includegraphics[scale=0.45]{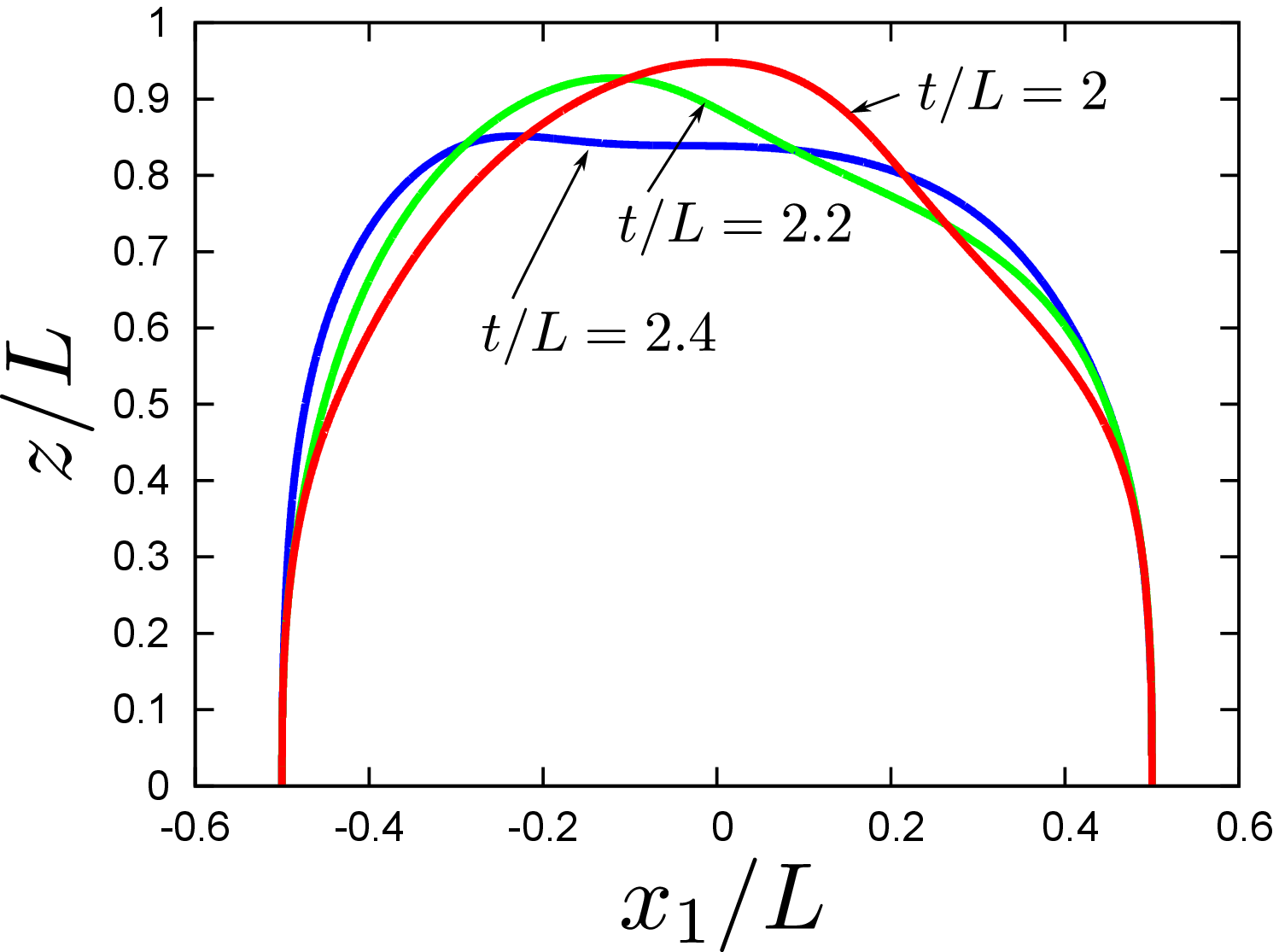}\label{ln_sn_early}
  }
  \subfigure[After the cusp formation]
  {\includegraphics[scale=0.45]{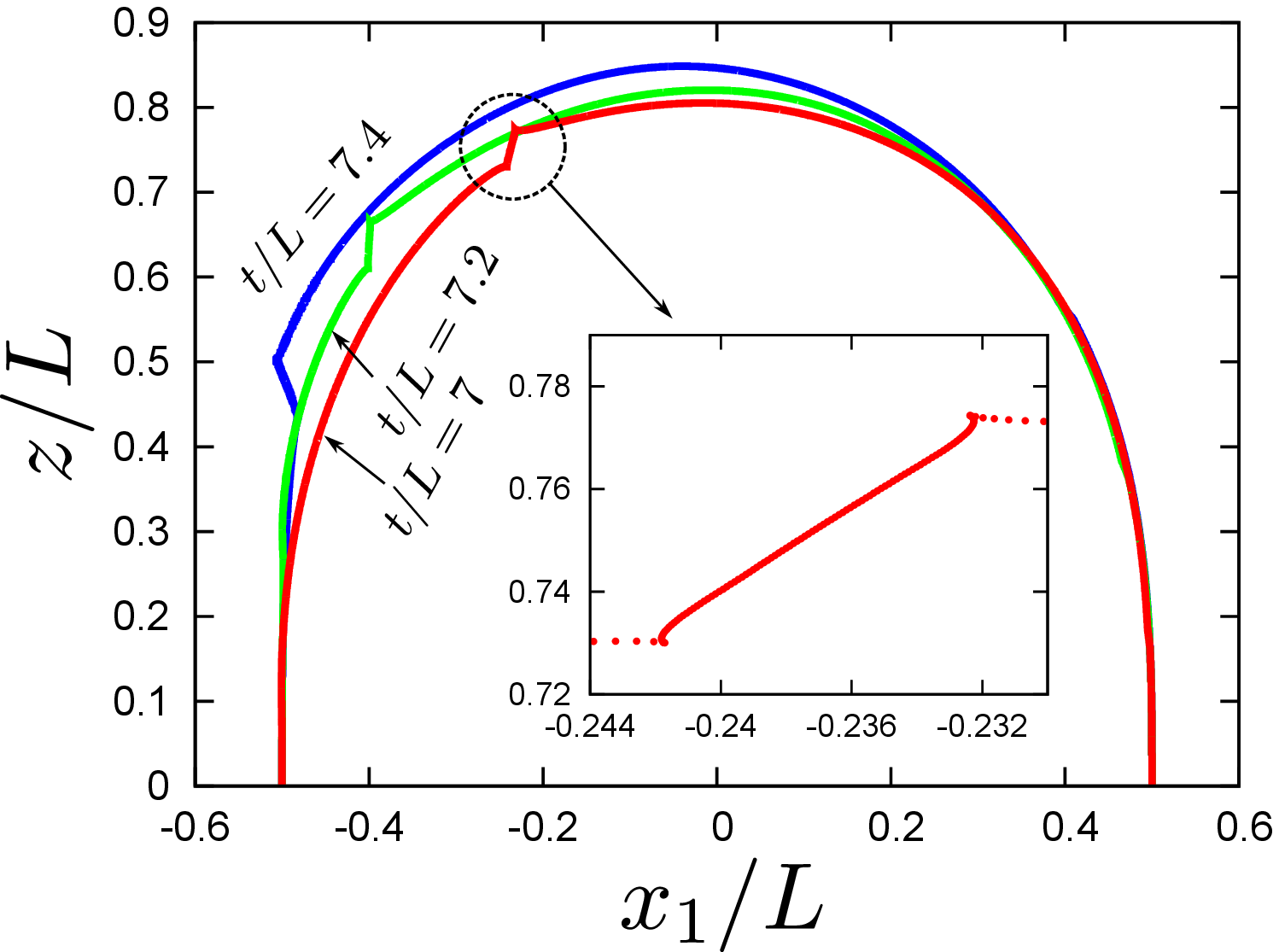} \label{ln_sn_cusp}
  }
  \caption{
Snapshots of the string time evolution for the longitudinal one-sided quench
with $\Delta t/L=2$, $\epsilon=0.03$.
\label{ln_snap}
}
\end{figure}

We start from the longitudinal
one-sided quench~(\ref{quench1}). In Fig.~\ref{ln_snap}, we show 
snapshots of the time evolution of the string for $\Delta t/L=2$ and $\epsilon=0.03$.
The left panel is just after the quench, $t/L=2, 2.2, 2.4$. We observe that 
perturbations are induced on the string by the quench.
Late time behavior is shown in the right panel for $t/L=7,7.2,7.4$,
where it is seen that cusps are formed on the string.
The time for the cusp formation is evaluated as $t/L \sim 5$ by using \eqref{cuspformJ},
while the snapshots are taken when it is easy to confirm the cusps visually.
Although the solution plotted in the target space $(z,x_1)$-plane looks singular on top of the cusps,
the fields $T(u,v)$, $Z(u,v)$ and $X_1(u,v)$ are indeed smooth on the $(u,v)$-plane.
Therefore, even after the formation of the cusps, the time evolution can continue
without a breakdown of numerical computations.
Physically, however, finite-$N_c$ effects can be important at cusp
singularities, and the time evolution after the cusp formation may be 
considered as unphysical without such corrections. 
We will discuss possible finite-$N_c$ effects at cusps in section~\ref{sec:sumdis}.
For the present, the dynamics of cusps are discussed without taking into account these corrections.
We find that the cusps appear as a pair. (The cusps at $t/L=7.0$ are magnified in the inset
of the right panel.) The propagating speeds of the two cusps are different, 
and they continue to propagate on the string.

\begin{figure}[t]
\centering
\subfigure[Cusp formation time]{
\includegraphics[scale=0.4]{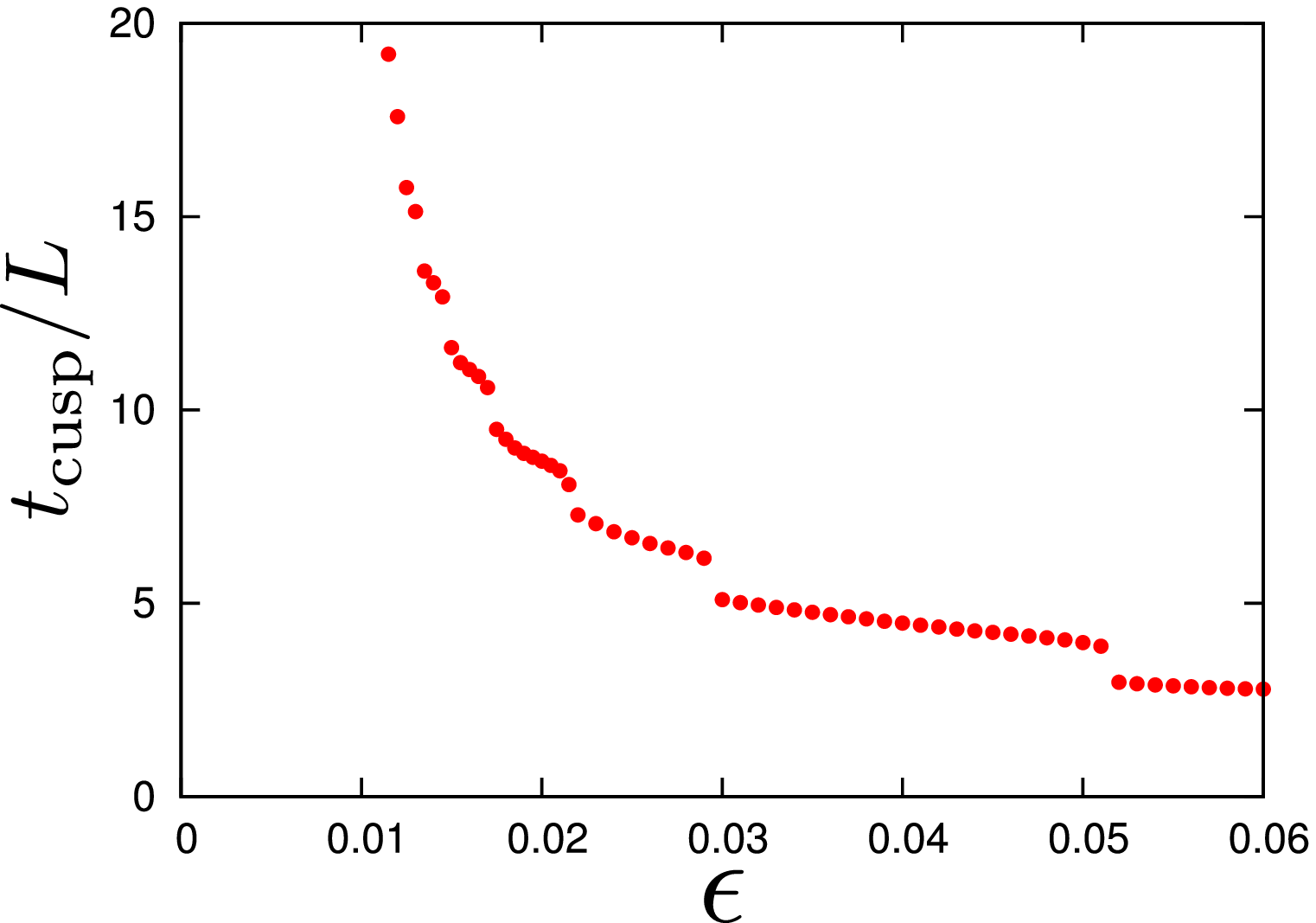}
}
\subfigure[Corresponding formation points]{
\includegraphics[scale=0.4]{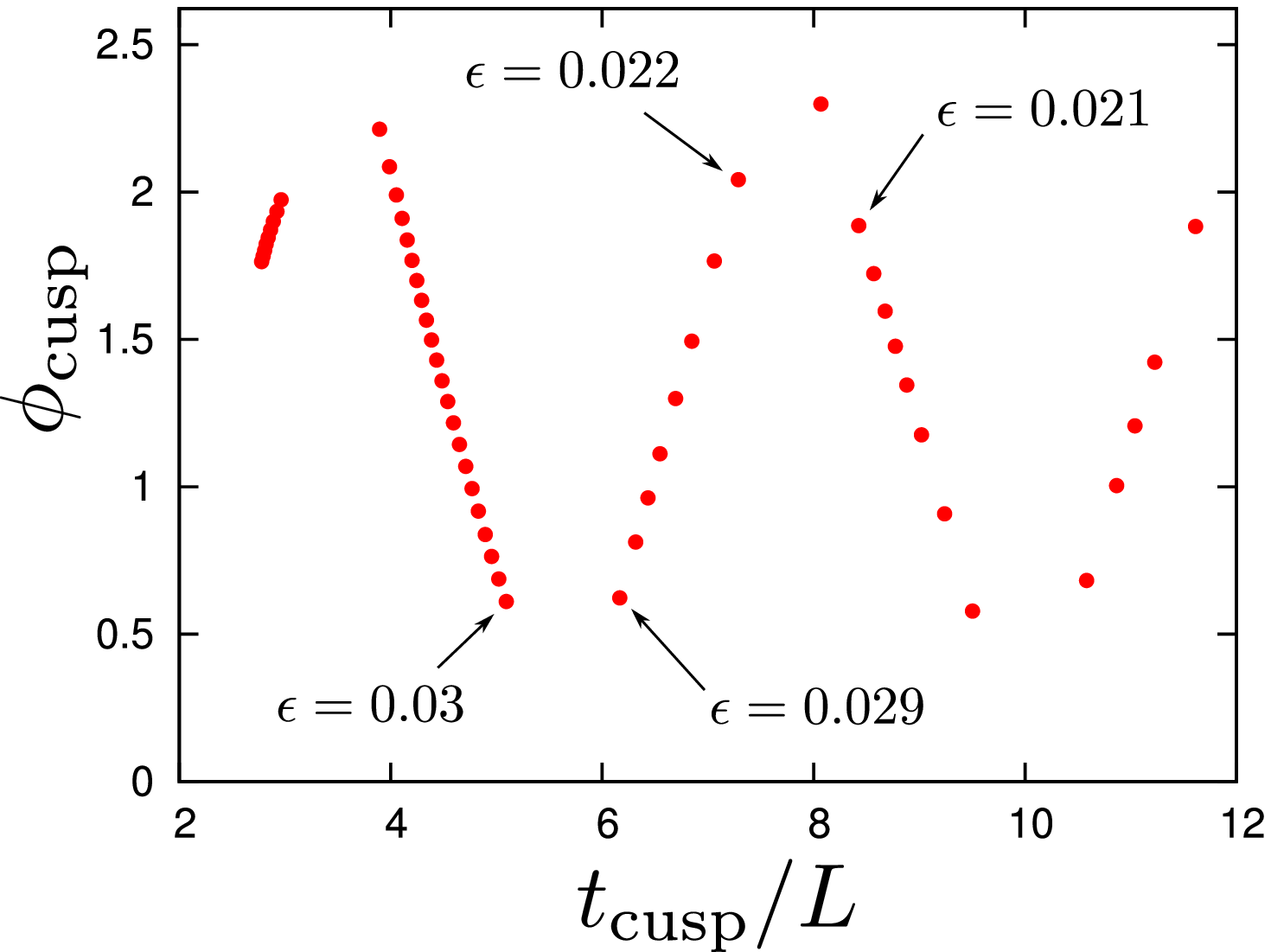}
}
\caption{Cusp formation for various $\epsilon$ when $\Delta t/L=2$. The formation
 points in (b) correspond to those at the same time in (a). In the
 plots, the points are computed at every 0.001 variation of $\epsilon$
 for $t_\mathrm{cusp}/L < 8$ and every $0.0005$ for $t_\mathrm{cusp}/L > 8$.}
\label{fig:cuspformation}
\end{figure}

Following this example, we survey cusp formation by changing quench parameters
using the method described in Section~\ref{sec:cusp_cond}.
Results are shown in Fig.~\ref{fig:cuspformation} when $\Delta t/L=2$ is fixed and $\epsilon$ is varied.
In the left panel, the times for the cusp formation for each $\epsilon$ are plotted,
and the corresponding locations in the $\phi$-coordinate are shown in the right panel.
The time for the cusp formation becomes longer as the
quench amplitude becomes smaller. It is also seen that the cusp formation
times are mildly discretized, as well as the locations of the formation away from the
boundary. This discretization indicates that the cusps might be 
difficult to form near the boundary.

The times and locations of the cusp formation tend to degenerate
in the very early time before the first reflection of the
initial perturbation wave at $X=-L/2$ around $t/L \sim 3.5$. 
Features in this region seem to be influenced by the largeness of the quench 
and would be different from late time dynamics.
Quenches with large amplitudes will be reported elsewhere~\cite{work_in_progress}.

\begin{figure}[t]
\centering
\includegraphics[scale=0.45]{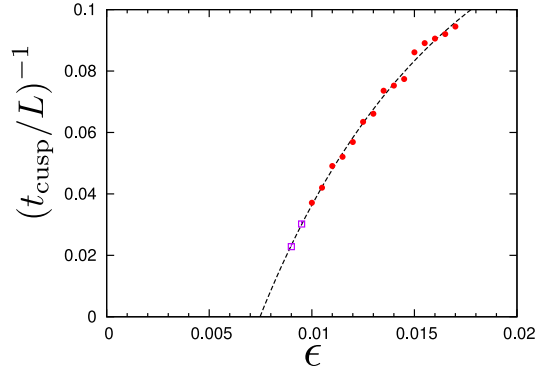}
\caption{Cusp formation in small $\epsilon$ for $\Delta t/L=2$. The dashed line is a fit of
 the red points. The points marked with the boxes are not used for the fit but are consistent with
 the fit curve. The fit curve reaches $(t_\mathrm{cusp}/L)^{-1}=0$ at 
$\epsilon \simeq 7.5 \times 10^{-3}$.}
\label{fig:cuspfit}
\end{figure}

The cusp formation time becomes longer as the amplitude gets smaller.
A natural question then is if the cusps can be formed for any small amplitude $\epsilon$.
As $\epsilon$ decreases, however, the variations in the fields are tinier and tinier, 
and eventually it becomes numerically difficult to use the cusp condition \eqref{cuspformJ} 
for finding the cusp formation. 
To obtain a reasonable inference, we fit results and extrapolate to smaller $\epsilon$. 
A result is given in Fig.~\ref{fig:cuspfit}, where we fit data points\footnote{
The points marked with the purple boxes were not derived from directly computing Eq.~\eqref{cuspformJ} 
because of difficulty in small $\epsilon$. We are, however, able to find a suspect of cusp formation 
by looking at the plot of the necessary condition \eqref{cuspneceJ}. 
We do not use these points for the fit, but they look consistent with the fit.
} at $\epsilon \ge 0.01$ and $t_\mathrm{cusp}/L \ge 10$ 
by a polynomial\footnote{
This fitting function is chosen by our intuition, and other choices for extrapolation could be utilized. 
For instance, the data can be also fit with $a \, \epsilon^2 + b \, \epsilon + c$, and 
a qualitatively consistent result can be obtained.
} $a \, \epsilon + b \, \epsilon^{1/2} + c$.
From the extrapolation, we find that there is a critical value $\epsilon_\mathrm{crit}$ below which the cusp formation time would be infinity. In Fig.~\ref{fig:cuspfit}, we obtain $\epsilon_\mathrm{crit} \simeq 7.5 \times 10^{-3}$.

\begin{figure}[t]
\centering
\includegraphics[scale=0.45]{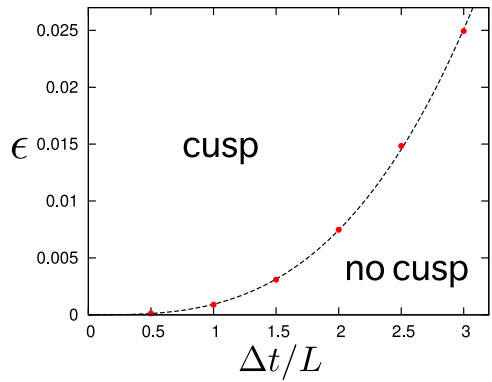}
\caption{The dynamical phase diagram for cusp formation. The red points are estimated values of critical $\epsilon$ at different $\Delta t/L$. The dashed curve is a fit given by $\epsilon = 9.3 \times 10^{-4} (\Delta t/L)^3$. Cusps are formed above this curve, but not below.}
\label{fig:cuspcritdiag}
\end{figure}

We repeat this procedure to estimate the critical $\epsilon$ for different $\Delta t/L$ and see how it changes. 
For each $\Delta t/L$, we fit data by $a \, \epsilon + b \, \epsilon^{1/2} + c$ and 
extrapolate to the limit of infinite cusp formation time to read off the value of critical $\epsilon$. 
Results are shown in Fig.~\ref{fig:cuspcritdiag}. 
We find that the critical value scales as $(\Delta t/L)^3$.
A fit of our results is $\epsilon_\mathrm{crit} = 9.3 \times 10^{-4} (\Delta t/L)^3$.
Note that this scaling may be altered if $\Delta t/L$ becomes very long and the wavelength of
the induced wave is comparable with the length of the hanging string. 
The critical amplitude in such a case, however, will be also big.
For this reason, we do not focus on larger $\Delta t/L$ in this paper.

The cusp formation found here is similar to the weakly turbulent
 instability in the global AdS~\cite{Bizon:2011gg}: 
Small fluctuations in that AdS propagate between the boundary and the
 center and, eventually, the perturbations collapse into a black hole after several bounces.
A difference between our cusp formation and that AdS instability is
their critical amplitude of initial perturbations.
The AdS instability occurs for arbitrary small initial
perturbations while 
our cusps are formed only for $\epsilon>\epsilon_\textrm{crit}> 0$.
By a perturbative study in Ref.~\cite{Bizon:2011gg}, 
it has been suggested that a commensurable spectrum is a necessary condition
for the collapse by arbitrary small perturbations.
As in Fig.~\ref{evef}, 
the linear spectrum of the string, $\omega_n$ and $\omega'_n$, 
are commensurable only for $n\to \infty$. 
Because of the non-commensurable spectrum of the string,
we need finite perturbations for the cusp formation.

\subsection{Energy spectrum in the non-linear theory}
\label{sec:Longi_En}
\begin{figure}
  \centering
  \subfigure[$\Delta t/L=2$, $\epsilon=0.005$ (no cusp)]
  {\includegraphics[scale=0.45]{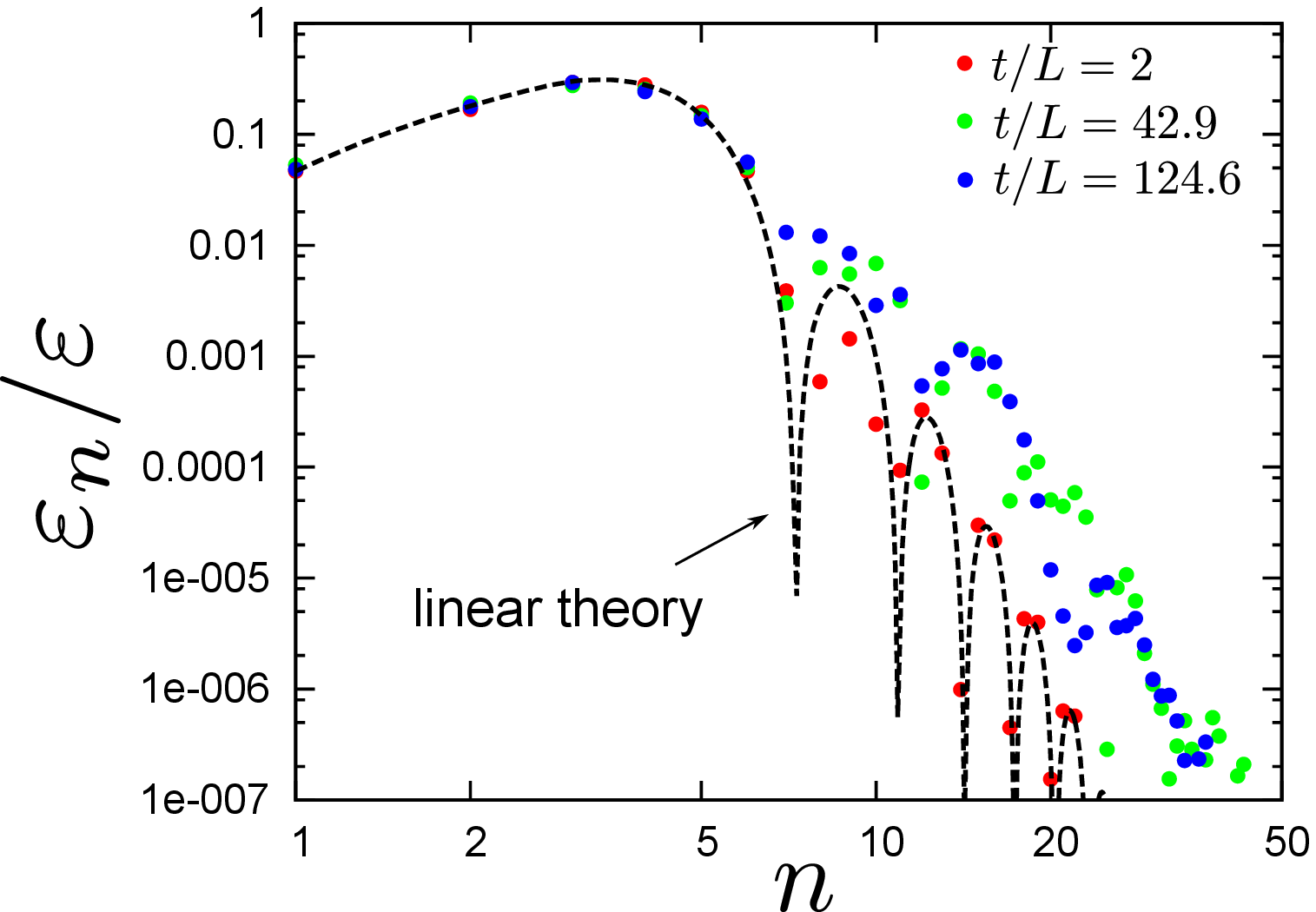}\label{dt=2_eps=0.005}
  }
  \subfigure[$\Delta t/L=2$, $\epsilon=0.01$ (cusp)]
  {\includegraphics[scale=0.45]{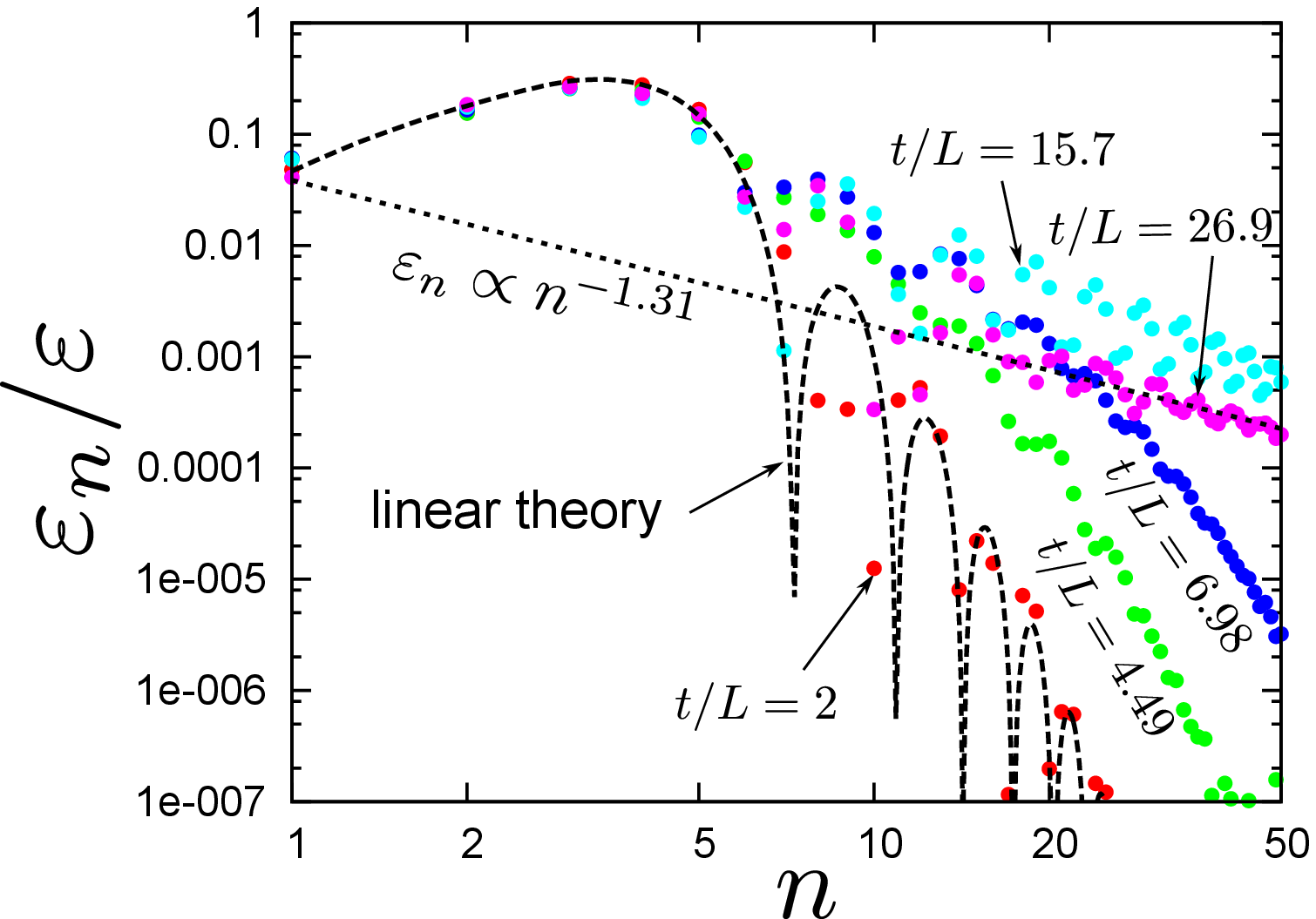}\label{dt=2_eps=0.01}
  }
  \subfigure[$\Delta t/L=2$, $\epsilon=0.03$ (cusp)]
  {\includegraphics[scale=0.45]{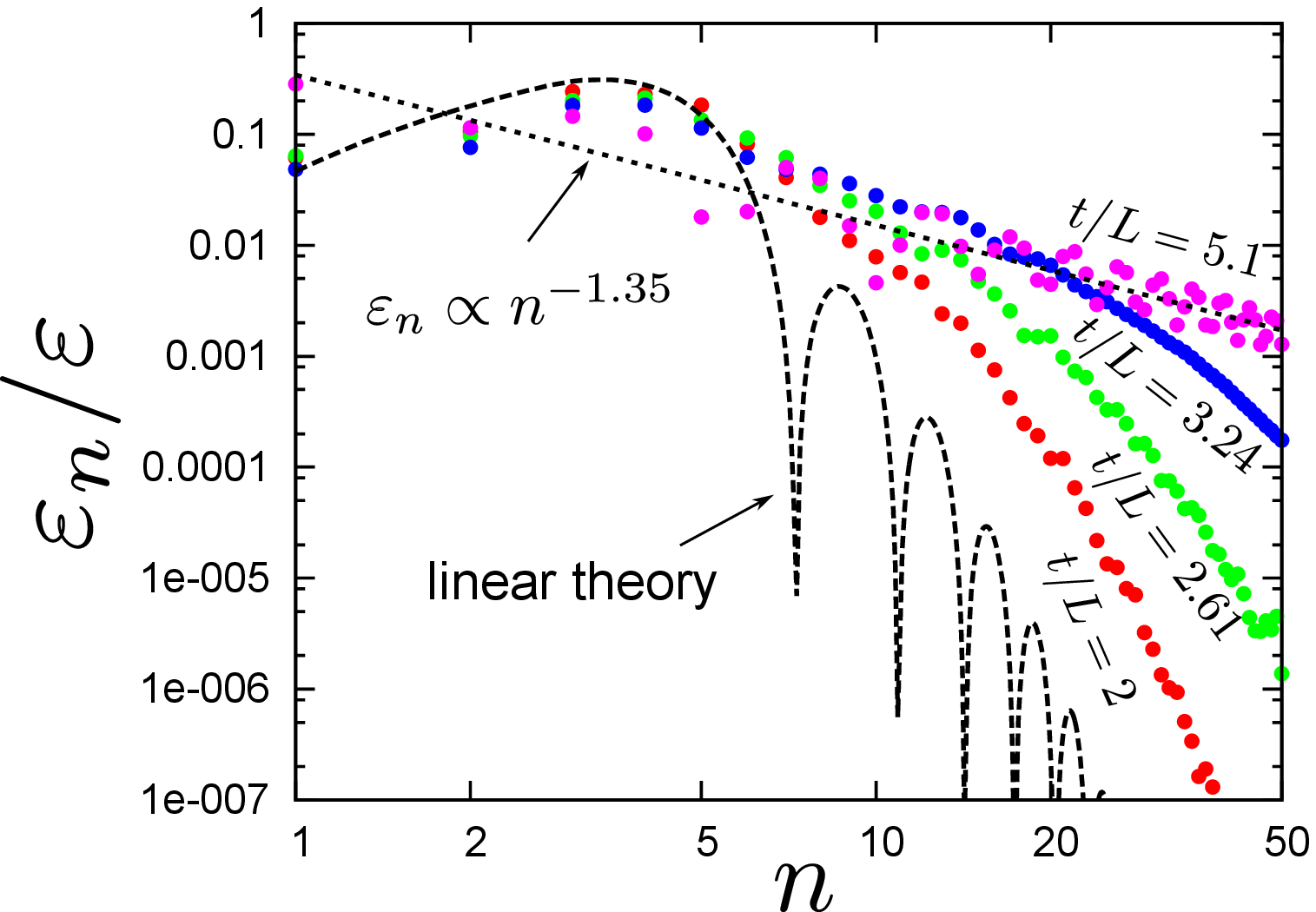}\label{dt=2_eps=0.03}
  }
  \subfigure[$\Delta t/L=4$, $\epsilon=0.07$ (cusp)]
  {\includegraphics[scale=0.45]{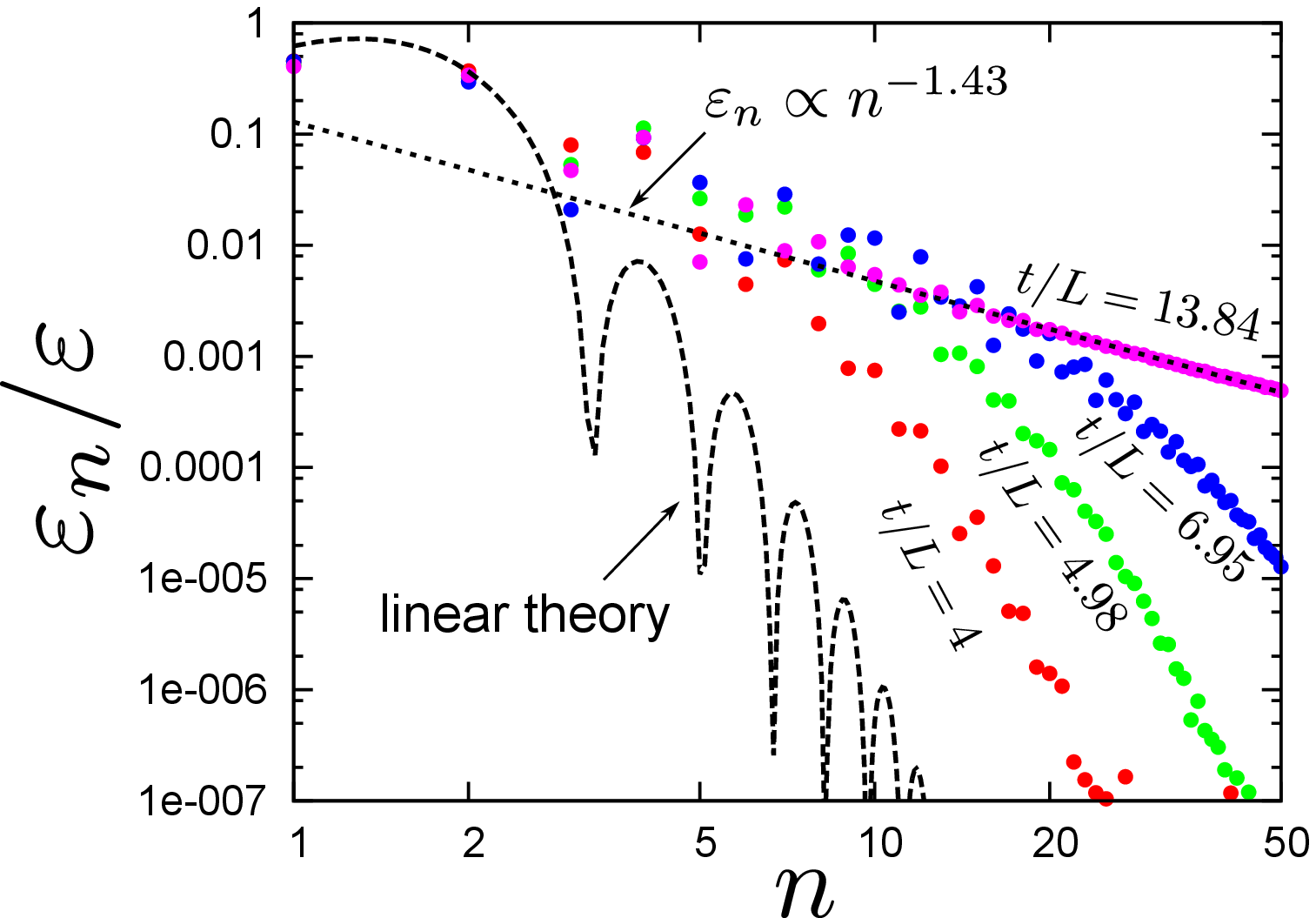}\label{dt=4_eps=0.07}
  }
  \caption{
Time dependence of the energy spectrum.
For the parameters (a), cusps do not form on the string.
For the parameters (b), (c) and (d), 
cusps are created on the string.
The dashed curves are the energy spectra
 computed in the linear theory.
Red points are spectra just after the quench.
Magenta points correspond to the time for cusp formation.
For (b), (c) and (d), 
the energy is transferred to higher modes as time
increases, and eventually power law spectra are reached.
In (b), light blue points indicate the time when the spectrum realizes the power law,
which lasts until the cusp formation.
We fit the magenta points by $\varepsilon\propto n^{-a}$, and the results are plotted 
with dotted lines.
\label{spec}
}
\end{figure}

Given the formation of cusps, we look into the time dependence
in the energy spectrum following the procedure described in Section~\ref{Enspec}.
For samples, we focus on the following four cases:
(a)~$\Delta t/L=2$, $\epsilon=0.005$, 
(b)~$\Delta t/L=2$, $\epsilon=0.01$,
(c)~$\Delta t/L=2$, $\epsilon=0.03$, and 
(d)~$\Delta t/L=4$, $\epsilon=0.07$. 
With the parameters (a), cusps do not form on the string, while 
cusps are created for (b), (c) and (d). 
In Fig.~\ref{spec}, we show the time dependence of the energy spectra for these parameters.
The dashed curves are the energy spectra computed in the linear theory; see Appendix~\ref{sec:Enspeclin} for the calculations.
Although the spectra are defined only for integer $n$, 
for visibility of the plot these results are generalized to continuous $n$.

High frequencies in the energy spectra are suppressed when the cusps are not formed.
In Fig.~\ref{dt=2_eps=0.005},
the spectrum can be well approximated by the linear theory just after the quench ($t/L=2$).
Although it slightly deviates from the linear theory as the time increases,
we do not find any remarkable change in the late time.

In contrast, the energy spectra show power law behaviors in the cases of cusp formation.
In Fig.~\ref{dt=2_eps=0.01}, although the spectrum can be well approximated by the linear theory
just after the quench, we see the growth of the spectrum in high frequencies as time passes,
apparently because of the nonlinearity in the time evolution equations.
In Fig.~\ref{dt=2_eps=0.03} and \ref{dt=4_eps=0.07}, the spectra deviate from those in the
linear theory even at $t=\Delta t$, and this indicates that the nonlinearity evolves  
even in the quenching time, $0<t<\Delta t$.
For these three cases, we find a direct energy cascade: 
The energy is transferred to higher $n$-modes during the time
evolution.
Eventually, power law spectra are observed, and these behaviors persist until the time of cusp
formation. (See magenta points.)\footnote{
After the cusp formation, the function $R(t,\phi)$ becomes multi-valued and
the energy spectrum is ill-defined.
Therefore, we only show spectra before the cusp formation.
}
In particular, as seen in~\ref{dt=2_eps=0.01}, the time for reaching the power law behavior can be rather earlier 
than that for the cusp formation,
and once realized, the behavior lasts until the cusps are formed.\footnote{
The string is smooth before the cusp formation, and the energy spectrum $\varepsilon_n$ must fall off faster than any power law
function as $n\to\infty$. 
Hence, it is indicated that the power law spectrum will be no longer maintained at $n \gg 50$, 
while many numerical efforts are necessary for computing the spectrum in such a region.
}
Hence, the cusp formation on the string can be regarded as a variation of weak
turbulence~\cite{Bizon:2011gg}.\footnote{The AdS weak turbulence was found to be characterized by a Kolmogorov-Zakharov scaling~\cite{deOliveira:2012dt}.}

We fit the power law spectra by $\varepsilon_n \propto n^{-a}$.
Just before the cusp formation, 
we obtain
$a=1.313\pm 0.171$, 
$a=1.354\pm 0.128$
and $a=1.432\pm 0.017$ for (b),
(c) and (d), respectively.
The exponents are distributed around $a\sim 1.4$.

\subsection{Time-dependence of the forces acting on the heavy quarks}

\begin{figure}
  \centering
  \subfigure[$0\leq t/L \leq 20$]
  {\includegraphics[scale=0.4]{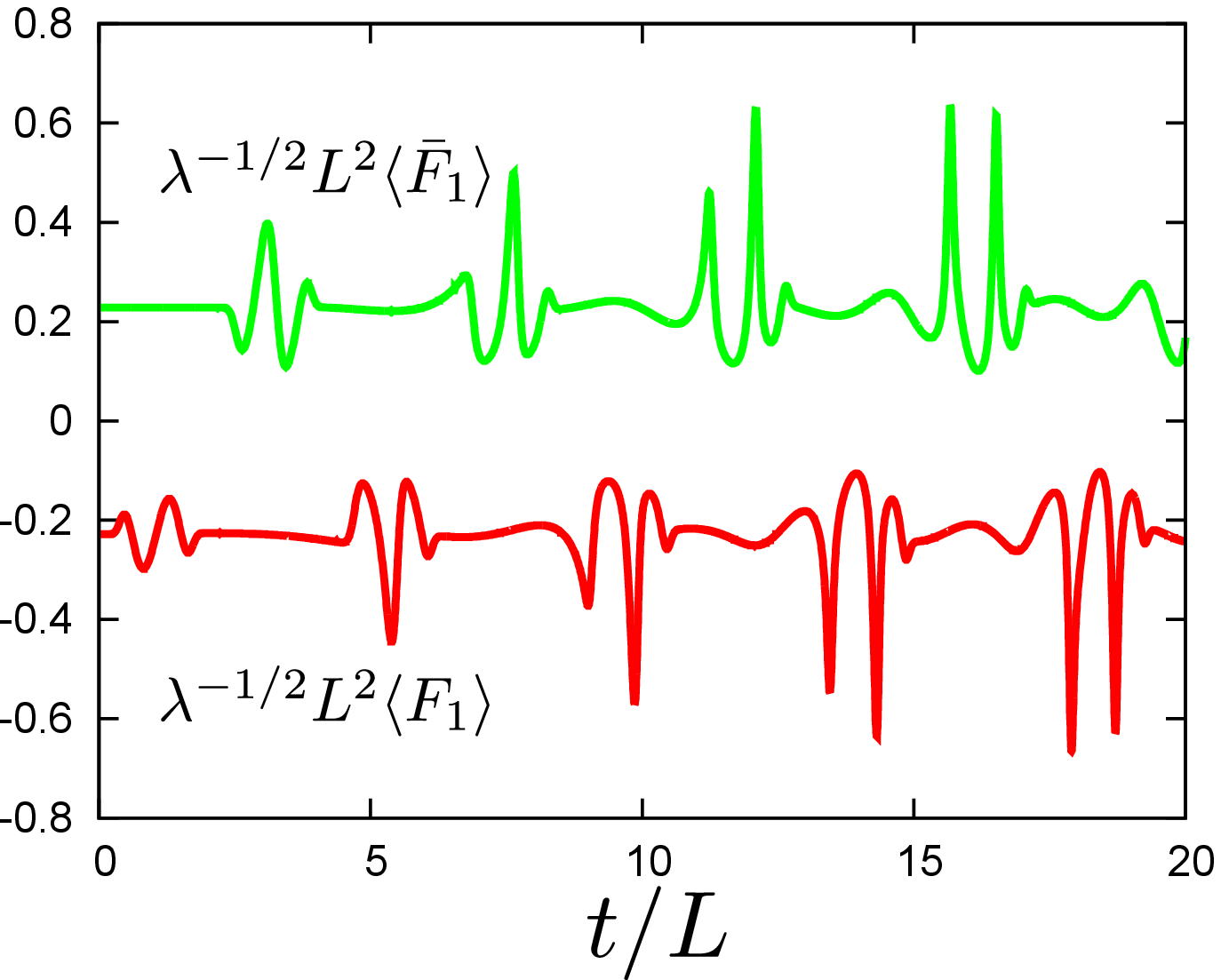}\label{FandFbar}
  }
  \subfigure[$0\leq t/L \leq 160$]
  {\includegraphics[scale=0.4]{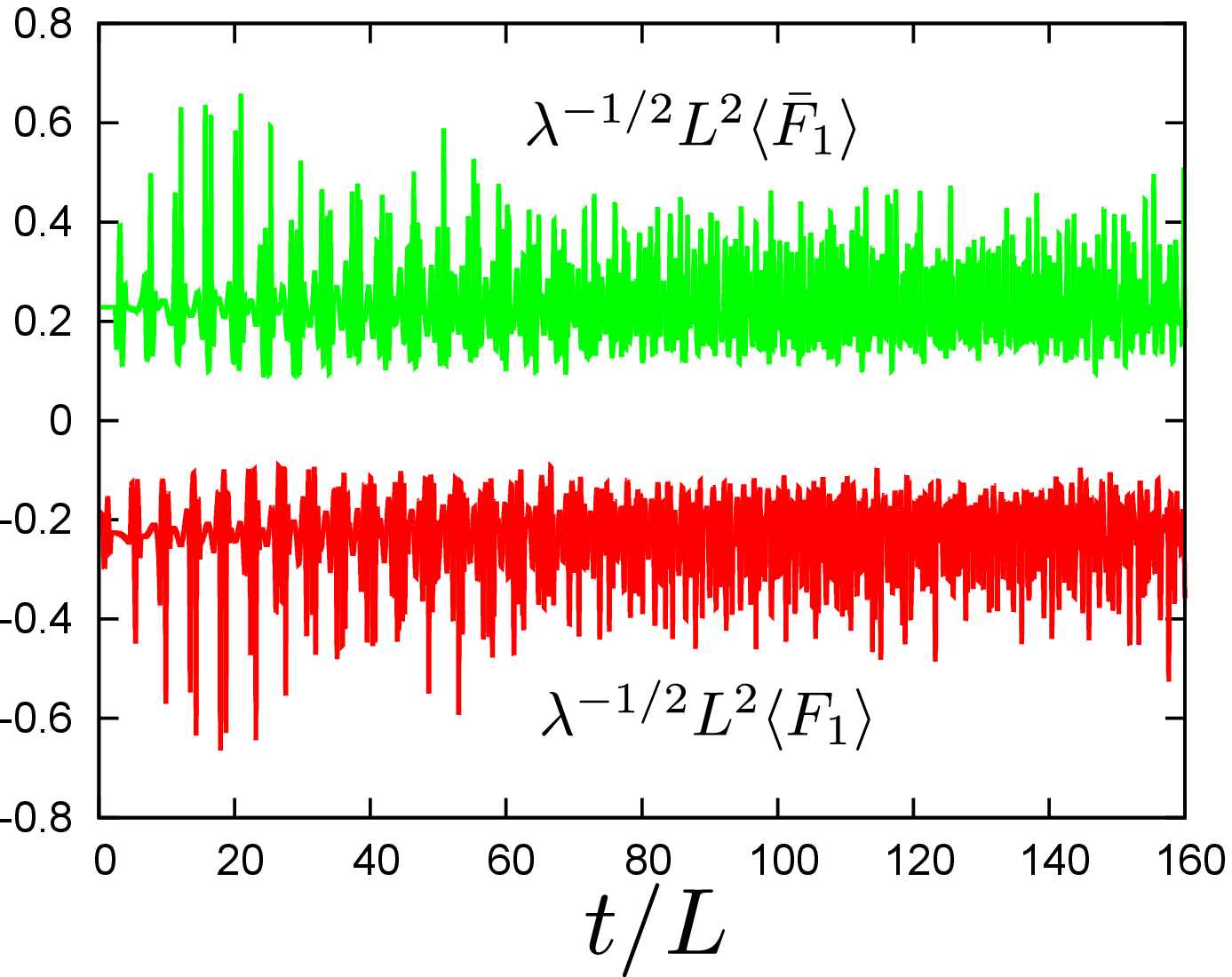} \label{FandFbar_long}
  }
  \caption{
Time dependence of the forces acting on the quark and the antiquark 
 for the longitudinal one-sided quench with $\epsilon=5.0\times 10^{-3}$ and
$\Delta t/L=2$. 
In this case, cusps do not form on the string.
\label{FFbar0}
}
\end{figure}

Following the procedure in Section~\ref{sec:force}, 
we compute the forces acting on the quark and antiquark as functions of
time. 
Since the string motion is now restricted in the $(z,x_1)$-plane, 
only the $x_1$-components of the forces are non-zero.
In Fig.~\ref{FFbar0}, we plot the forces $\langle F_1(t) \rangle$ and 
$\langle \bar{F}_1(t) \rangle$ for $\epsilon=0.005$ and
$\Delta t/L=2$. For these parameters, cusps do not form on the string.
Figure~\ref{FandFbar} is for the early stage in the time evolution ($0\leq t/L \leq 20$), and 
Fig.~\ref{FandFbar_long} for a long period ($0\leq t/L \leq 160$).
In Fig.~\ref{FandFbar},
pulse-like oscillations are repeated at regular time intervals.
The period corresponds to the timescale for the fluctuation induced by the quench~(\ref{quench1})
to go back and forth between the two boundaries.
Because of the dispersive spectrum in the linear perturbation, 
the initially localized oscillations tend to spread as time increases
as in Fig.~\ref{FandFbar_long}, and there is no way for them to converge again.
We can also see that $\langle F_1(t) \rangle<0$ and $\langle \bar{F}_1(t) \rangle>0$
throughout 
the time evolution, and this implies that the force between the quark and antiquark is always attractive.

\begin{figure}
  \centering
  \subfigure[$0\leq t/L \leq 13$]
  {\includegraphics[scale=0.4]{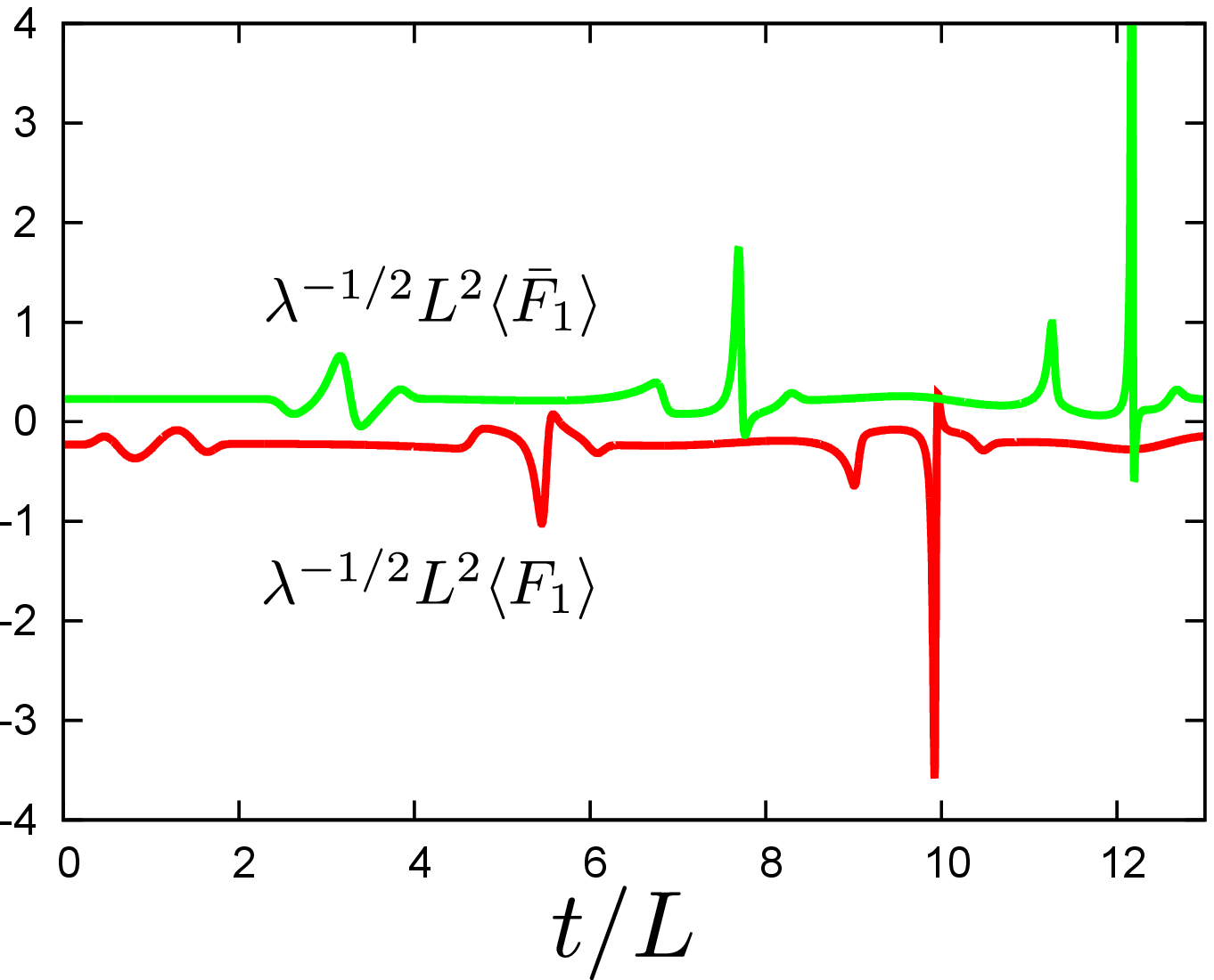}\label{Fcusp}
  }
  \subfigure[$0\leq t/L \leq 30$]
  {\includegraphics[scale=0.4]{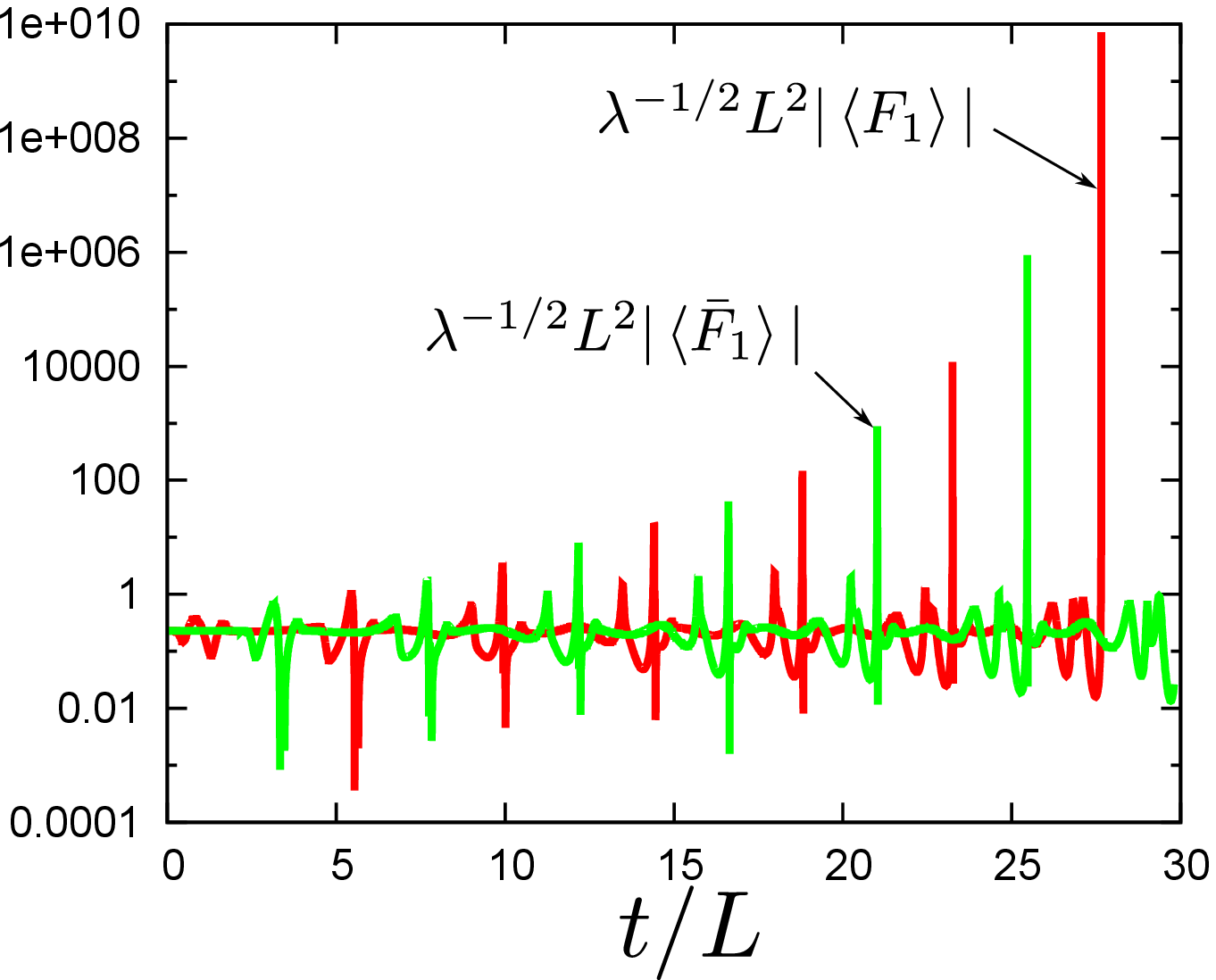} \label{Fcusp_long}
  }
  \caption{
Time dependence of the forces acting on the quark and the antiquark 
 for the longitudinal one-sided quench with $\epsilon=1.0\times 10^{-2}$ and
$\Delta t/L=2$. For these parameters, cusps form on the string.
Pulse-like oscillations in the forces are
getting amplified as time increases and diverge eventually.
\label{Fcusp0}
}
\end{figure}

For the parameters where the cusps form on the string, 
the time evolution of the forces is different from the previous example.
In Fig.~\ref{Fcusp0}, we show $\langle F_1(t) \rangle$ and 
$\langle \bar{F}_1(t) \rangle$ for $\epsilon=1.0\times 10^{-2}$ and
$\Delta t/L=2$, where
figures (a) and (b) are for $0\leq t/L \leq 13$ and 
for $0\leq t/L \leq 30$, respectively.
In figure (b), we take the absolute values of the forces, 
and plot in the log scale in the vertical axis.
We find that the pulse-like oscillations 
are getting sharp and amplified as the time increases.
The forces can change the sign because of their large oscillations,
and this implies that the force between the quark and antiquark can be repulsive temporarily.
Eventually, when a cusp arrives at the boundary after the cusp formation, 
the force diverges.

A mathematical explanation why the forces diverge after the cusp
formation is given as follows. Near the AdS boundary, 
the time dependent solution is well approximated by Eq.~(\ref{expand}).
The conditions for the cusp are then given by 
$\partial_\sigma Z=0$ and $\partial_\sigma \bm{X}=0$ as discussed in Section~\ref{sec:cusp_cond}. 
The latter condition is automatically satisfied near the boundary because
of $\partial_\sigma \bm{X} \sim \sigma$, while
the former gives $\dot{t}_0=0$.
In Eq.~(\ref{Force2}), the denominator has $\dot{t}_0$, 
and it appears that the numerator does not cancel the zero in the denominator.
Hence, it is natural that the force diverges when the cusp arrives at the boundary.

\subsection{$\mathbb{Z}_2$-symmetric quench}

When the two endpoints of the string are simultaneously quenched in the opposite directions with the same amplitude, there is a new contribution from the one-sided quench that the propagating waves collide at the $\mathbb{Z}_2$-symmetric point $X=0$, and cusps are expected to form on the collision. This case is equivalent to impose the Neumann boundary condition at $X=0$. Such a condition is typically imposed in probe D-brane embeddings. We would like to emphasize that understanding the difference between this case and the case without the Neumann condition would be important for distinguishing mechanisms for cusp formation whether the cusps are formed spontaneously as discussed in previous sections or formed with a help of the collisions or the Neumann condition.

\begin{figure}[t]
\centering
\subfigure[Cusp formation times]{
\includegraphics[scale=0.4]{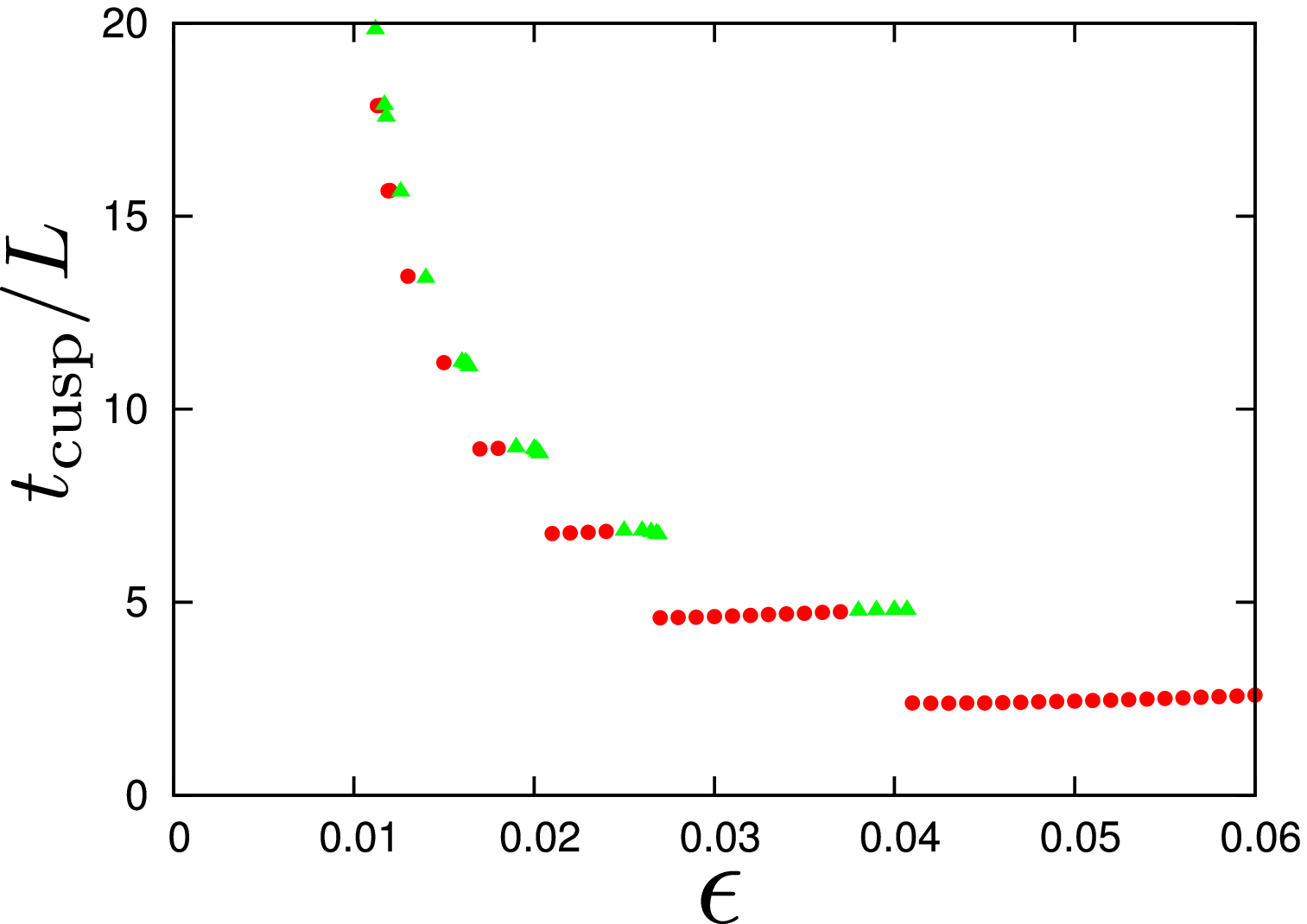}\label{fig:cuspZ2T}
}
\subfigure[Corresponding formation points]{
\includegraphics[scale=0.4]{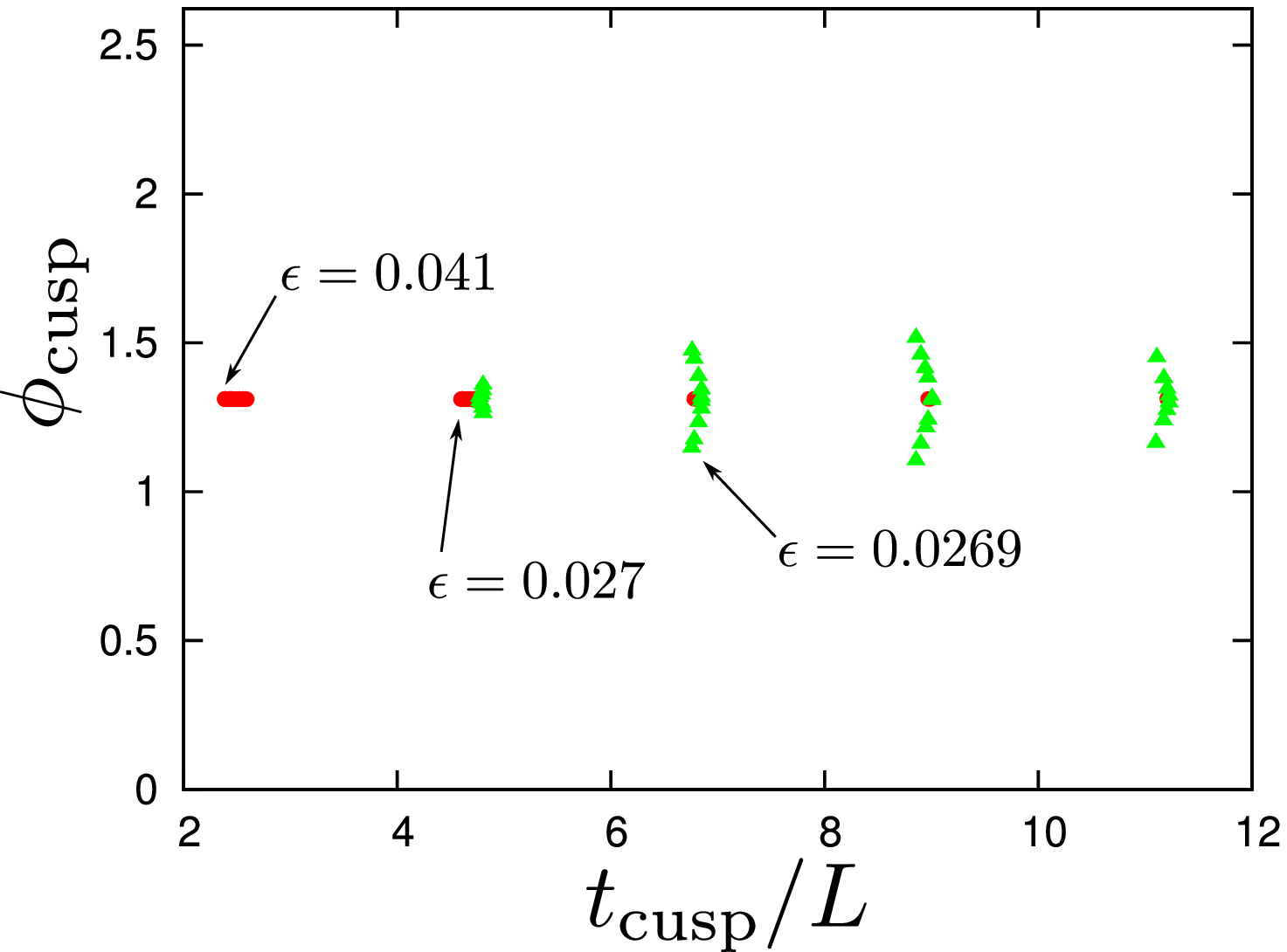}\label{fig:cuspZ2phi}
}
\caption{Cusp formation times for the $\mathbb{Z}_2$-symmetric quench with $\Delta t=2$. Cusp creation by wave collisions around $\phi=\beta_0/2$ are plotted with red dots, while events apart from that point and in the same way as the one-sided quench are plotted with green triangles.}
\label{fig:cuspZ2}
\end{figure}

We investigate the cusp formation by computing the condition
\eqref{cuspformJ}. Results are shown in Fig.~\ref{fig:cuspZ2} for
varying $\epsilon$ with $\Delta t/L=2$ fixed. The left panel shows the times
for \eqref{cuspformJ} to be satisfied for the first time, and the
corresponding $\phi$-coordinates are plotted in the right panel.
We find that there are two kinds of cusp formation and wave collisions at the $\mathbb{Z}_2$-symmetric point:
One is cusp formation by the wave collision, and the first cusp formation in this case is marked with red points in the plots, since in this case these cusps disappear once the colliding waves pass.
The other is that cusps are formed on the propagating waves in the same way as in the one-sided quench, and the cusp formation for this case is marked with green triangles.
The times for the cusp formation are clearly discretized, and the locations are concentrated to $\phi=\beta_0/2$. 
The small change in the formation time is because the propagating speed of the wave slightly varies for different $\epsilon$.

Closely looking at the results in evaluating \eqref{cuspformJ}, we find that even numbers of cusps are created for the first case (red points). In the bulk coordinates, the cusp formation is seen at very close to $\phi=\beta_0/2$ and one might naively think that only one cusp was generated at a point. In the worldvolume results, however, when the waves collide at $\phi=\beta_0/2$, we find that a pair of cusps whose orientations are opposite are created, and then these cusps pair-annihilate shortly when the waves pass by. Hence, the cusps created by the collision do not propagate away from that point, and therefore these instantaneous cusps are not observed in the force at the boundary. In this case, other formation of cusps in the same way as that in the one-sided quench also happens afterward.

In the cases of the green triangles in Fig.~\ref{fig:cuspZ2}, the cusps are formed slightly before the collision point, and these cusps subsequently collide at the $\mathbb{Z}_2$-symmetric point. In fact, it is seen in Fig.~\ref{fig:cuspZ2T} that these points go ahead of the cusp formation by the collisions. These cusps then continue traveling on the string, inferring that the waves are already magnified enough for forming cusps.

\begin{figure}[t]
\centering
\includegraphics[scale=0.45]{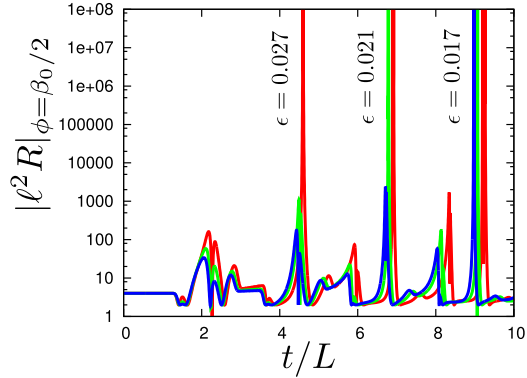}
\caption{Ricci scalar evaluated at $\phi=\beta_0/2$ when $\Delta t/L=2$.}
\label{fig:Z2Ricci}
\end{figure}

In the $\mathbb{Z}_2$-symmetric case, it is convenient to evaluate the time evolution of the worldsheet Ricci scalar at $\phi=\beta_0/2$ since many events of the cusp formation occur there. The Ricci scalar is given by
\begin{align}
R = \frac{2(\gamma_{uv,u} \gamma_{uv,v} - \gamma_{uv} \gamma_{uv,uv})}{\gamma_{uv}^3} \ .
\end{align}
For the static configuration, this monotonically changes from $R=-2/\ell^2$ at $\phi=0$ to $R=-4/\ell^2$ at $\phi=\beta_0/2$, and because of this coordinate dependence it might be desirable to compare the Ricci scalar at a fixed $\phi$. The Ricci scalar diverges on top of a cusp, and the cusp formation condition \eqref{cuspformJ} is consistent with the divergence of the Ricci scalar since $\gamma_{uv}$ in the denominator becomes zero when \eqref{cuspformJ} is satisfied. Practically, as we use discretized computations, the Ricci scalar does not exactly become infinity, while at least it becomes huge. Results of the Ricci scalar representing the cusp formation at second, third, and forth collisions are shown in Fig.~\ref{fig:Z2Ricci}. In these results, the absolute value of the Ricci scalar suddenly becomes huge at the cusp formation, and come back to of order one when the waves pass. In our setup, the numerical evolution does not breakdown after the Ricci scalar diverges in contrast to the D3/D7 case~\cite{Hashimoto:2014yza,Hashimoto:2014xta,Hashimoto:2014dda}, although finite-$N_c$ corrections may have to be taken into account.

\section{Results for the transverse linear quench}
\label{sec:transv}

\subsection{Cusp formation}
\label{sec:tr_lin_cus}

\begin{figure}
  \centering
  \subfigure[Just after the quench]
  {\includegraphics[scale=0.45]{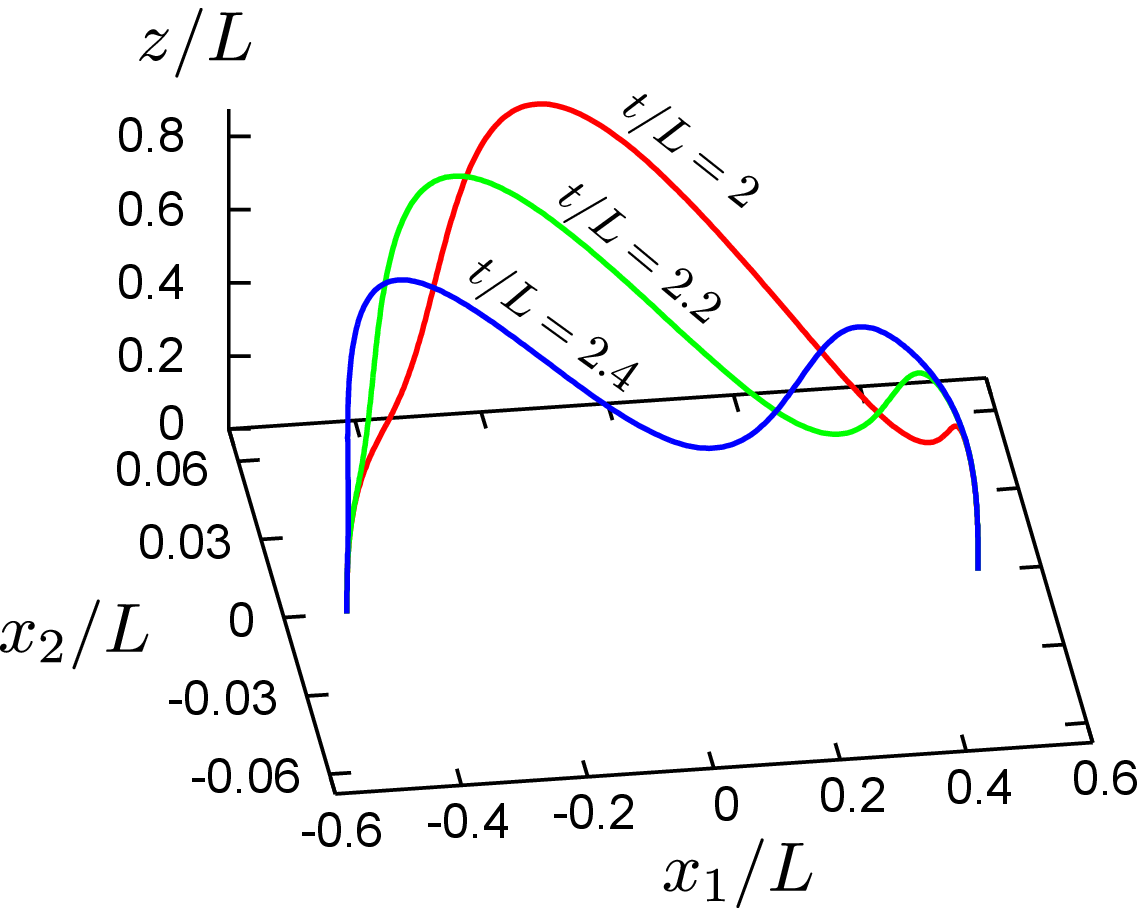}\label{tr_sn_early}
  }
  \subfigure[After the cusp formation]
  {\includegraphics[scale=0.45]{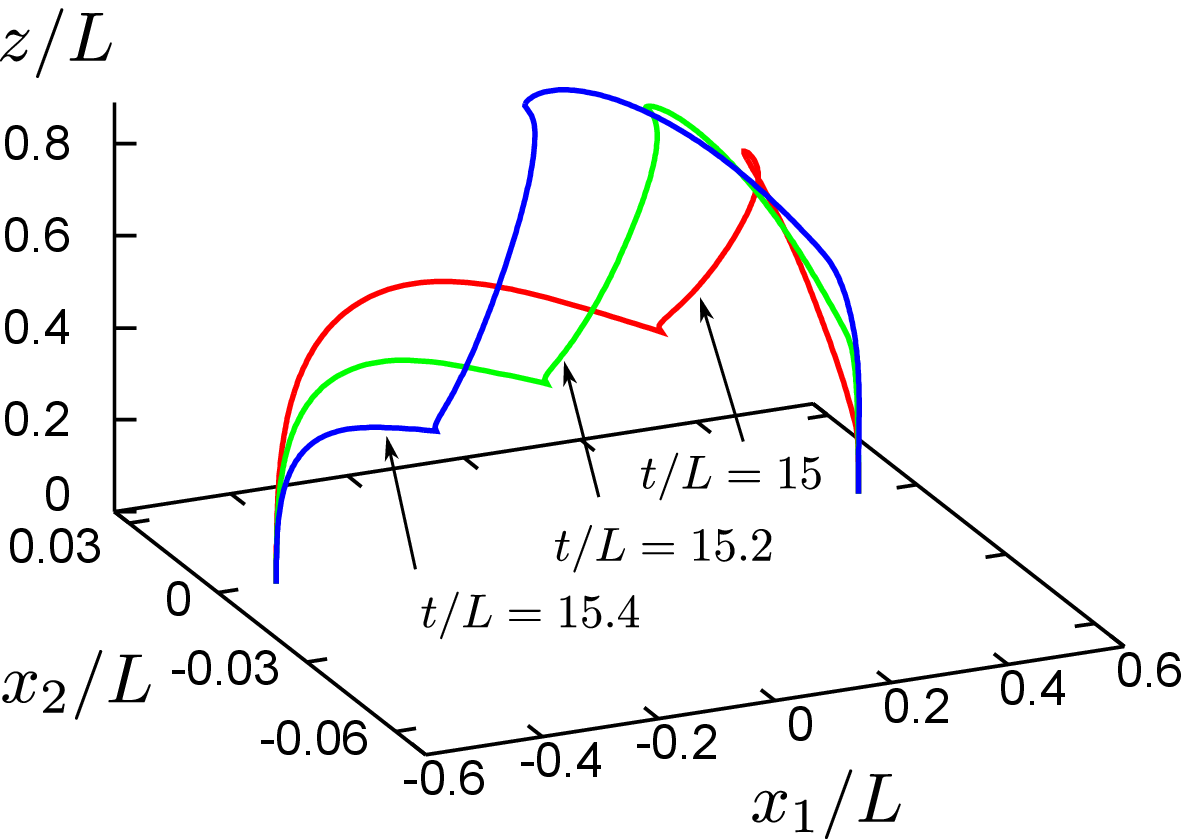} \label{tr_sn_cusp}
  }
  \caption{
Snapshots of the string in the transverse linear quench.
We set the parameters as $\Delta t/L=2$, $\epsilon=0.03$.
\label{tr_snap}
}
\end{figure}

In this section, we study the string dynamics induced by the
transverse linear quench~(\ref{quench3}).
With such a quench, the string moves in the (3+1)-dimensions spanned by $(t,z,x_1,x_2)$.
Snapshots of string configurations under a quench with parameters $\Delta t/L=2$ and $\epsilon=0.03$ are shown in Fig.~\ref{tr_snap}.
The left panel is just after the quench, $t/L=2, \, 2.2, \, 2.4$, where
we do not find cusps. 
However, in the right panel for late time $t/L=15, \, 15.2,\, 15.4$,
we find cusps on the string.
This demonstrates that cusps can form in the
transverse linear quench.
For quenches with smaller amplitudes ($\epsilon\lesssim 0.02$), we did not
find cusp formation for the period we computed the time evolution.

\subsection{Energy spectrum in the non-linear theory}

\begin{figure}
\begin{center}
\includegraphics[scale=0.5]{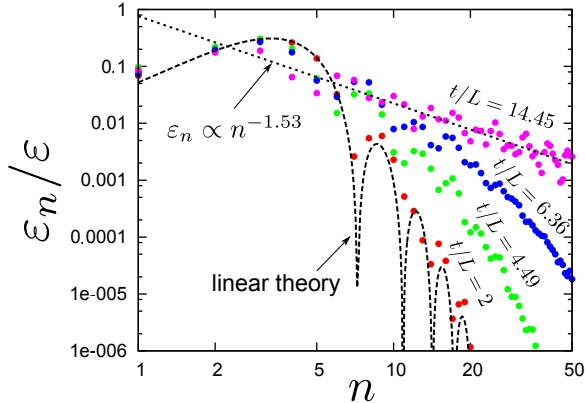}
\end{center}
\caption{
Time dependence of the energy spectrum in the transverse linear quench with
 $\Delta t/L=2$ and $\epsilon=0.03$.
Magenta points correspond to the time for the cusp formation.
We fit them by $\varepsilon\propto n^{-a}$ and show the result
by a dotted line.
}
 \label{spec2}
\end{figure}

We expect to see nonlinear origin for the cusp formation in the energy spectrum
also in the case of the transverse linear quench.
In Fig.~\ref{spec2}, we show
the time dependence of the energy spectrum when the parameters are $\Delta t/L=2$ and $\epsilon=0.03$.
The time for the cusp formation is $t/L=14.45$ for these parameters.
Red points are just after the quench, and
magenta points correspond to the time at cusp formation.
We find the direct energy cascade as in Section~\ref{sec:Longi_En},
and eventually the spectrum obeys a power law until the time of cusp
formation. 
Thus, also in the transverse linear
quench, we find the turbulent behavior toward the cusp formation.
Fitting the spectrum by $\varepsilon\propto n^{-a}$, 
we obtain $a= -1.533\pm 0.133$. 
The exponent has similar value as that for the longitudinal quench.

\subsection{Time-dependence of forces acting on the heavy quarks}

\begin{figure}
  \centering
  \subfigure[$\Delta t/L=2$, $\epsilon=0.01$ (no cusp)]
  {\includegraphics[scale=0.45]{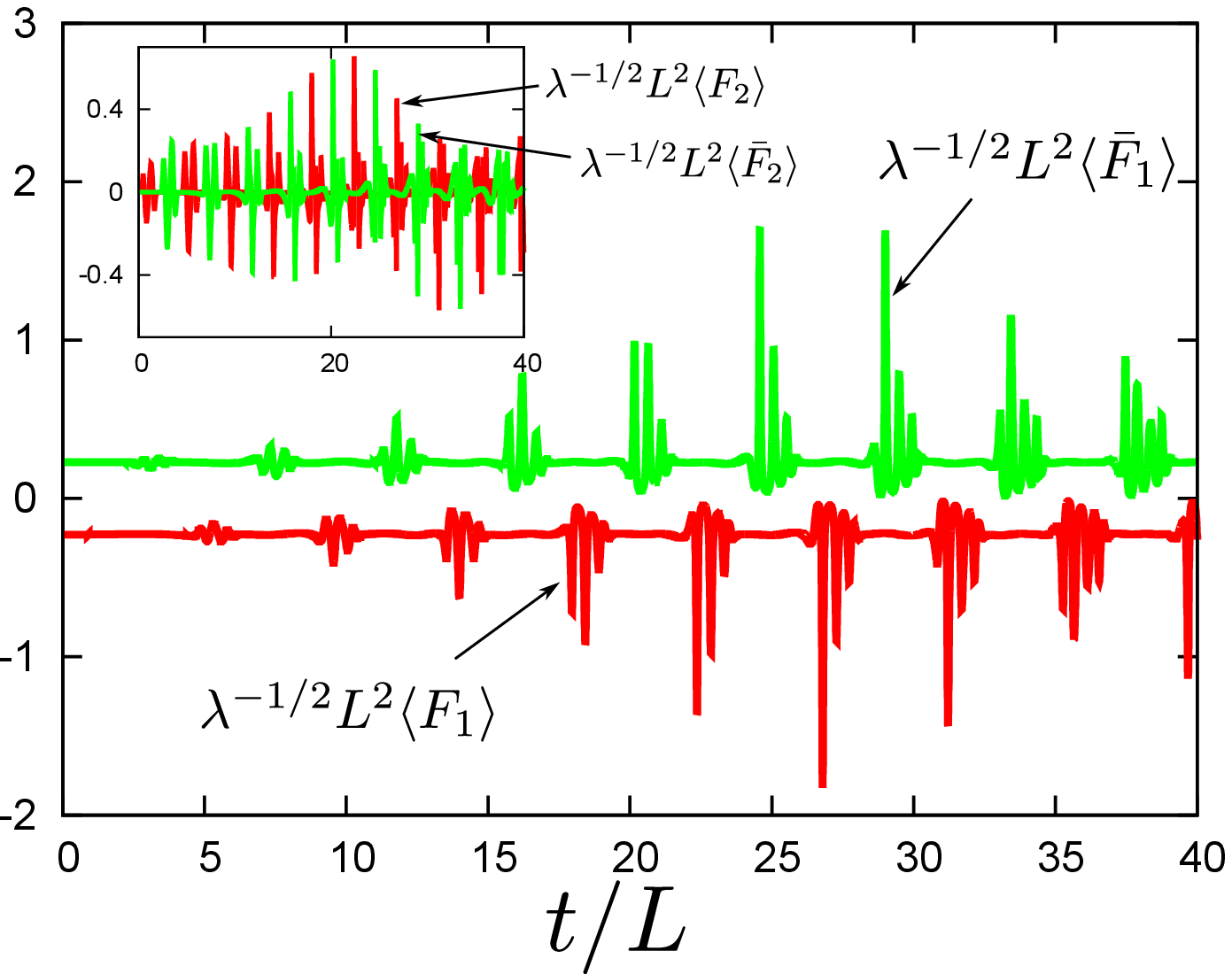}\label{F001_trans}
  }
  \subfigure[$\Delta t/L=2$, $\epsilon=0.03$ (cusp)]
  {\includegraphics[scale=0.45]{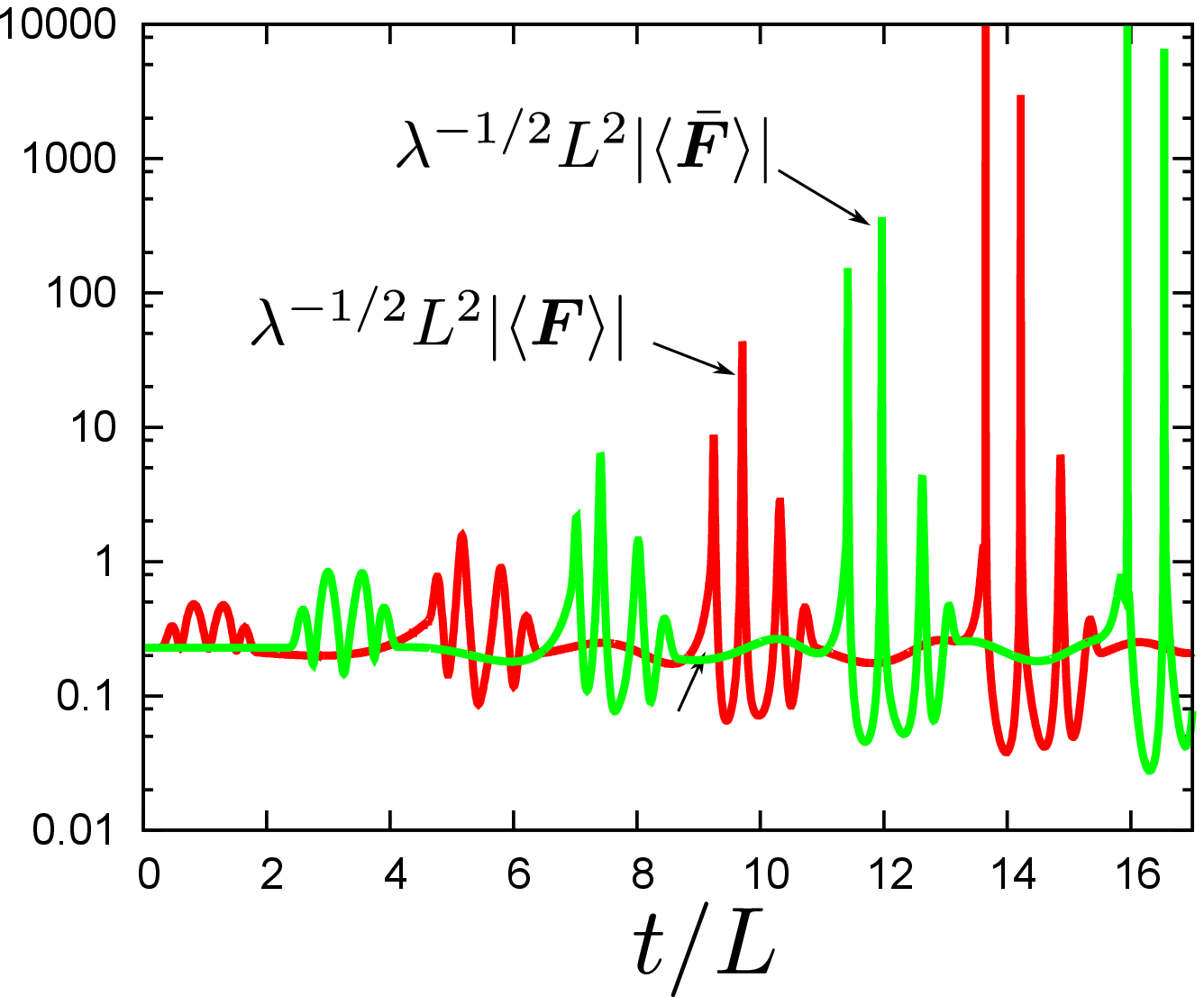} \label{F003_trans}
  }
  \caption{
Forces acting on the quark and the antiquark in the transverse linear quench. 
We fix the time scale of the quench as $\Delta t/L=2$.
Left and right panels are 
for $\epsilon=0.01$ and $\epsilon=0.03$, respectively.
In the right panel, we take the absolute values of the forces, and the vertical
 axis is log scale.
\label{F_trans}
}
\end{figure}

We turn to the time-dependence of the forces acting on the quark and the antiquark
 in the transverse linear quench.
Since the motion of the string is in the $(z,x_1,x_2)$-space, 
the $x_1$- and $x_2$-components of the forces can be non-zero.
In Fig.~\ref{F001_trans}, we show $\langle \bm{F} \rangle$ and 
$\langle \bar{\bm{F}} \rangle$ as functions of time for $\Delta t/L=2$
and $\epsilon=0.01$, with which cusps do not appear on the string.
Similar to the longitudinal quench, 
pulse-like oscillations are repeated at intervals.
Although there are sharp peaks, 
they are always $\mathcal{O}(1)$ in units of $\lambda^{-1/2}L^2$.
We also find that $\langle F_1(t) \rangle<0$ and $\langle \bar{F}_1(t) \rangle>0$.
Thus, the force between quarks is always attractive.

In Fig.~\ref{F003_trans}, we show the absolute values of the forces for 
$\Delta t/L=2$ and $\epsilon=0.03$, where cusps are formed
as seen in Section~\ref{sec:tr_lin_cus}.
The pulses are getting sharp and amplified as time increases, and
eventually after the cusp formation, 
the forces diverge when the cusps arrive at the boundaries.
We also monitored $\dot{t}_0$, which is in the
denominator in Eq.~(\ref{Force2}), at the boundaries, 
and found that it is consistent with zero at $t/L\simeq 13.6$ and $15.9$ at
$\bm{x}=\bm{x}_q$ and $\bm{x}_{\bar{q}}$, respectively.

\section{Results for the transverse circular quench}
\label{sec:trancirc}

\begin{figure}
  \centering
  \subfigure[Direct energy cascade]
  {\includegraphics[scale=0.45]{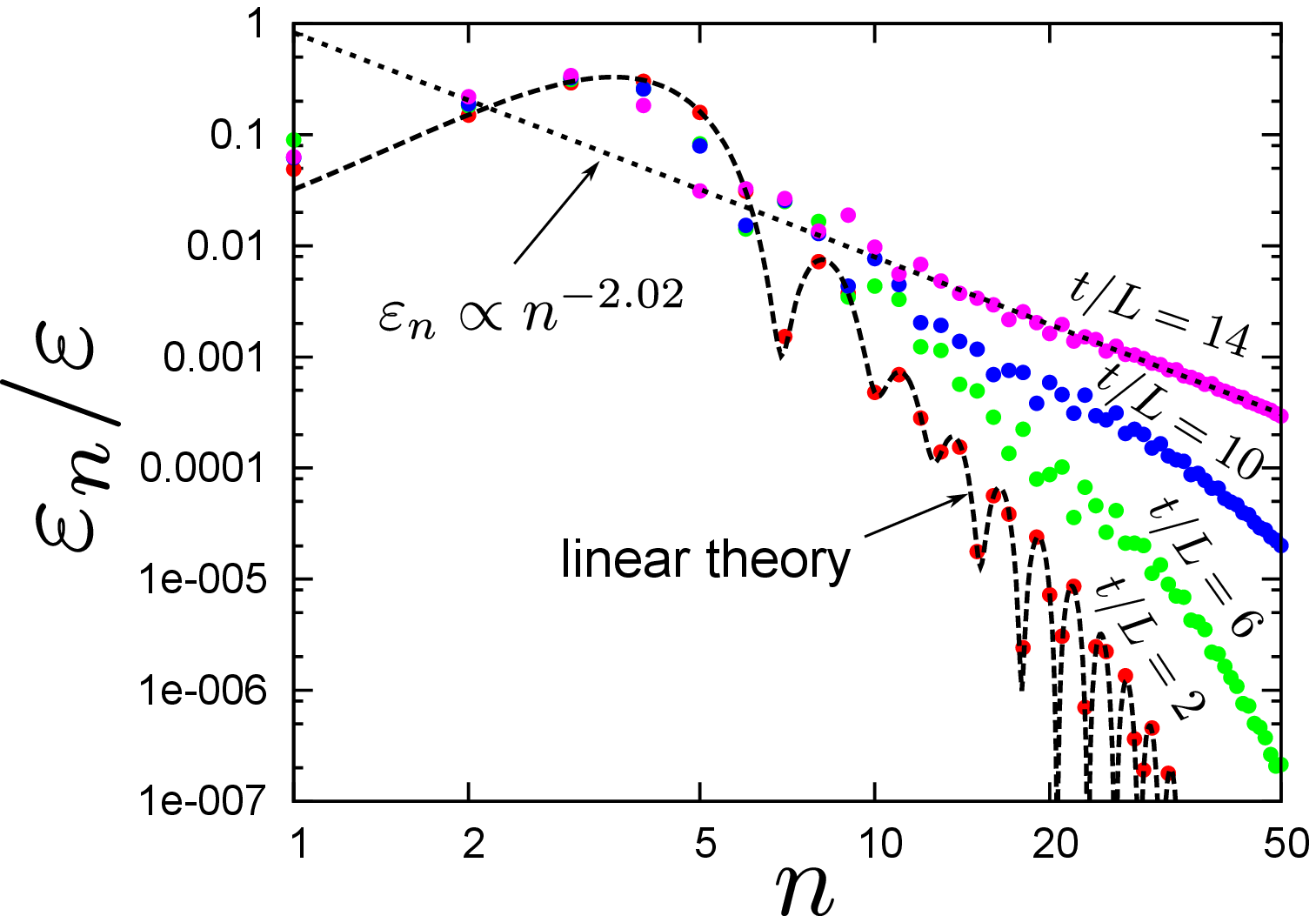}\label{circ1}
  }
  \subfigure[Inverse energy cascade]
  {\includegraphics[scale=0.45]{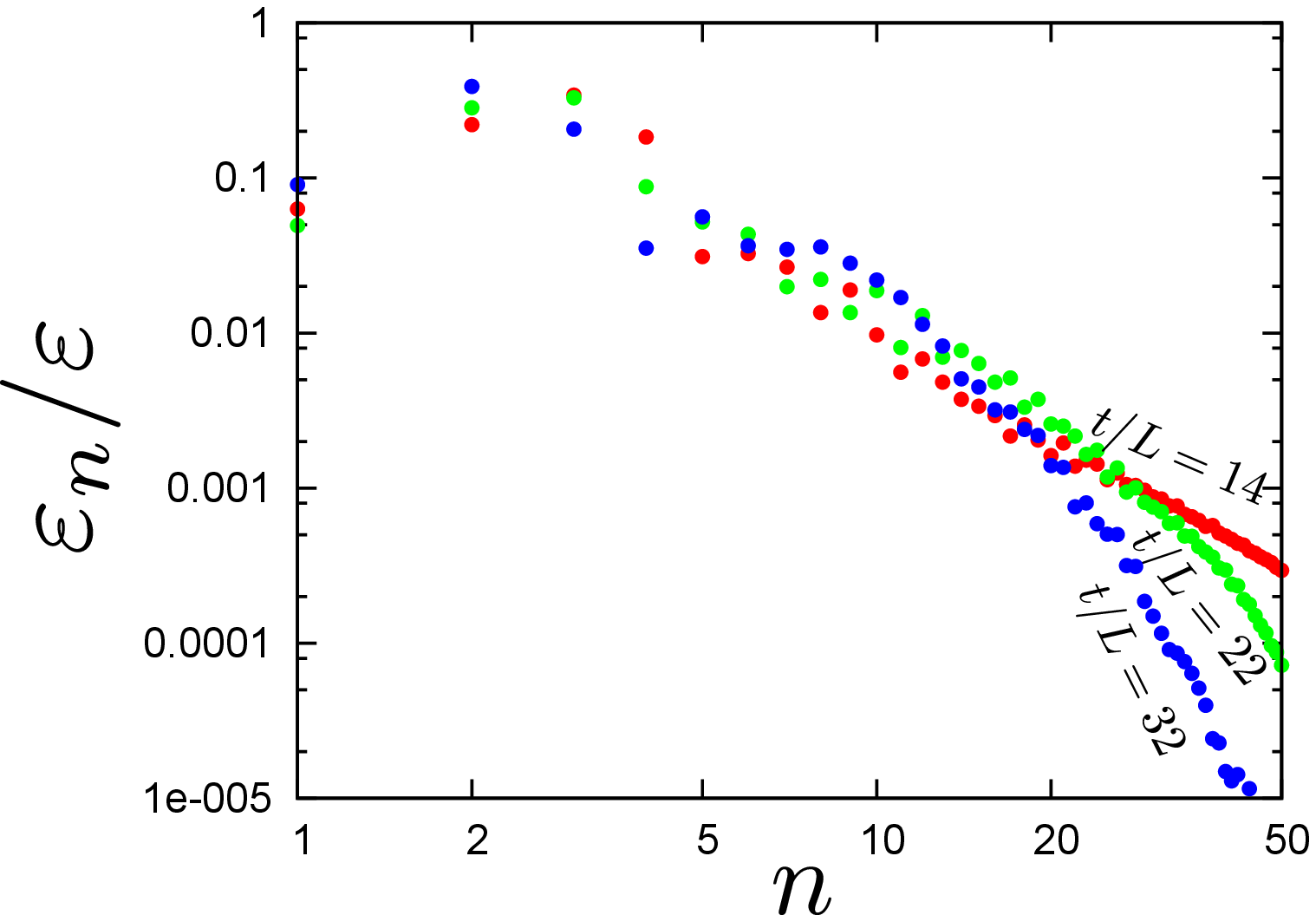} \label{circ2}
  }
  \caption{
Time dependence of the energy spectrum for the transverse circular quench with
 $\Delta t/L=2$ and $\epsilon=0.02$.
\label{circ}
}
\end{figure}

Finally, we study the string dynamics induced by the
transverse circular quench~(\ref{quench4}).
The string moves in all (4+1)-dimensions spanned by $(t,z,x_1,x_2,x_3)$.
For the transverse circular quench, we did not find any cusp formation
at least for modest parameters: around at $\epsilon\sim 0.01$ and $\Delta t/L\sim 1$.
Nevertheless, we found an interesting behavior in the energy spectrum.
In Fig.~\ref{circ}, we show the time dependence of the energy spectrum
for parameters $\Delta t/L=2$ and $\epsilon=0.02$.
In the early time evolution until $t/L\lesssim 14$, 
there is a direct energy cascade: the energy is transferred from
large to small scales, and eventually the spectrum obeys a power law
at $t/L\sim 14$.
Fitting the numerical data at $t/L=14$, we obtain 
$\varepsilon_n\propto n^{-2.019\pm 0.029}$.
For $t/L\gtrsim 14$, however, we find that 
this turns into an inverse energy cascade: 
The energy is transferred to the large scale.
Thus, in contrast to the previous low dimensional cases, 
the power law once realized at an intermediate time $t/L=14$ is not maintained in the late time.

\begin{figure}
  \centering
  \subfigure[$(x_1,x_2,z)$-space]
  {\includegraphics[scale=0.5]{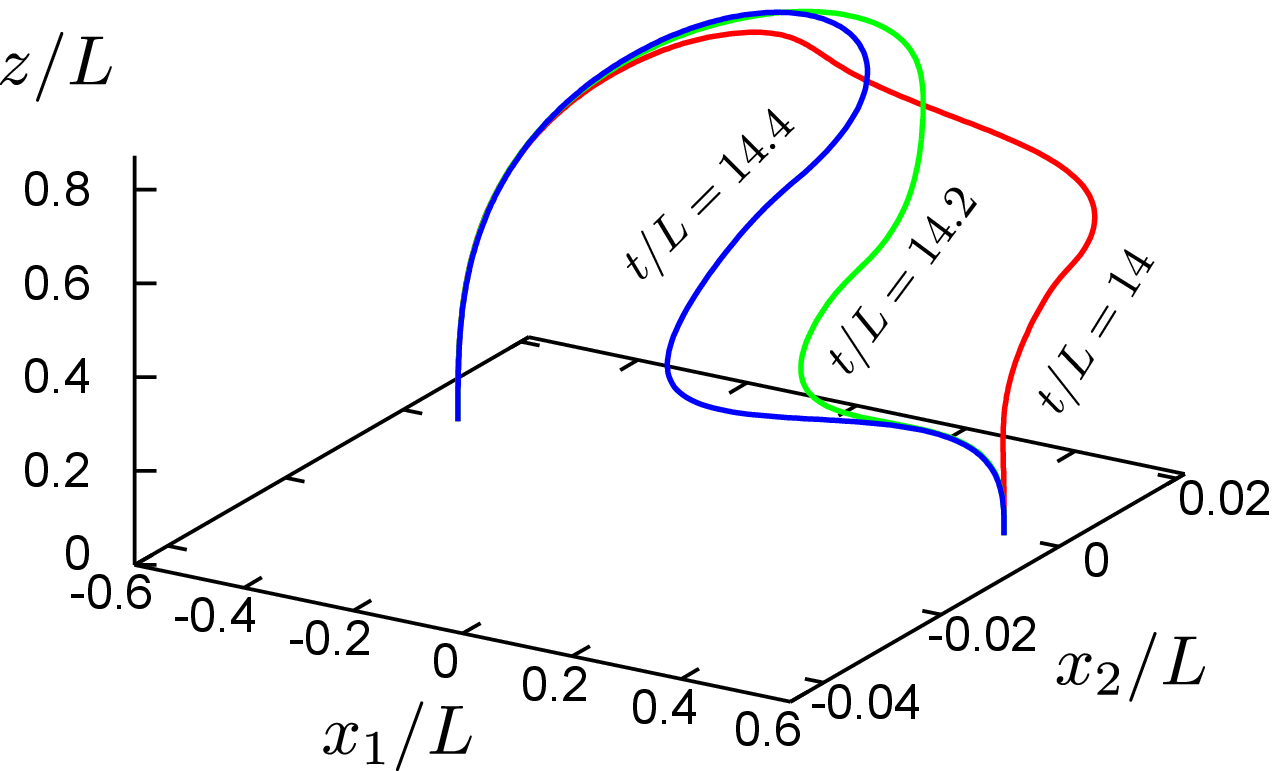}\label{x1x2}
  }
  \subfigure[$(x_1,x_3,z)$-space]
  {\includegraphics[scale=0.5]{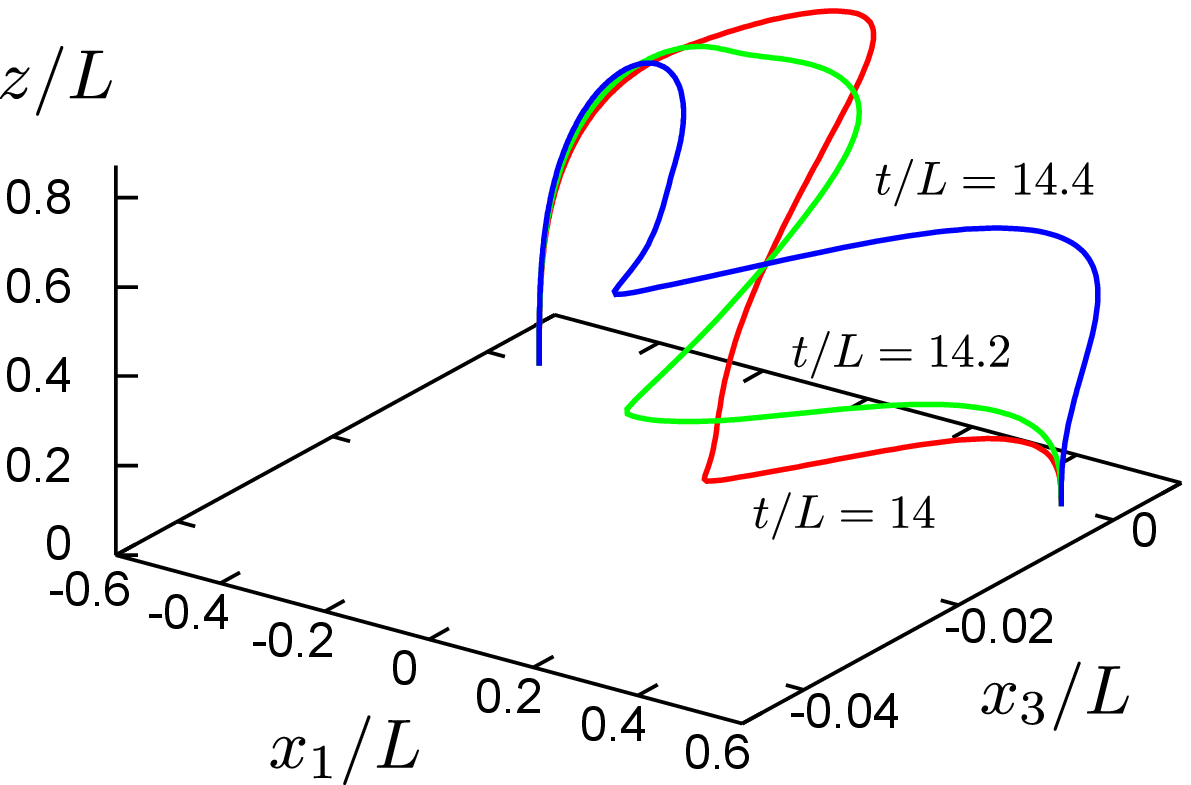} \label{x1x3}
  }
  \caption{
Snapshots of the string in the transverse circular quench with
 $\Delta t/L=2$ and $\epsilon=0.02$.
In the left and right figures, we project the profiles of the string into
 $(x_1,x_2,z)$- and  $(x_1,x_3,z)$-spaces, respectively.
\label{snap4d}
}
\end{figure}

In Fig.~\ref{snap4d}, we show 
snapshots of string configurations around the ``turning point'' of the
energy cascade: $t/L=14,14.2,14.4$.
Since the string motion is in the $(4+1)$-dimensions, 
we project the string profile into $(x_1,x_2,z)$- and  
$(x_1,x_3,z)$-spaces.
Although cusp-like points can be seen in the right figure, 
these are not real cusps: 
We find that although roots of $J_Z$, $J_{X_2}$
and $J_{X_3}$ become close at $t/L\simeq 14$,
$J_{X_1}$ is not zero at the point. (See Eq.~(\ref{cuspformJ}).)
However, the perturbation variable $\hat{\chi}_3$ defined in
Eq.~(\ref{pertvar}) becomes cuspy, namely, its energy is transferred to the
small scale. Hence the direct energy cascade appears in $t/L\lesssim 14$.
After $t/L\simeq 14$, the cuspy shape gets loose 
because of the dispersive spectrum in the linear perturbation.

\begin{figure}
\begin{center}
\includegraphics[scale=0.5]{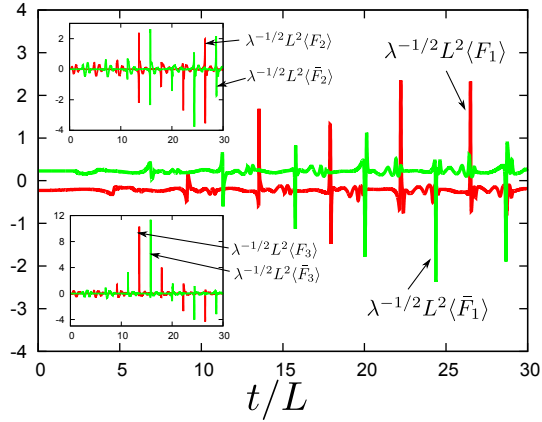}
\end{center}
\caption{
Time dependence of the forces acting on the quark and the antiquark 
 for the transverse circular quench with 
$\epsilon=0.02$ and
$\Delta t/L=2$. 
}
 \label{Force_circ}
\end{figure}

In Fig.~\ref{Force_circ}, we show the time dependence of the forces acting
on the quark and the antiquark for $\epsilon=0.02$ and
$\Delta t/L=2$. We do not find the divergence of the forces.
However, the forces are magnified until $t/L\simeq 14$ and can be repulsive at some time intervals
($\langle F_1 \rangle > 0$ and $\langle \bar{F}_1 \rangle < 0$) even
though there is no cusp formation. The magnitude of the forces in the late time 
is not as big as that period, reflecting the looseness in the cuspy shape.

\section{Summary and Discussion}
\label{sec:sumdis}

We studied nonlinear dynamics of the flux tube
between an external quark-antiquark pair in $\mathcal{N}=4$ SYM theory
using the AdS/CFT duality.
We numerically computed the time evolution of the string in AdS dual to the flux tube
when we perturbed the positions of the string endpoints to induce string motions.
We considered four kinds of quenches that were chosen to represent typical string motions:
(i)~longitudinal one-sided quench,
(ii)~longitudinal $\mathbb{Z}_2$-symmetric quench,
(iii)~transverse linear quench, and
(iv)~transverse circular quench. (See Eqs.~(\ref{quench1}-\ref{quench4})
and Fig.~\ref{quench_ponchi}.)
For (i)-(iii), we found cusp formation on the string.
In the time evolution of the energy spectrum, 
we observed the weak turbulence, that is, 
the energy was transferred to the small scale,
 and the energy
spectrum eventually obeyed a power law until the time of the cusp formation.
The cusp formation occurred only when the amplitude of the quench was larger
than a critical value, $\epsilon>\epsilon_\textrm{crit}$,
and the dependence of its magnitude on the quench duration $\Delta t$ 
was given by a simple form $\epsilon_\textrm{crit} \propto (\Delta t/L)^3$ in small $\Delta t/L$.
When the cusps arrived at the AdS boundary, we observed the divergence of 
the force between the quark pair.
For (iv), we found no cusp formation.
Nevertheless, we observed a direct energy cascade and the power law spectrum for a while.
However, in late time the direct cascade turned into an inverse energy cascade, 
where the energy was transferred to the large scale.
There was no divergence of the force between the quark pair.

How can we understand the weak turbulence of the string in view of
gauge theory?
Eigen normal modes $e_n$ of the fundamental string studied in
section~\ref{sec:lin} can be regarded as 
the excited states $|n\rangle$ 
of the flux tube in the gauge theory side.
Hence, the fluctuating string solution, such as
$R(t,\phi)=z_0 + \sum_n c_n(t)e_n(\phi)$, 
corresponds to 
\begin{equation}
 |\psi\rangle = |0\rangle +\sum_{n=1}^\infty c_n(t) |n\rangle\ 
\end{equation}
in the boundary theory, 
where $|0\rangle$ is the ground state.
The weak turbulence implies that $|c_n(t)|$ 
with $n\gg 1$ tends to increase as a function of time.
Therefore, in the late time, 
the probability of observing highly excited states 
is high compared to the linear theory.
Similar phenomenon has been found in the D3/D7 system dual to
$\mathcal{N}=2$ supersymmetric QCD~\cite{Hashimoto:2014yza,Hashimoto:2014xta,Hashimoto:2014dda}, where
a direct energy cascade was found in the fluctuations on the D7-brane and
regarded as production of many heavy mesons in the SQCD.
In that paper, this phenomenon was referred as ``turbulent meson condensation''.
Although the endpoints of the flux tube we considered are regarded as 
nondynamical and infinitely heavy quarks, 
the string turbulence found in this paper 
would be regarded as the microscopic picture 
of the turbulent meson condensation.

We found cusp formation when the motion of the string is restricted 
in $(2+1)$- and $(3+1)$-dimensions. 
The divergence of the forces acting on the quarks is accompanied by the cusp
singularities, where we expect that finite-$N_c$ effects will become important.
These will contain quantum effects of the string,
and such effects may resolve the cusp singularities and the divergence of the forces.
Nevertheless, the cusp formation in the classical sense can give us 
observable effects:
Finite-$N_c$ effects will also appear as gravitational backreactions.
If these are taken into account, a strong gravitational wave will be emitted
at the onset of the cusp formation.\footnote{
In the asymptotically flat spacetime,
gravitational self-interaction of cosmic strings 
has been perturbatively studied in Ref.~\cite{Quashnock:1990wv}.
This work suggests that cusps survive the backreaction.
}
For cosmic strings in flat spacetime, 
gravitational wave bursts from cusps have been studied in
Ref.~\cite{Damour:2001bk}, and 
it has been found that their spectra obey a power law in the high-frequency regime.
It would be nice to compute the gravitational waves from the
string in AdS$_5$
and find the power law spectrum. The description in the dual field theory 
may be the power law spectrum in gluon jets from the flux tube.

When the motion of the string is in
$(4+1)$-dimensions, we did not find cusp formation.
Hence, in AdS$_5$ spacetime, 
the cusp formation on the string is not a general phenomenon but accidental one. 
This implies that the dual phenomenon to the cusp formation is not ubiquitous 
in the $(3+1)$-dimensional boundary field theory.
However, if we consider the many-body system of quark-antiquark pairs,
it is possible that the time evolution of some flux tubes happens to be restricted
in lower dimensional spaces
approximately,\footnote{We thank Claude Warnick for pointing out this argument.}
and such flux tubes would be able to emit the gluon jets with the approximately power law spectrum, which
is characteristic to the cusp formation.
Besides, gravitational wave bursts in the presence of extra dimensions were 
discussed in \cite{O'Callaghan:2010ww,O'Callaghan:2010sy}. Even though
real cusps are not formed, there may be some gluon emission from cuspy shapes.

There are some future directions in our work.
In this paper, we only considered modest values for the amplitude of the quench,
$\epsilon\sim 0.01$.
For a large value of $\epsilon$, we expect that 
the string can even plunge into the Poincare 
horizon because of the strong perturbation.
This will demonstrate a non-equilibrium process of breaking of the flux tube.
It is also straightforward to take into account finite temperature effects in this process.
It will be also interesting to consider the string motions in confined geometries~\cite{Witten:1998zw}.
In the theories dual to these backgrounds, the quark-antiquark potential is linear,
and the presence of such potential may affect conditions for cusp formation.
Studying nonlinear string dynamics in such geometries may give new insights into 
understanding the QCD flux tubes and non-equilibrium processes in realistic QCD.

Closed strings rotating in AdS and having cusps were constructed in
Ref.~\cite{Kruczenski:2004wg}. Although our dynamical cusp formation on
an open string is different from the existence of cusps in those steady
solutions, it may be interesting to obtain useful information from such
configurations. In \cite{Hashimoto:2014yza}, the universal exponent in
the power law was deduced from a stationary solution called critical
embedding in the D3/D7-brane system in the presence of a constant
electric field, and results in time dependent computations obeyed that
universal value. In our setup, we do not have a corresponding static cuspy configuration, but our power law exponents, distributed around 1.4, may be naturally understood from cuspy stationary strings in AdS.

%In non-linear systems, there may be underlying
%chaos. In~\cite{Zayas:2010fs}, closed strings moving in AdS background
%were studied from the viewpoint of chaos. It may be interesting if
%chaos is seen also in the motion of open strings such as those
%considered in this paper.

Ultimately, it will be important to understand the mechanism relevant for 
the the turbulent behavior.
For the integrable Wilson loops such as those in AdS$_5 \times$ S$^5$,
the turbulent behavior may be studied with the techniques of integrability.

In non-linear systems, there may be underlying
chaos. In~\cite{Zayas:2010fs}, closed strings moving in Schwarzschild-AdS background
were studied from the viewpoint of chaos. 
It may be interesting if
chaos is seen also in the motion of open strings and the turbulent behavior is understood,
particularly in non-integrable situations.

\acknowledgments

The authors thank Constantin Bachas, Koji Hashimoto, Shunichiro
Kinoshita, Shin Nakamura,
Vasilis Niarchos, So Matsuura, Giuseppe Policastro, Harvey Reall, Christopher Rosen and Claude
Warnick for valuable discussions and comments.
The work of T.I.~was supported in part by European Union's Seventh Framework Programme under grant agreements (FP7-REGPOT-2012-2013-1) no 316165, the EU program ``Thales" MIS 375734 and was also cofinanced by the European Union (European Social Fund, ESF) and Greek national funds through the Operational Program ``Education and Lifelong Learning" of the National Strategic Reference Framework (NSRF) under ``Funding of proposals that have received a positive evaluation in the 3rd and 4th Call of ERC Grant Schemes''.

\appendix

\section{Numerical methods}
\label{sec:numerical}
In this appendix, we explain our numerical method for solving the equations of
motion of the string~(\ref{evol}).
Basic ideas are explained in Appendix A in \cite{Ishii:2014paa}.

We found that the original form of the equations of motion~(\ref{evol}) is numerically unstable. 
To stabilize the numerical evolution, it is effective to use the constraints~(\ref{CON}). 
From them, we have
\begin{equation}
 T_{,u}=\sqrt{Z_{,u}^2+\bm{X}_{,u}^2}\ ,\qquad
T_{,v}=\sqrt{Z_{,v}^2+\bm{X}_{,v}^2}\ ,
\end{equation}
where we choose the positive signs for the square roots since we take $\partial_u$ and
$\partial_v$ as future directed vectors.
Eliminating $T_{,u}$ and $T_{,v}$ from Eq.~(\ref{evol}), we obtain
\begin{equation}
\begin{split}
&T_{,uv}=\frac{1}{Z}[(Z_{,u}^2+\bm{X}_{,u}^2)^{1/2}Z_{,v}+(Z_{,v}^2+\bm{X}_{,v}^2)^{1/2}Z_{,u}]\ ,\\
&Z_{,uv}=\frac{1}{Z}[(Z_{,u}^2+\bm{X}_{,u}^2)^{1/2}(Z_{,v}^2+\bm{X}_{,v}^2)^{1/2}+Z_{,u}Z_{,v}-\bm{X}_{,u}\cdot
 \bm{X}_{,v}]\ ,\\
&\bm{X}_{,uv}=\frac{1}{Z}(\bm{X}_{,u}Z_{,v}+\bm{X}_{,v}Z_{,u})\ .
\end{split}
\label{evol_stable}
\end{equation}
The evolution equations in these expressions are found numerically stable.

\begin{figure}[t]
\centering
\includegraphics[scale=0.5]{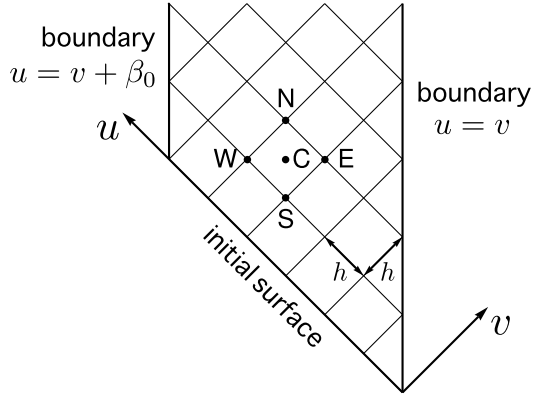}
\caption{Discretization of the world volume.} 
\label{ponchi_Fnews}
\end{figure}

To numerically solve \eqref{evol_stable}, we discretize the world volume $(u,v)$-coordinates with the grid spacing $h$
as shown in Fig.~\ref{ponchi_Fnews}. Let us denote the fields 
$(T, \,Z, \, \mathbf{X})$ by $\Psi$. At a point $C$ apart from the boundary, the fields and
their derivatives are discretized with second-order
accuracy as
\begin{align}
\Psi_{,uv} |_C &= \frac{\Psi_N - \Psi_E - \Psi_W + \Psi_S}{h^2}\ , \quad
\Psi_{,u} |_C = \frac{\Psi_N - \Psi_E + \Psi_W - \Psi_S}{2 h}\ , \nonumber \\
\Psi_{,v} |_C&= \frac{\Psi_N + \Psi_E - \Psi_W - \Psi_S}{2 h}\ , \quad
\Psi |_C = \frac{\Psi_E + \Psi_W}{2}\ .
\end{align}
Discretization error is $\mathcal{O}(h^2)$.
Substituting these into the evolution equations~(\ref{evol_stable}), we obtain nonlinear
equations to determine $\Psi_N$ by using known data of $\Psi_E$, $\Psi_W$,
and $\Psi_S$. We use the Newton-Raphson method for solving the coupled
nonlinear equations.

The equations for the boundary time evolution~\eqref{bdry_T_evo} become coupled
nonlinear equations of $T_N$ and $X_N$. ($Z_N=0$ is trivially imposed.)
The form at $\sigma=0$ is
\begin{equation}
T_N = T_S + \sqrt{4Z_W^2 + (\mathbf{X}_N-\mathbf{X}_S)^2} \ , \quad \mathbf{X}_N = \bm{x}_q(T_N) \ ,
\label{disc_bdry_TN_nonlinear}
\end{equation}
where we used a relation for $Z$, $Z_E = -Z_W$, derived from
the boundary condition $Z_{,uv}=0$. At the other boundary
$\sigma=\beta_0$, $Z_W$ and $\bm{x}_q$ in
\eqref{disc_bdry_TN_nonlinear} are replaced with $Z_E$ and
$\bm{x}_{\bar{q}}$. These equations are also solved by using the
Newton-Raphson method.

\section{Error analysis}
\label{sec:err}

\begin{figure}
  \centering
  \subfigure[Longitudinal]
  {\includegraphics[scale=0.32]{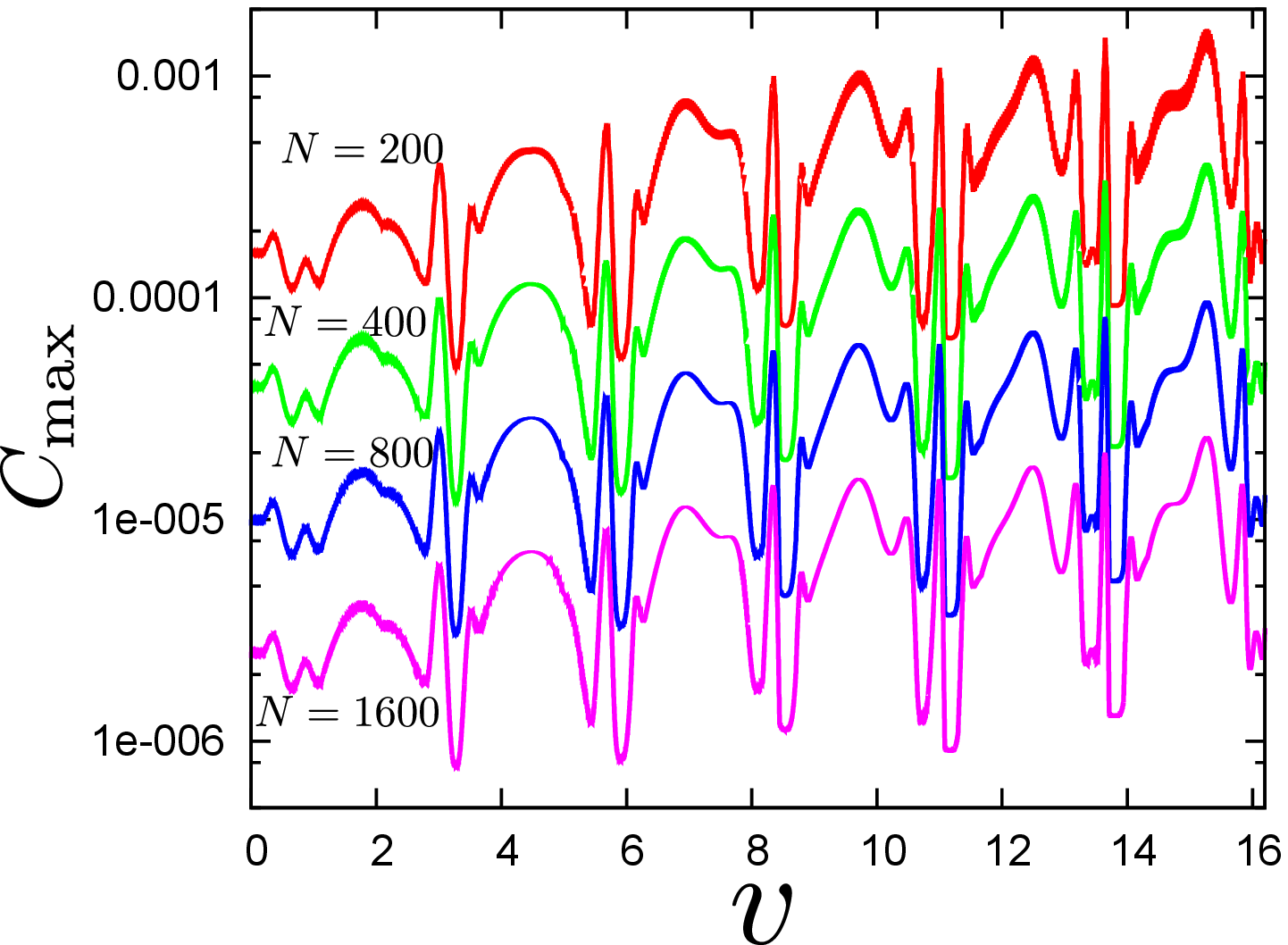}\label{cmax_L}
  }
  \subfigure[Transverse linear]
  {\includegraphics[scale=0.32]{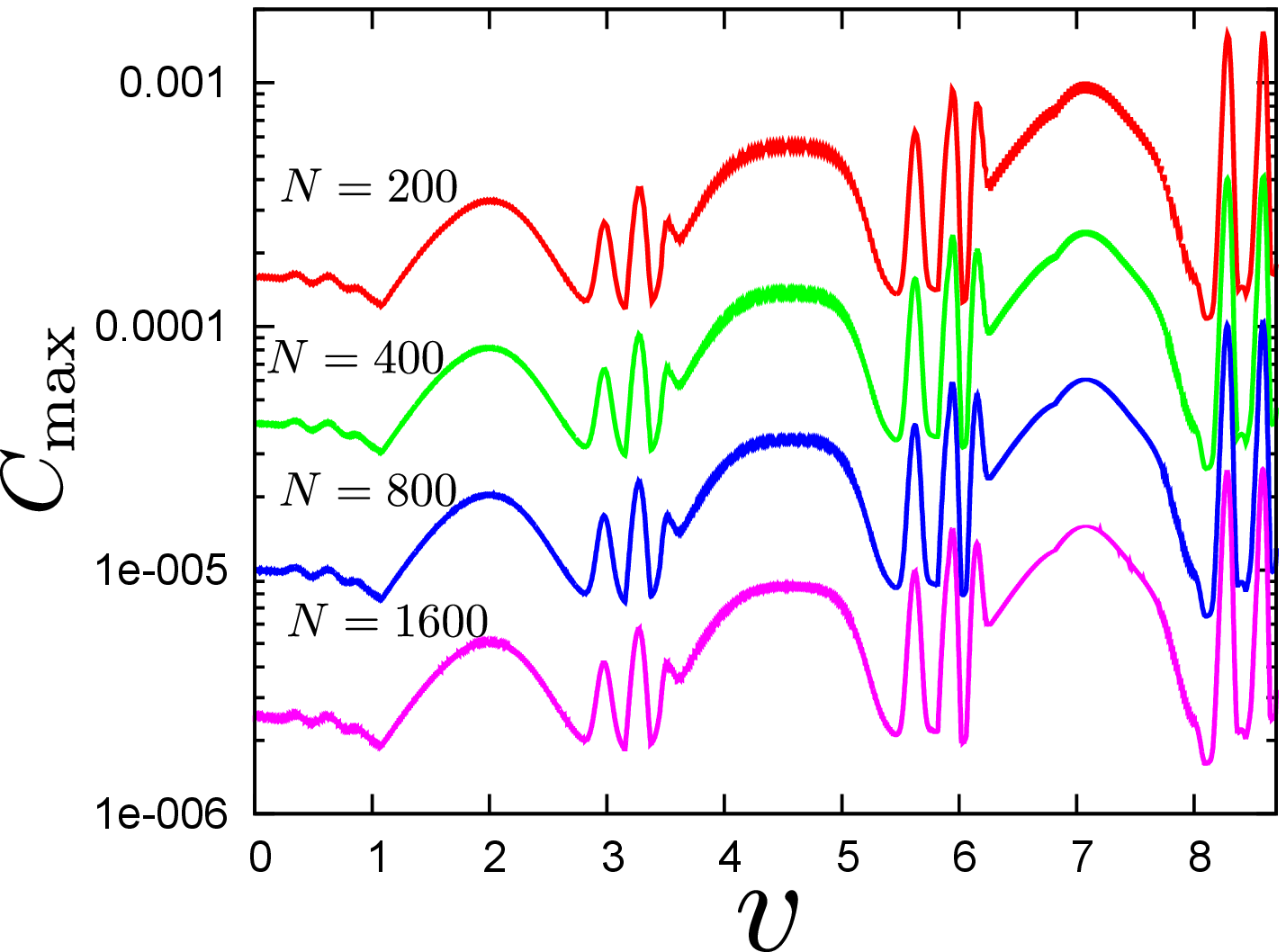} \label{cmax_TL}
  }
  \subfigure[Transverse circular]
  {\includegraphics[scale=0.32]{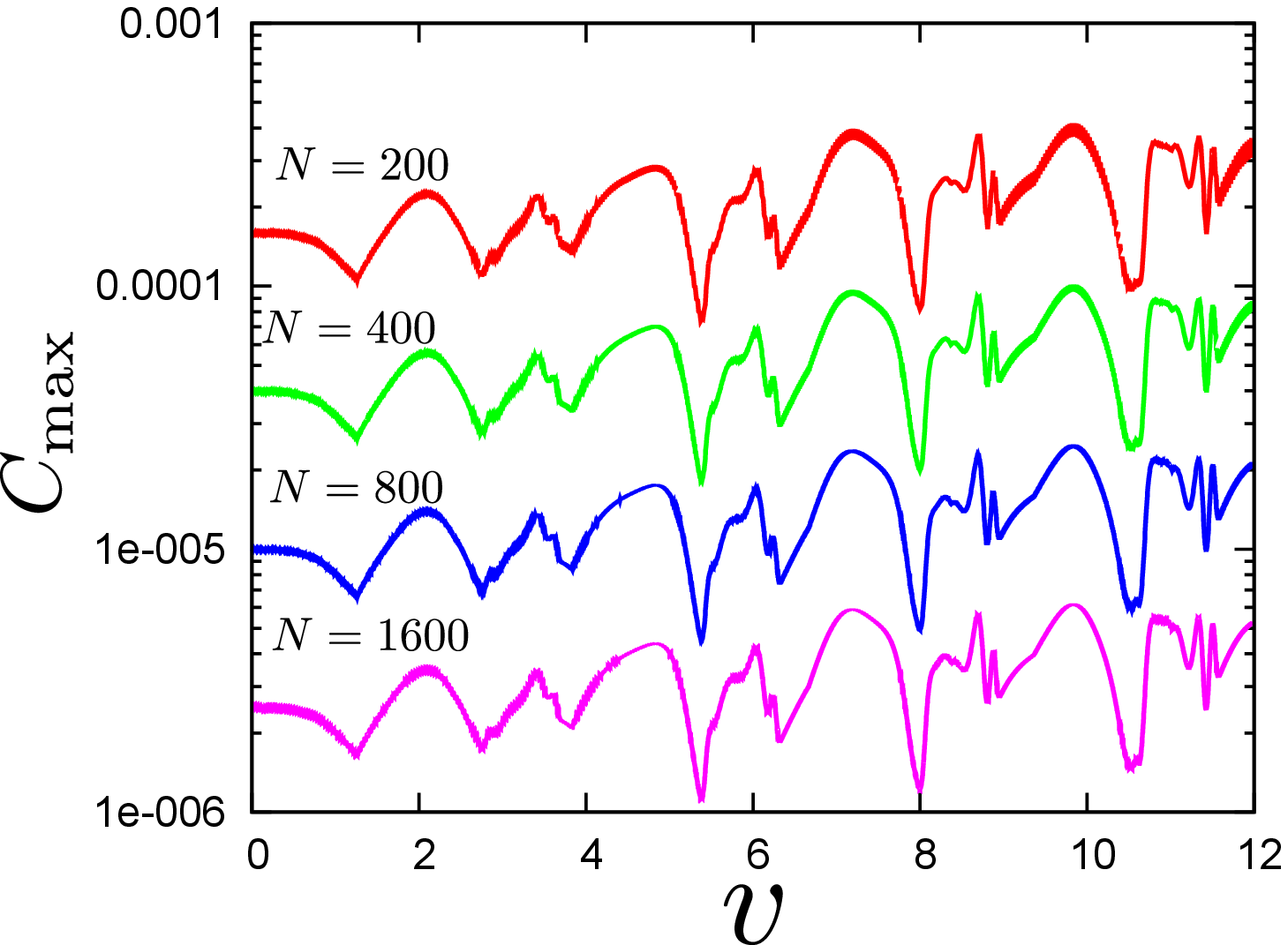} \label{cmax_TC}
  }
  \caption{
Constraint violation for several resolutions, $N=200, \, 400, \, 800, \, 1600$.
(a)~Longitudinal one-sided quench with $\Delta t/L=2$ and $\epsilon=0.01$.
(b)~Transverse linear quench with $\Delta t/L=2$ and $\epsilon=0.03$.
(c)~Transverse circular quench with $\Delta t/L=2$ and $\epsilon=0.02$.
For (a) and (b), 
the right ends of the figures correspond to the times for the cusp formation.
\label{Cmax}
}
\end{figure}

In this section, we estimate errors in our numerical calculations.
We define 
\begin{equation}
\tilde{C}_1=\frac{1}{L^2}(-T_{,u}^2+Z_{,u}^2+\bm{X}_{,u}^2)\ ,\qquad
\tilde{C}_2=\frac{1}{L^2}(-T_{,v}^2+Z_{,v}^2+\bm{X}_{,v}^2)\ .
\end{equation}
These constraints should be zero for exact solutions.
Hence, these can be nice indicators of our numerical errors.
For visibility of the constraint violation, we introduce 
\begin{equation}
 C_\textrm{max}(v)=\max_{\textrm{fixed } v}(|\tilde{C}_1|,|\tilde{C}_2|)\ ,
\end{equation}
where we take the maximum value when we vary $u$ on a fixed $v$ surface.
We also choose the bigger of the two constraints, $|\tilde{C}_1|$ and $|\tilde{C}_2|$.
Introducing an integer $N$ such that the mesh size is given by
$h=\beta_0/N$,
we plot $C_\textrm{max}(v)$ for several values of $N$ in Fig.~\ref{Cmax}. 
We see that the constraint violation is small 
($C_\textrm{max}\sim 10^{-3}$ even for $N=200$) and behaves as
$C_\textrm{max}\propto 1/N^2$. 
This is consistent with the fact that 
our numerical method has the second order accuracy.
In this paper, we mainly set $N=800$. Then, the constraint violation is $\mathcal{O}(10^{-4})$.

\section{Energy spectrum in the linear theory}
\label{sec:Enspeclin}

In this appendix, we derive the energy spectrum induced in the linear theory of section~\ref{sec:lin} when a quench is added on the boundary.
The equations of motion for the perturbation variables $(\chi_1, \, \chi_2, \, \chi_3)$ are given in Eq.~(\ref{chieq}).
Here, we focus on the longitudinal mode $\chi_1$ for simplicity
and denote it as $\chi_1=\chi$.
Application to transverse modes is straightforward.
For the quench, we consider the following boundary conditions for $\chi$:
\begin{equation}
 \chi(t,\phi=0)=\chi_b(t)\ ,\quad
 \chi(t,\phi=\beta_0)=0\ ,
\end{equation}
where $\chi_b(t)$ is the quench function assumed to have a compact support at
$0<t<\Delta t$. We also assume that the solution is trivial before the quench,
$\chi(t \le 0,\phi)=0$.

We firstly consider a time independent solution to (\ref{chieq}),
\begin{equation}
 \mathcal{H}S(\phi)=0\ .
\label{HS}
\end{equation}
A solution is given by 
\begin{equation}
 S(\phi)=-\frac{1}{A}\int^\phi_{\beta_0}\frac{d\phi'}{h(\phi')}\ ,
\end{equation}
where $A\equiv \int^{\beta_0}_0 d\phi'/h(\phi')$ is a constant for normalization.
Near the boundaries the function $S$ behaves as
\begin{equation}
S=1-\frac{1}{3A\Gamma_0^2}\phi^3 + \cdots \quad (\phi\sim
 0)\ , \quad
S=\frac{1}{3A\Gamma_0^2}(\beta_0-\phi)^3 + \cdots \quad (\phi\sim \beta_0) \ .
\end{equation}

Let us introduce the quench $\chi_b(t)$. 
Using the function $S$, we define $\tilde{\chi}$ as
\begin{equation}
 \chi(t,\phi)=\tilde{\chi}(t,\phi)+\chi_b(t)S(\phi)\ .
\end{equation}
The new variable $\tilde{\chi}$ satisfies trivial boundary conditions, 
$\tilde{\chi}(t,\phi=0)=\tilde{\chi}(t,\phi=\beta_0)=0$.
The equation of motion for $\tilde{\chi}$ becomes
\begin{equation}
 (\partial_t^2 + \mathcal{H})\tilde{\chi}=-\ddot{\chi}_b(t)S(\phi)\ ,
\label{tilchieq}
\end{equation}
where we used Eq.~(\ref{HS}).

To solve the equation, we consider a Green's equation:
\begin{equation}
 (\partial_t^2 +
  \mathcal{H})G(t,t';\phi,\phi')=\delta(t-t')\delta(\phi-\phi')\ ,
\end{equation}
where $G$ is the Green's function.
By using $G$, a special solution to
Eq.~(\ref{tilchieq}) can be written in the form
\begin{equation}
 \tilde{\chi}=-\int^\infty_{-\infty}dt' \int^{\beta_0}_0 d\phi'
  G(t,t';\phi,\phi')\ddot{\chi}_b(t')S(\phi')\ .
\label{sp_sol}
\end{equation}
Operating $\int^{t'+\epsilon}_{t'-\epsilon} dt$ on both sides of the Green's
equation and taking the limit of $\epsilon\to 0$, we obtain the junction
condition as
\begin{equation}
 \partial_t G|^{t=t'+0}_{t=t'-0}=\delta(\phi-\phi')\ .
\label{jnk}
\end{equation}
For $t<t'$, we assume the Green's function is trivial: $G=0$. 
For $t>t'$, the Green's function can be written as
\begin{equation}
 G=\sum_n[a_n \sin\omega_n(t-t')+b_n \cos\omega_n(t-t')]e_n(\phi)\ ,
\end{equation}
From the continuity of $G$ at $t=t'$ we have $b_n=0$, and then
from the junction condition~(\ref{jnk}), we find that $a_n$ satisfies
\begin{equation}
 \sum_n a_n \omega_n e_n(\phi)=\delta(\phi-\phi')\ .
\end{equation}
Operating $(e_n,*)$ to the above equation, we obtain
\begin{equation}
 a_n=\frac{1}{\omega_n}\gamma(\phi')e_n(\phi')\ .
\end{equation}
Thus, the Green's function can be written as
\begin{equation}
G(t,t';\phi,\phi')=
\begin{cases}
    0 &  (t<t')\ ,\\
    \sum_n
 \omega_n^{-1}\sin\omega_n(t-t')\, \gamma(\phi')e_n(\phi')e_n(\phi) &
 (t>t')\ .
\end{cases}
\end{equation}
Since the Green's function is zero at $t<t'$, 
the special solution obtained from Eq.~(\ref{sp_sol}) is also zero before
the quench, $t<0$, and this is nothing but the solution we are looking for.
After the quench $t>T$, the solution becomes
\begin{equation}
\begin{split}
 \chi&=-\sum_n \omega_n^{-1}S_n e_n(\phi) \int^T_0dt'
  \ddot{\chi}_0(t')\sin\omega_n(t-t')\\
&=\sum_n \omega_n S_n e_n(\phi) \int^T_0dt'
  \chi_b(t')\sin\omega_n(t-t')\ ,
\end{split}
\label{chilinthsol}
\end{equation}
where $S_n\equiv (S,e_n)$. Note that $\tilde{\chi}=\chi$ after the quench.
At the second equality, we integrated by parts twice. 

It is then straightforward to compute the energy spectrum.
The mode coefficient $c_n=(\chi,e_n)$ is computed by using \eqref{chilinthsol} as
\begin{equation}
 c_n(t)=\omega_n S_n \int^T_0dt'
  \chi_b(t')\sin\omega_n(t-t')\ ,
\end{equation}
and then from Eq.~(\ref{eps_n})
the energy spectrum for the longitudinal quench in the linear theory becomes
\begin{equation}
\varepsilon_n
=\frac{\sqrt{\lambda} z_0}{4\pi}
 \left[\dot{c}_n^2+\omega_n^2 c_n^2\right]
=\frac{\sqrt{\lambda} z_0 \omega_n^4 S_n^2}{4\pi}
|\hat{\chi}(\omega_n)|^2
\ .
\end{equation}
where we define $\hat{\chi}(\omega)=\int dt \chi_b(t)e^{-i\omega t}$.
The energy spectrum does not depend on $t$ as we expect.
Taking into account the transverse modes, we obtain the energy spectrum as
\begin{equation}
\varepsilon_n
=\frac{\sqrt{\lambda} z_0}{4\pi}\left[
\omega_n^4 S_n^2
|\hat{\chi}(\omega_n)|^2
+\sum_{i=2,3}\omega_n'{}^4 S_n'{}^2
|\hat{\chi}_i(\omega_n)|^2
\right]
\ ,
\label{lin_spec_full}
\end{equation}
where 
\begin{equation}
\hat{\chi}_i(\omega)=\int dt \chi_{b}^{i}(t,\phi=0)e^{-i\omega t}\ ,\quad
S'=\frac{g(\phi)+\Gamma_0}{2\Gamma_0}\ ,\quad
S_n'=(S',e_n)'\ ,
\end{equation}
and $\chi_b^i$ are the quench functions for $\chi_i$. The total energy is $\varepsilon = \sum_{n=1}^{\infty} \varepsilon_n$.

The spectrum~(\ref{lin_spec_full}) is defined only for
integer $n$. However, in Figs.~\ref{spec}, \ref{spec2} and \ref{circ}, 
we generalize $\varepsilon_n$ to a continuous number by
interpolating $\omega_n$, $S_n$, $\omega_n'$, and $S_n'$ and regarding
them as function of continuous number $n$ for visibility.

\section{Forces acting on the quark and the antiquark}
\label{app:force}

In this appendix, we derive the formula for the forces acting on the
quark endpoints \eqref{Force2}.
We denote the on-shell Nambu-Goto action as $S[\bm{x}_q,\bm{x}_{\bar{q}}]$, where 
$\bm{x}_q$ and $\bm{x}_{\bar{q}}$ are the locations of the string endpoints regarded as 
the quark and the antiquark, respectively, at the
AdS boundary: $\bm{X}(t,z\to 0) = \bm{x}_q, \bm{x}_{\bar{q}}$.
The on-shell action relates to the partition function of the boundary
theory as
\begin{equation}
 Z_\textrm{CFT}[\bm{x}_q,\bm{x}_{\bar{q}}]=e^{iS[\bm{x}_q,\bm{x}_{\bar{q}}]}\ .
\label{ZeS}
\end{equation}
In the field theory, the partition function is written as
\begin{equation}
 Z_\textrm{CFT}[\bm{x}_q,\bm{x}_{\bar{q}}]=\int \mathcal{D}\phi\,
  \exp
\left(iS_\textrm{SYM}[\phi]+iS_q[\phi(\bm{x}_q),\bm{x}_q]+iS_{\bar{q}}[\phi(\bm{x}_{\bar{q}}),\bm{x}_{\bar{q}}]\right)\ ,
\end{equation}
where $\phi$ represents the set of the fields in $\mathcal{N}=4$ super
Yang-Mills theory and $S_\textrm{SYM}$ is its action.
$S_q$ and $S_{\bar{q}}$ denote the actions for the quark and the antiquark, respectively.
We regard $\bm{x}_q(t)$ and $\bm{x}_{\bar{q}}(t)$ as external fields. 
The quark action is schematically written as
\begin{equation}
 S_q[\phi(\bm{x}_q),\bm{x}_q]=\int dt \left[
 -m\sqrt{1-\bm{v}^2}+L_\textrm{int}[\phi(\bm{x}_q),\bm{x}_q]
\right]\ ,
\end{equation}
where $m$ is the quark mass and $L_\textrm{int}$ corresponds to the interaction term
with SYM fields, and the velocity of the quark is introduced as
$\bm{v}\equiv d\bm{x}_q/dt$.
We define the force acting on the quark as
\begin{equation}
 \bm{F} =\frac{\delta}{\delta \bm{x}_q} \int dt
  L_\textrm{int}[\phi(\bm{x}_q),\bm{x}_q] \ .
\end{equation}
The variation of the on-shell Nambu-Goto action is then related to the field theory terms as
\begin{equation}
 \frac{\delta S}{\delta \bm{x}_q}=-i\frac{\delta}{\delta \bm{x}_q}\ln Z_\textrm{CFT}
=-m(\gamma \bm{v})^\cdot+\langle \bm{F}\rangle\ ,
\label{dSdx1}
\end{equation}
where 
$\langle \bm{F} \rangle=Z^{-1}\int\mathcal{D}\phi\bm{F}e^{iS_\textrm{SYM}+iS_q+iS_{\bar{q}}}$,
$\gamma=1/\sqrt{1-\bm{v}^2}$, and the dot in the upper right of the parentheses denotes ${}^\cdot\equiv d/dt$.
In the first equality, we used the 
AdS/CFT duality~(\ref{ZeS}).

Now, we evaluate $\delta S/\delta \bm{x}_q$ in the gravity side.
For this purpose, it is convenient to use target space coordinates
$(t,z)$ as the world sheet coordinates.
Then, the position of the string is specified by $\bm{x}=\bm{X}(t,z)$,
and the Nambu-Goto action becomes
\begin{equation}
 S=-\frac{\sqrt{\lambda}}{2\pi}\int dt dz \,
  \frac{1}{z^2}\big[(1-\dot{\bm{X}}^2)(1+\bm{X}'{}^2)+(\dot{\bm{X}}\cdot\bm{X}')^2\big]^{1/2}\ .
\end{equation}
The equations of motion are given by
\begin{equation}
 \left[
\frac{-(1+\bm{X}'{}^2)\dot{\bm{X}}+(\dot{\bm{X}}\cdot\bm{X}')\bm{X}'}{z^2\sqrt{\xi}}
\right]^\cdot
+ \left[
\frac{(1-\dot{\bm{X}}^2)\bm{X}{}'+(\dot{\bm{X}}\cdot\bm{X}')\dot{\bm{X}}}{z^2\sqrt{\xi}}
\right]'=0\ ,
\end{equation}
where ${}^\cdot\equiv \partial/\partial t$, ${}'\equiv \partial/\partial z$,
and $\xi\equiv(1-\dot{\bm{X}}^2)(1+\bm{X}'{}^2)+(\dot{\bm{X}}\cdot\bm{X}')^2$.
Solving these near the AdS boundary, we obtain
\begin{equation}
 \bm{X}=\bm{x}_q(t)-\frac{1}{2}\gamma^2 \bm{a}z^2+\bm{x}_3 z^3
  +\mathcal{O}(z^4)\ ,
\label{Xasymp}
\end{equation}
where $\bm{a}=d^2\bm{x}_q/dt^2$.
Let us consider the variation of the on-shell
action with respect to one of the endpoints of the string, $\bm{x}_q$, and
the other endpoint is fixed, $\delta \bm{x}_{\bar{q}}=0$.
By defining Lagrangian as $S=\int d\tau d\sigma \mathcal{L}$, 
the variation
of the action becomes
\begin{equation}
\begin{split}
\delta S[\bm{x}_q,\bm{x}_{\bar{q}}]
&=\int dt dz \left[
\delta(\partial_a \bm{X})\cdot \frac{\partial \mathcal{L}}{\partial(\partial_a \bm{X})}
+\delta \bm{X} \cdot \frac{\partial \mathcal{L}}{\partial \bm{X}}\right]\\
&=\int dt dz \,
\partial_a \left(\delta \bm{X} \cdot \frac{\partial
 \mathcal{L}}{\partial(\partial_a \bm{X})}\right) 
=-\int dt \, \left.
 \delta \bm{X}\cdot \frac{\partial \mathcal{L}}{\partial \bm{X}'}\right|_{\bm{x}=\bm{x}_q,z=\epsilon}\ ,
\end{split}
\label{delS}
\end{equation}
where we take the cutoff at $z=\epsilon$.
At the second equality, we used Euler-Lagrange equation, 
$\partial \mathcal{L}/\partial \bm{X}=\partial_a[\partial \mathcal{L}/\partial (\partial_a \bm{X})]$.
There is no contribution from the other boundary 
$\bm{X}=\bm{x}_{\bar{q}}$ and $z=\epsilon$ since $\delta \bm{X}=0$ there. 
Substituting Eq.~(\ref{Xasymp}) into Eq.~(\ref{delS}), we obtain
\begin{equation}
\delta S[\bm{x}_q,\bm{x}_{\bar{q}}]
=\int dt \,\delta \bm{x}_q\cdot \left(-\frac{\sqrt{\lambda}}{2\pi \epsilon}(\gamma \bm{v})^\cdot
+\frac{3\sqrt{\lambda}}{2\pi\gamma}(\bm{x}_3+\gamma^2(\bm{v}\cdot\bm{x}_3)\bm{v})\right)\ . 
\end{equation}
Hence, the upshot for $\delta S/\delta \bm{x}_q$ is
\begin{equation}
 \frac{\delta S}{\delta \bm{x}_q}=-\frac{\sqrt{\lambda}}{2\pi \epsilon}(\gamma \bm{v})^\cdot
+\frac{3\sqrt{\lambda}}{2\pi\gamma}(\bm{x}_3+\gamma^2(\bm{v}\cdot\bm{x}_3)\bm{v})\ .
\label{dSdx2}
\end{equation}
Comparing above expression with Eq.~(\ref{dSdx1}),
we can see that the first term in (\ref{dSdx2}) corresponds to a diverging quark mass $m\sim 1/\epsilon$.
This is a natural consequence since we are considering an infinitely extended string.
Setting $m=\sqrt{\lambda}/(2\pi \epsilon)$, we obtain 
the force acting on the quark from Eqs.~(\ref{dSdx1}) and
(\ref{dSdx2}) as
\begin{equation}
 \langle F_i(t) \rangle = 
\frac{\sqrt{\lambda}}{4\pi\gamma}(\delta_{ij}+\gamma^2v_iv_j)\partial_z^3
X_j|_{z=0}\ ,
\label{Fexp}
\end{equation}
where we replaced $\bm{x}_3$ with $\partial_z^3 \bm{X}|_{z=0}/6$.
Note that there can be a finite difference between $m$ and 
$\sqrt{\lambda}/(2\pi \epsilon)$, and
an extra-term proportional to $(\gamma \bm{v})^{\cdot}$ may appear in Eq.~(\ref{Fexp}).
However, it can be eliminated by adding a local counter term 
proportional to $\sqrt{1-\bm{v}^2}$
in the quark action $S_p$.
The same formula can be applied to the force acting on the antiquark at $\bm{x}_{\bar{q}}$.

\bibliography{bunkenF1}

\providecommand{\href}[2]{#2}\begingroup\raggedright\begin{thebibliography}{10}

\bibitem{Maldacena:1997re}
J.~M. Maldacena, {\it {The Large N limit of superconformal field theories and
  supergravity}},  {\em Int.J.Theor.Phys.} {\bf 38} (1999) 1113--1133,
  [\href{http://arxiv.org/abs/hep-th/9711200}{{\tt hep-th/9711200}}].

\bibitem{Gubser:1998bc}
S.~Gubser, I.~R. Klebanov, and A.~M. Polyakov, {\it {Gauge theory correlators
  from noncritical string theory}},  {\em Phys.Lett.} {\bf B428} (1998)
  105--114, [\href{http://arxiv.org/abs/hep-th/9802109}{{\tt hep-th/9802109}}].

\bibitem{Witten:1998qj}
E.~Witten, {\it {Anti-de Sitter space and holography}},  {\em
  Adv.Theor.Math.Phys.} {\bf 2} (1998) 253--291,
  [\href{http://arxiv.org/abs/hep-th/9802150}{{\tt hep-th/9802150}}].

\bibitem{Chesler:2008hg}
P.~M. Chesler and L.~G. Yaffe, {\it {Horizon formation and far-from-equilibrium
  isotropization in supersymmetric Yang-Mills plasma}},  {\em Phys.Rev.Lett.}
  {\bf 102} (2009) 211601, [\href{http://arxiv.org/abs/0812.2053}{{\tt
  arXiv:0812.2053}}].

\bibitem{Chesler:2009cy}
P.~M. Chesler and L.~G. Yaffe, {\it {Boost invariant flow, black hole
  formation, and far-from-equilibrium dynamics in N = 4 supersymmetric
  Yang-Mills theory}},  {\em Phys.Rev.} {\bf D82} (2010) 026006,
  [\href{http://arxiv.org/abs/0906.4426}{{\tt arXiv:0906.4426}}].

\bibitem{Chesler:2010bi}
P.~M. Chesler and L.~G. Yaffe, {\it {Holography and colliding gravitational
  shock waves in asymptotically AdS$_5$ spacetime}},  {\em Phys.Rev.Lett.} {\bf
  106} (2011) 021601, [\href{http://arxiv.org/abs/1011.3562}{{\tt
  arXiv:1011.3562}}].

\bibitem{Chesler:2013lia}
P.~M. Chesler and L.~G. Yaffe, {\it {Numerical solution of gravitational
  dynamics in asymptotically anti-de Sitter spacetimes}},  {\em JHEP} {\bf
  1407} (2014) 086, [\href{http://arxiv.org/abs/1309.1439}{{\tt
  arXiv:1309.1439}}].

\bibitem{Chesler:2015wra}
P.~M. Chesler and L.~G. Yaffe, {\it {Holography and off-center collisions of
  localized shock waves}},  \href{http://arxiv.org/abs/1501.04644}{{\tt
  arXiv:1501.04644}}.

\bibitem{Ishii:2014paa}
T.~Ishii, S.~Kinoshita, K.~Murata, and N.~Tanahashi, {\it {Dynamical Meson
  Melting in Holography}},  {\em JHEP} {\bf 1404} (2014) 099,
  [\href{http://arxiv.org/abs/1401.5106}{{\tt arXiv:1401.5106}}].

\bibitem{Hashimoto:2014yza}
K.~Hashimoto, S.~Kinoshita, K.~Murata, and T.~Oka, {\it {Electric Field Quench
  in AdS/CFT}},  {\em JHEP} {\bf 1409} (2014) 126,
  [\href{http://arxiv.org/abs/1407.0798}{{\tt arXiv:1407.0798}}].

\bibitem{Hashimoto:2014xta}
K.~Hashimoto, S.~Kinoshita, K.~Murata, and T.~Oka, {\it {Turbulent meson
  condensation in quark deconfinement}},
  \href{http://arxiv.org/abs/1408.6293}{{\tt arXiv:1408.6293}}.

\bibitem{Hashimoto:2014dda}
K.~Hashimoto, S.~Kinoshita, K.~Murata, and T.~Oka, {\it {Meson turbulence at
  quark deconfinement from AdS/CFT}},
  \href{http://arxiv.org/abs/1412.4964}{{\tt arXiv:1412.4964}}.

\bibitem{Ali-Akbari:2015bha}
M.~Ali-Akbari, F.~Charmchi, A.~Davody, H.~Ebrahim, and L.~Shahkarami, {\it
  {Time-Dependent Meson Melting in External Magnetic Field}},
  \href{http://arxiv.org/abs/1503.04439}{{\tt arXiv:1503.04439}}.

\bibitem{Bizon:2011gg}
P.~Bizon and A.~Rostworowski, {\it {On weakly turbulent instability of anti-de
  Sitter space}},  {\em Phys.Rev.Lett.} {\bf 107} (2011) 031102,
  [\href{http://arxiv.org/abs/1104.3702}{{\tt arXiv:1104.3702}}].

\bibitem{Rey:1998ik}
S.-J. Rey and J.-T. Yee, {\it {Macroscopic strings as heavy quarks in large N
  gauge theory and anti-de Sitter supergravity}},  {\em Eur.Phys.J.} {\bf C22}
  (2001) 379--394, [\href{http://arxiv.org/abs/hep-th/9803001}{{\tt
  hep-th/9803001}}].

\bibitem{Maldacena:1998im}
J.~M. Maldacena, {\it {Wilson loops in large N field theories}},  {\em
  Phys.Rev.Lett.} {\bf 80} (1998) 4859--4862,
  [\href{http://arxiv.org/abs/hep-th/9803002}{{\tt hep-th/9803002}}].

\bibitem{Witten:1998zw}
E.~Witten, {\it {Anti-de Sitter space, thermal phase transition, and
  confinement in gauge theories}},  {\em Adv.Theor.Math.Phys.} {\bf 2} (1998)
  505--532, [\href{http://arxiv.org/abs/hep-th/9803131}{{\tt hep-th/9803131}}].

\bibitem{Rey:1998bq}
S.-J. Rey, S.~Theisen, and J.-T. Yee, {\it {Wilson-Polyakov loop at finite
  temperature in large N gauge theory and anti-de Sitter supergravity}},  {\em
  Nucl.Phys.} {\bf B527} (1998) 171--186,
  [\href{http://arxiv.org/abs/hep-th/9803135}{{\tt hep-th/9803135}}].

\bibitem{Brandhuber:1998bs}
A.~Brandhuber, N.~Itzhaki, J.~Sonnenschein, and S.~Yankielowicz, {\it {Wilson
  loops in the large N limit at finite temperature}},  {\em Phys.Lett.} {\bf
  B434} (1998) 36--40, [\href{http://arxiv.org/abs/hep-th/9803137}{{\tt
  hep-th/9803137}}].

\bibitem{Herzog:2006gh}
C.~Herzog, A.~Karch, P.~Kovtun, C.~Kozcaz, and L.~Yaffe, {\it {Energy loss of a
  heavy quark moving through N=4 supersymmetric Yang-Mills plasma}},  {\em
  JHEP} {\bf 0607} (2006) 013, [\href{http://arxiv.org/abs/hep-th/0605158}{{\tt
  hep-th/0605158}}].

\bibitem{Gubser:2006bz}
S.~S. Gubser, {\it {Drag force in AdS/CFT}},  {\em Phys.Rev.} {\bf D74} (2006)
  126005, [\href{http://arxiv.org/abs/hep-th/0605182}{{\tt hep-th/0605182}}].

\bibitem{CasalderreySolana:2006rq}
J.~Casalderrey-Solana and D.~Teaney, {\it {Heavy quark diffusion in strongly
  coupled N=4 Yang-Mills}},  {\em Phys.Rev.} {\bf D74} (2006) 085012,
  [\href{http://arxiv.org/abs/hep-ph/0605199}{{\tt hep-ph/0605199}}].

\bibitem{Gubser:2006nz}
S.~S. Gubser, {\it {Momentum fluctuations of heavy quarks in the gauge-string
  duality}},  {\em Nucl.Phys.} {\bf B790} (2008) 175--199,
  [\href{http://arxiv.org/abs/hep-th/0612143}{{\tt hep-th/0612143}}].

\bibitem{CasalderreySolana:2007qw}
J.~Casalderrey-Solana and D.~Teaney, {\it {Transverse Momentum Broadening of a
  Fast Quark in a N=4 Yang Mills Plasma}},  {\em JHEP} {\bf 0704} (2007) 039,
  [\href{http://arxiv.org/abs/hep-th/0701123}{{\tt hep-th/0701123}}].

\bibitem{Liu:2006ug}
H.~Liu, K.~Rajagopal, and U.~A. Wiedemann, {\it {Calculating the jet quenching
  parameter from AdS/CFT}},  {\em Phys.Rev.Lett.} {\bf 97} (2006) 182301,
  [\href{http://arxiv.org/abs/hep-ph/0605178}{{\tt hep-ph/0605178}}].

\bibitem{Liu:2006he}
H.~Liu, K.~Rajagopal, and U.~A. Wiedemann, {\it {Wilson loops in heavy ion
  collisions and their calculation in AdS/CFT}},  {\em JHEP} {\bf 0703} (2007)
  066, [\href{http://arxiv.org/abs/hep-ph/0612168}{{\tt hep-ph/0612168}}].

\bibitem{Liu:2006nn}
H.~Liu, K.~Rajagopal, and U.~A. Wiedemann, {\it {An AdS/CFT Calculation of
  Screening in a Hot Wind}},  {\em Phys.Rev.Lett.} {\bf 98} (2007) 182301,
  [\href{http://arxiv.org/abs/hep-ph/0607062}{{\tt hep-ph/0607062}}].

\bibitem{Chernicoff:2006hi}
M.~Chernicoff, J.~A. Garcia, and A.~Guijosa, {\it {The Energy of a Moving
  Quark-Antiquark Pair in an N=4 SYM Plasma}},  {\em JHEP} {\bf 0609} (2006)
  068, [\href{http://arxiv.org/abs/hep-th/0607089}{{\tt hep-th/0607089}}].

\bibitem{Balasubramanian:2010ce}
V.~Balasubramanian, A.~Bernamonti, J.~de~Boer, N.~Copland, B.~Craps, et~al.,
  {\it {Thermalization of Strongly Coupled Field Theories}},  {\em
  Phys.Rev.Lett.} {\bf 106} (2011) 191601,
  [\href{http://arxiv.org/abs/1012.4753}{{\tt arXiv:1012.4753}}].

\bibitem{Balasubramanian:2011ur}
V.~Balasubramanian, A.~Bernamonti, J.~de~Boer, N.~Copland, B.~Craps, et~al.,
  {\it {Holographic Thermalization}},  {\em Phys.Rev.} {\bf D84} (2011) 026010,
  [\href{http://arxiv.org/abs/1103.2683}{{\tt arXiv:1103.2683}}].

\bibitem{Mikhailov:2003er}
A.~Mikhailov, {\it {Nonlinear waves in AdS / CFT correspondence}},
  \href{http://arxiv.org/abs/hep-th/0305196}{{\tt hep-th/0305196}}.

\bibitem{Turok:1984cn}
N.~Turok, {\it {Grand Unified Strings and Galaxy Formation}},  {\em Nucl.Phys.}
  {\bf B242} (1984) 520.

\bibitem{Davis:2008kg}
A.-C. Davis, W.~Nelson, S.~Rajamanoharan, and M.~Sakellariadou, {\it {Cusps on
  cosmic superstrings with junctions}},  {\em JCAP} {\bf 0811} (2008) 022,
  [\href{http://arxiv.org/abs/0809.2263}{{\tt arXiv:0809.2263}}].

\bibitem{Garcia:2012gw}
J.~A. Garcia, A.~Guijosa, and E.~J. Pulido, {\it {No Line on the Horizon: On
  Uniform Acceleration and Gluonic Fields at Strong Coupling}},  {\em JHEP}
  {\bf 1301} (2013) 096, [\href{http://arxiv.org/abs/1210.4175}{{\tt
  arXiv:1210.4175}}].

\bibitem{Callan:1999ki}
J.~Callan, Curtis~G. and A.~Guijosa, {\it {Undulating strings and gauge theory
  waves}},  {\em Nucl.Phys.} {\bf B565} (2000) 157--175,
  [\href{http://arxiv.org/abs/hep-th/9906153}{{\tt hep-th/9906153}}].

\bibitem{Klebanov:2006jj}
I.~R. Klebanov, J.~M. Maldacena, and I.~Thorn, Charles~B., {\it {Dynamics of
  flux tubes in large N gauge theories}},  {\em JHEP} {\bf 0604} (2006) 024,
  [\href{http://arxiv.org/abs/hep-th/0602255}{{\tt hep-th/0602255}}].

\bibitem{Avramis:2006nv}
S.~D. Avramis, K.~Sfetsos, and K.~Siampos, {\it {Stability of strings dual to
  flux tubes between static quarks in N = 4 SYM}},  {\em Nucl.Phys.} {\bf B769}
  (2007) 44--78, [\href{http://arxiv.org/abs/hep-th/0612139}{{\tt
  hep-th/0612139}}].

\bibitem{Arias:2009me}
R.~E. Arias and G.~A. Silva, {\it {Wilson loops stability in the gauge/string
  correspondence}},  {\em JHEP} {\bf 1001} (2010) 023,
  [\href{http://arxiv.org/abs/0911.0662}{{\tt arXiv:0911.0662}}].

\bibitem{work_in_progress}
T.~Ishii and K.~Murata, {\it {work in progress}}, .

\bibitem{deOliveira:2012dt}
H.~de~Oliveira, L.~A. Pando~Zayas, and E.~Rodrigues, {\it {A
  Kolmogorov-Zakharov Spectrum in AdS Gravitational Collapse}},  {\em
  Phys.Rev.Lett.} {\bf 111} (2013), no.~5 051101,
  [\href{http://arxiv.org/abs/1209.2369}{{\tt arXiv:1209.2369}}].

\bibitem{Quashnock:1990wv}
J.~M. Quashnock and D.~N. Spergel, {\it {Gravitational Selfinteractions of
  Cosmic Strings}},  {\em Phys.Rev.} {\bf D42} (1990) 2505--2520.

\bibitem{Damour:2001bk}
T.~Damour and A.~Vilenkin, {\it {Gravitational wave bursts from cusps and kinks
  on cosmic strings}},  {\em Phys.Rev.} {\bf D64} (2001) 064008,
  [\href{http://arxiv.org/abs/gr-qc/0104026}{{\tt gr-qc/0104026}}].

\bibitem{O'Callaghan:2010ww}
E.~O'Callaghan, S.~Chadburn, G.~Geshnizjani, R.~Gregory, and I.~Zavala, {\it
  {Effect of Extra Dimensions on Gravitational Waves from Cosmic Strings}},
  {\em Phys.Rev.Lett.} {\bf 105} (2010) 081602,
  [\href{http://arxiv.org/abs/1003.4395}{{\tt arXiv:1003.4395}}].

\bibitem{O'Callaghan:2010sy}
E.~O'Callaghan, S.~Chadburn, G.~Geshnizjani, R.~Gregory, and I.~Zavala, {\it
  {The effect of extra dimensions on gravity wave bursts from cosmic string
  cusps}},  {\em JCAP} {\bf 1009} (2010) 013,
  [\href{http://arxiv.org/abs/1005.3220}{{\tt arXiv:1005.3220}}].

\bibitem{Kruczenski:2004wg}
M.~Kruczenski, {\it {Spiky strings and single trace operators in gauge
  theories}},  {\em JHEP} {\bf 0508} (2005) 014,
  [\href{http://arxiv.org/abs/hep-th/0410226}{{\tt hep-th/0410226}}].

\bibitem{Zayas:2010fs}
L.~A. Pando~Zayas and C.~A. Terrero-Escalante, {\it {Chaos in the Gauge /
  Gravity Correspondence}},  {\em JHEP} {\bf 1009} (2010) 094,
  [\href{http://arxiv.org/abs/1007.0277}{{\tt arXiv:1007.0277}}].

\end{thebibliography}\endgroup

\end{document}